\documentclass[12pt,reqno]{gtpart}
\usepackage{times,euscript}
\usepackage[utf8]{inputenc}
\usepackage{hyperref}
 
\usepackage{amsmath}
\usepackage{amsthm}
\usepackage{amsfonts}
\usepackage{amssymb}
\usepackage{epic}
\usepackage{eepic,epsfig}
\usepackage{times} 
\usepackage{graphics}
\usepackage{epsfig,multicol,bbm}

\usepackage{xcolor}
\usepackage{graphicx}

\theoremstyle{plain}

\theoremstyle{remark}
\newtheorem{rem}{Remark}

\newcommand{\sst}{\scriptscriptstyle}

\renewcommand{\theequation}{\thesection.\arabic{equation}} 

\newcommand{\ra}{\to}

\newcommand{\fsl}{{\mathfrak s}{\mathfrak l}}

\def\beq{\begin{equation}}
\def\ee{\end{equation}}
\def\bea{\begin{eqnarray}}
\def\eea{\end{eqnarray}}

\def\cW{\mathcal{W}}

\def\cG{\mathcal{G}}

\newcommand{\al}{\alpha}
\newcommand{\1}{\one}
\newcommand{\2}{\two}
\newcommand{\3}{\three}

\newcommand{\be}{\beta}
\newcommand{\ga}{\gamma}
\newcommand{\Ga}{\Gamma}
\newcommand{\de}{\delta}

\newcommand{\ep}{\epsilon}
\newcommand{\la}{\lambda}

\newcommand{\si}{\sigma}

\newcommand{\vf}{\varphi}

\newcommand{\pa}{\partial}

\newcommand{\CB}{{\mathcal B}}
\newcommand{\CC}{{\mathcal C}}
\newcommand{\CD}{{\mathcal D}}
\newcommand{\CE}{{\mathcal E}}
\newcommand{\CF}{{\mathcal F}}
\newcommand{\CG}{{\mathcal G}}
\newcommand{\CH}{{\mathcal H}}
\newcommand{\CI}{{\mathcal I}}
\newcommand{\CL}{{\mathcal L}}
\newcommand{\CM}{{\mathcal M}}
\newcommand{\CN}{{\mathcal N}}
\newcommand{\CO}{{\mathcal O}}  

\newcommand{\CR}{{\mathcal R}}
\newcommand{\CS}{{\mathcal S}}
\newcommand{\CT}{{\mathcal T}}

\newcommand{\CV}{{\mathcal V}}

\newcommand{\CZ}{{\mathcal Z}}

\newcommand{\SA}{{\mathsf A}}
\newcommand{\SB}{{\mathsf B}}

\newcommand{\SH}{{\mathsf H}}

\renewcommand{\sf}{{\mathsf f}}
\renewcommand{\sh}{{\mathsf h}}

\newcommand{\mst}{{\mathsf t}}

\newcommand{\sw}{{\mathsf w}}

\newcommand{\one}{{\mathfrak 1}}
\newcommand{\two}{{\mathfrak 2}}
\newcommand{\three}{{\mathfrak 3}}

\newcommand{\BR}{{\mathbb R}}

\newcommand{\BC}{{\mathbb C}}

\newcommand{\BZ}{{\mathbb Z}}

\newcommand{\rf}[1]{(\ref{#1})}

\begin{document}\thispagestyle{empty}
\title{From quantum curves\\ to\\ topological string partition functions}
\author{Ioana Coman, Elli Pomoni, J\"org Teschner}
\address{
{\sc IC, EP, JT:} DESY theory, 
Notkestrasse 85,
20607 Hamburg,
Germany; 
{\sc JT:} Department of Mathematics, 
University of Hamburg, 
Bundesstrasse 55,
20146 Hamburg, Germany,
E-mail: joerg.teschner@desy.de;
{\sc IC:} Institute of Physics, 
University of Amsterdam,
1098 XH Amsterdam, the Netherlands.}
\maketitle

\begin{quote}
\begin{center}{\bf Abstract}\end{center}
{\small
This paper describes the reconstruction of the topological string partition function for 
certain local Calabi-Yau (CY) manifolds from 
the quantum curve, an ordinary  differential equation obtained by quantising their defining equations.  
Quantum curves are characterised as solutions to  a Riemann-Hilbert problem.
The isomonodromic tau-functions associated to these Riemann-Hilbert 
problems admit a family of natural normalisations labelled by 
the chambers in the extended K\"ahler 
moduli space of the local CY under consideration.
The corresponding isomonodromic tau-functions admit 
a series expansion of generalised theta series type from which one can extract 
the topological string partition functions for each chamber.}
\end{quote}

\begin{quote} 
\hspace{-0.83cm}{\fontfamily{cmss}\selectfont \small DESY-18-198 }
\end{quote}

\setcounter{tocdepth}{2}
\tableofcontents
\section{Introduction}

Topological string theory on Calabi-Yau (CY) manifolds is a subject which has attracted considerable interest both from theoretical physics 
and from mathematics.  From the point of view of physics, it can provide non-perturbative information on various 
string compactifications with possible applications to supersymmetric field theories
and  black hole physics. 
The subject is mathematically related to various curve counting invariants and to the phenomenon of mirror symmetry. 
A very fruitful interplay between mathematics and physics on this subject has emerged, with duality conjectures motivated 
by arguments from theoretical physics suggesting profound and unexpected relations between different parts of mathematics, and mathematical research providing the groundwork for making the ideas from physics sufficiently 
precise for extracting the relevant predictions, and understanding the theoretical foundations.

A key object in topological string theory is the topological string partition function, mathematically defined as the 
generating function of the Gromov-Witten invariants.  
String-theoretic duality conjectures suggest that this partition function
is related to the generating functions of the enumerative invariants associated with the names 
Donaldson-Thomas and Gopakumar-Vafa, respectively. These interpretations do not easily lead to a 
conceptual characterisation of the topological string partition functions as mathematical objects of their own right,
as the relevant generating functions are without further input only defined in the sense of formal series.
Various alternative characterisations have been proposed, including matrix models, topological recursion, 
Chern-Simons theory and the quantisation of the moduli spaces of geometric structures of the relevant families of 
CY  manifolds.
 
These approaches all have their virtues and drawbacks, as usual, and it seems to us that there is still room 
for an improvement of our understanding of the 
topological string partition functions as mathematical objects of their own right. Our paper is an attempt to 
improve our understanding of the topological string partition functions for a certain class of local CY manifolds.
The manifolds $Y$ of our interest can be locally described by equations of the form 
\begin{equation}
uv-R(x,y)=0,
\end{equation}
where $x$ and $y$ are local coordinates for the cotangent bundle $T^*C$ of a given Riemann surface 
$C$ such that the equation $R(x,y)=0$ defines a covering of $C$. This class of 
local CY manifolds 
will be referred to as class $\Sigma$. The local CY in this class are relevant  for the description 
of the $\CN=2$, $d=4$-supersymmetric field theories of class $\CS$ \cite{Ga,GMN09}
within string theory by geometric engineering \cite{KKV,KMV}, see \cite{DDDHP} or \cite[Section 3]{Sm}  
for more details
on the geometry of local CY of class $\Sigma$.
Theories of class $\CS$ are labelled by the data $(C,\mathfrak{g})$, with $C$ being a possibly
punctured Riemann surface, and $\mathfrak{g}$ a Lie algebra of ADE-type. Our goal is to give a non-perturbative
definition of the topological string partition functions for local CY of class $\Sigma$. A subset 
of the local CY of class $\Sigma$ can be represented by certain limits of toric
CY, but such a description does not seem to be known for all CY of class $\Sigma$.

As main example we will consider the case $C=C_{0,4}$, the Riemann sphere with four punctures,
and $R(x,y)=y^2-q(x)$, $q(x)(dx)^2$ being a quadratic differential on $C$ with regular singularities at the 
punctures. It corresponds to an $A_1$-theory of class $\CS$ often referred to as the $SU(2)$, $N_f=4$ theory.
The generalisation to the 
cases $C=C_{0,n}$ is absolutely straightforward, and the cases where $C$ has higher genus or 
$q$ has irregular singularities are certainly within reach.   
We believe that the resulting picture has a high potential for further generalisations. 
Covers of higher order corresponding to $A_n$-theories of class $\CS$, 
for example, can be an interesting next step.

The approach taken here is inspired by the previous work described in
\cite{N,OP,LMN,NO,ADKMV}, indicating a deep interplay between 
topological string partition functions, free fermions on algebraic curves, 
and the theory of classically integrable hierarchies. 
Our approach can be seen in particular as a concrete realisation of some  ideas 
discussed in \cite{ADKMV} suggesting that a 
non-commutative deformation
of the  curve $\Sigma$, often referred to as ``quantum curve'', can be used to 
characterise the topological string partition functions. 
It seems to us, however, that these ideas have not been realised concretely 
for the local CY of class $\Sigma$ yet.
We will here offer a precise definition of the 
quantum curves for the cases of our interest, and explain how the quantum curve can be used
to define the topological string partition functions. 

Another source of inspiration
for us were  the 
works \cite{DHSV,DHS} where it has been argued on the basis of string dualities that
there exists a dual description for the topological string in terms of a system of D4 and D6 branes intersecting
along the surface $\Sigma$. 
It can can be argued that the topological string partition functions get represented by the 
partition functions $Z_{\rm ff}$ of the massless chiral open strings stretching between D4 and D6 branes, 
defining a system of free fermions on the intersection $\Sigma$.
Having a nonzero value of the topological 
string coupling $\la$ corresponds to turning on a B-field on the D6-branes. The effect 
of the B-field can be described in terms of  
a non-commutative deformation of $\Sigma$.
In \cite{DHS} it has been proposed  that in the case of local CY of class $\Sigma$
it is possible  to describe the relevant deformation of $\Sigma$
by a differential equation, 
or equivalently a $\CD$-module, on the underlying base curve $C$.
A generalisation of the Krichever correspondence \cite{Kr77a,Kr77b} is proposed in 
\cite{DHS} leading to a construction of the relevant free fermion partitions $Z_{\rm ff}(\xi,t;\la)$
as Fredholm determinants of certain operators build from the 
solutions of the differential equation defining the quantum curve, where $t$ denotes a collection of
parameters for the complex structures of $\Sigma$, while $\xi$ is a tuple of chemical potentials 
for the free fermion charges. 
This line of thought leads to the prediction that the 
topological string partition function $Z_{\rm top}(t,\la)$ is related to
$Z_{\rm ff}(\xi,t;\la)$  by an expansion of the form
\begin{align}\label{thetaseries0}
Z_{\rm ff}(\xi,t;\la)
=\sum_{p\in H^2(Y,\BZ)}e^{p\xi}Z_{\rm top}(t+p\la,\la).
\end{align}
We will in the following refer to series of the form \rf{thetaseries0} as generalised theta 
series.\footnote{This terminology can be motivated in two ways. Weighted sums over functions with arguments
shifted by lattice translations are sometimes called theta series in the mathematical 
literature. In \cite{CLT} it is shown, on the other hand, that ordinary theta functions can be recovered from 
$Z_{\rm ff}(\xi,t;\la)$ in the limit $\la\ra 0$. We may therefore regard the partition functions
$Z_{\rm ff}(\xi,t;\la)$
as deformations of ordinary theta functions.}
This would lead to an elegant mathematical characterisation of the topological string partition function
whenever one knows how to define the partition functions of free fermionic field theories on the relevant non-commutative 
surfaces, and how exactly to extract the topological string partition functions from these objects.
The program suggested in \cite{DHSV,DHS} has been realised in some basic examples. Our goal here
is to realise it in a case that is sufficiently rich to indicate what needs to be done to generalise this
approach to  much wider classes of cases.

We will observe two main issues that need to be addressed. 
It will, on the one hand, be crucial in our approach to allow certain 
quantum corrections to the equation of the quantum curve represented by 
terms of higher order in $\la$. The quantum corrections turn out to be
determined  by the integrable structures of the problem. We will furthermore 
observe that the issue of normalisation of the solutions plays a crucial role: Different normalisations
for the solutions yield different partition functions. It turns out that there exist  
distinguished choices of normalisation which are mathematically very natural, and 
lead to the definition of  functions which coincide with the results of topological vertex calculations.
The impatient reader may jump to Section \ref{sec:summ} 
for a slightly more precise summary of 
our results.

In the context of Donaldson-Thomas theory for toric CY there is an interesting approach to the emergence
of the quantum curve \cite{O09}, revealing the origin of the integrable structures of the topological 
string \cite{OR}. Our goals are different. We use the quantum curve as a key ingredient
in a precise description of the topological string partition functions as  {\it analytic objects}.
The results can be described as products of certain
Fredholm determinants with explicit meromorphic functions. 
Other approaches to the reconstruction of the topological string 
partition functions from the quantum curve have been proposed in \cite{ACDKV,GS,GHM,MS}\footnote{The approach of 
\cite{GHM} considers Fredholm determinants constructed from the quantum curves of toric CY. However, 
the relation to the Fredholm determinants appearing in our paper is not clear to us.}.

The precise relation between free fermion partition functions and topological string partition
functions established in this paper can be seen as a prediction of the duality conjectures
used in  \cite{DHSV,DHS}. From a mathematical point of view one may find this relation quite
non-obvious.
One may, in particular,  regard our results as a rather non-trivial quantitative check of the 
string duality conjectures 
predicting such relations.
We'd ultimately hope that learning to define the topological string partition function non-perturbatively 
may provide the groundwork for a
mathematical understanding of various string dualities.

\subsection{Overview}\label{overview}

Our goal is to define and calculate the topological string partition functions for 
the families $Y_{z,u}$ of local CY,
$
vw-R(x,y)=0,
$
where $\Sigma=\Sigma_{z,u}$ is the double cover of a Riemann surface $C$
defined by the equation $R(x,y)=0$, where
$R(x,y)=q(x)-y^2$,
$q(x)(dx)^2$ being a quadratic differential on $C$.
This will be fully worked out in the case $C=C_{0,4}$, which is prototypical enough 
to serve as a guideline for the case of general $C$. The solution will be described in the following steps.
Section 2  summarises  the relevant features of the geometry of the 
family $Y_{z,u}$ of local CY, and of their mirror manifolds $X_{z,u}$
which can be described as certain limits of a family of toric CY. 
We then introduce the differential equations defining the quantum curves
in Section 3. The following Section 4 associates a free fermion partition function to these differential equations.
We demonstrate that the free fermion partition function $\CZ_{\rm ff}$ is proportional to the isomonodromic tau-function
$\CT(\mu;z)$ for the case at hand. 
Section 5 explains how to obtain series expansions for the isomonodromic tau-functions. The form of these
expansions depends on the chosen parameterisation for the monodromy data characterising the quantum curves
by the Riemann-Hilbert correspondence.
{In  Section 6 it is observed that one may 
obtain series expansions having the required form \rf{thetaseries0} depending on a proper  
choice of coordinates for the moduli space of quantum curves.  
The expansion coefficients are compared to the topological string partition functions computed
using the topological vertex 
in Section 7. The partition functions differ from chamber to chamber in the extend K\"ahler moduli space.
Agreement with the expansion coefficients of tau-functions holds if one picks the 
coordinates defining the theta series expansions in a way that depends on the 
chamber under consideration. In Section 8 it is finally observed that the same assignment 
of coordinates to chambers in the moduli space is obtained by applying a construction
called abelianisation in the literature \cite{HN}. Using the coordinates provided by abelianisation 
to define  theta series expansions of the isomonodromic tau-functions 
automatically yields expansion coefficients given by the topological string partition functions 
for each chamber under consideration. 
}


{This paper is the first part of a series of papers devoted to this subject. It
has been revised after the appearance of the second part \cite{CLT} in order to minimize
overlap, and to add some clarifications.  Some material from a previous version of this 
paper has been moved to \cite{CLT}.
}

\newpage

\section{A family of local CY}\label{sec:curves}

\setcounter{equation}{0}

In this section we will discuss the relevant geometric features of the 
families of local CY-manifolds studied in the paper. 
As algebraic varieties one may define the manifolds $Y$ by equations of the form
\begin{equation}\label{CYeqn}
vw-R(x,y)=0,
\end{equation}
where $R(x,y)$ is a polynomial in two variables. Important geometric features 
of $Y$ are encoded in the curve $\Sigma$ defined by the equation $R(x,y)=0$.
Families of curves $\Sigma$ define families $Y\equiv Y_{\Sigma}$ of local CY via \rf{CYeqn}.

\subsection{Curves}\label{Curves}

We will mainly focus our attention on the  family $Y_{u,z}\equiv Y_{\Sigma_{u,z}}$ of local CY 
associated to the family of curves $\Sigma_{u,z}$ defined as 
\begin{equation}\label{nf4-curve}
\begin{aligned}
&\Sigma_{u,z}=
\big\{\,(x,y)\in T^*C\,;\, y^2 = q(x)\,\big\},\\
&q(x)=\frac{a_1^2}{x^2}  + \frac{a_2^2}{(x-z)^2} + \frac{a_3^2}{(x-1)^2} + 
\frac{\kappa}{x(1-x)}+ \frac{z(z-1)}{x(x-1)} \frac{u}{(x-z)},
\end{aligned}
\end{equation}
with
$\kappa = a_1^2 + a_2^2 + a_3^2 - a^2_{4}$. It has a complex two-dimensional moduli space 
parameterised by the complex variables $z$ and $u$. We will see below that the defining equation 
for $\Sigma_{u,z}$ can be brought into the form $R(x,v)=0$ with a polynomial $R(x,v)$
by a change of coordinates $v=v(x,y)$.
The curve $\Sigma_{u,z}$ is a two-fold covering
of the four-punctured sphere $C_z\equiv C_{0,4}=\mathbb{P}^1\setminus\{0,z,1,\infty\}$. The variable 
$u$ determines how $\Sigma$ covers the base curve $C_z$, in particular the positions of the four branch 
points. 

The description simplifies in a useful way in the limit $z\ra 0$ corresponding to a degeneration 
of the base curve $C_z$. Let $\ga_{s}$ be the cycle on $C_z$ that is pinched when $z\ra 0$, and let 
$\hat{\ga}_{s}$ be a lift of $\ga_{s}$ to $\Sigma_{u,z}$ which is odd under the involution exchanging the sheets.
We will be interested in degenerations keeping the period of the canonical differential $ydx$ along 
$\hat{\ga}_{s}$ finite for $z\ra 0$. This will be the case if we consider families $(z,u_z)$ such that 
$u_z=\frac{1}{z}(a^2-a_1^2-a_2^2)$, with $a\in\BC$ finite. Indeed,
setting $u=\frac{1}{z}(a^2-a_1^2-a_2^2)$ in \rf{nf4-curve}, it is straightforward to see
that the region on $\Sigma_{u,z}$ with $x=\CO(1)$ for $z\ra 0$ can be approximately 
represented by the branched cover $\Sigma_{\rm out}$ of $C_{0,3}=\mathbb{P}^1\setminus\{0,1,\infty\}$
defined by the equation
\begin{equation}\label{C03out}
y^2=\frac{x^2 a_4^2-x(a^2+a_4^2-a_3^2)+a^2}{x^2(x-1)^2}.
\end{equation}
From \rf{C03out} is easy to see that the integral $\int_{\hat{\ga}_s}ydx$ is proportional to $a$, as required.

The region in $\Sigma_{u,z}$ with $x=tz$, with $t$ finite when $z\ra 0$, may be represented
as another branched cover $\Sigma_{\rm in}$ of $C_{0,3}$, defined by
\begin{equation}\label{C03curve}
(zy)^2=\frac{t^2 a^2-t(a^2+a_1^2-a_2^2)+a^2_1}{t^2(t-1)^2}.
\end{equation}
We see that $\Sigma_{u,z}$ degenerates into the union of $\Sigma_{\rm out}$ and 
$\Sigma_{\rm in}$ for $z\ra 0$. The parameter $a$ determining the behaviour of 
the parameter  $u$ in the degeneration 
of $\Sigma_{u,z}$ is found to describe the singular behaviour
at the points of $\Sigma_{\rm out}$ and 
$\Sigma_{\rm in}$ corresponding to the double point on $\Sigma_{u,z}$ arising in the degeneration.

\subsection{Four-dimensional limit and local mirror symmetry}\label{sec:geoeng}

\begin{figure}[t]
 \begin{center}
  \includegraphics[width=70mm,clip]{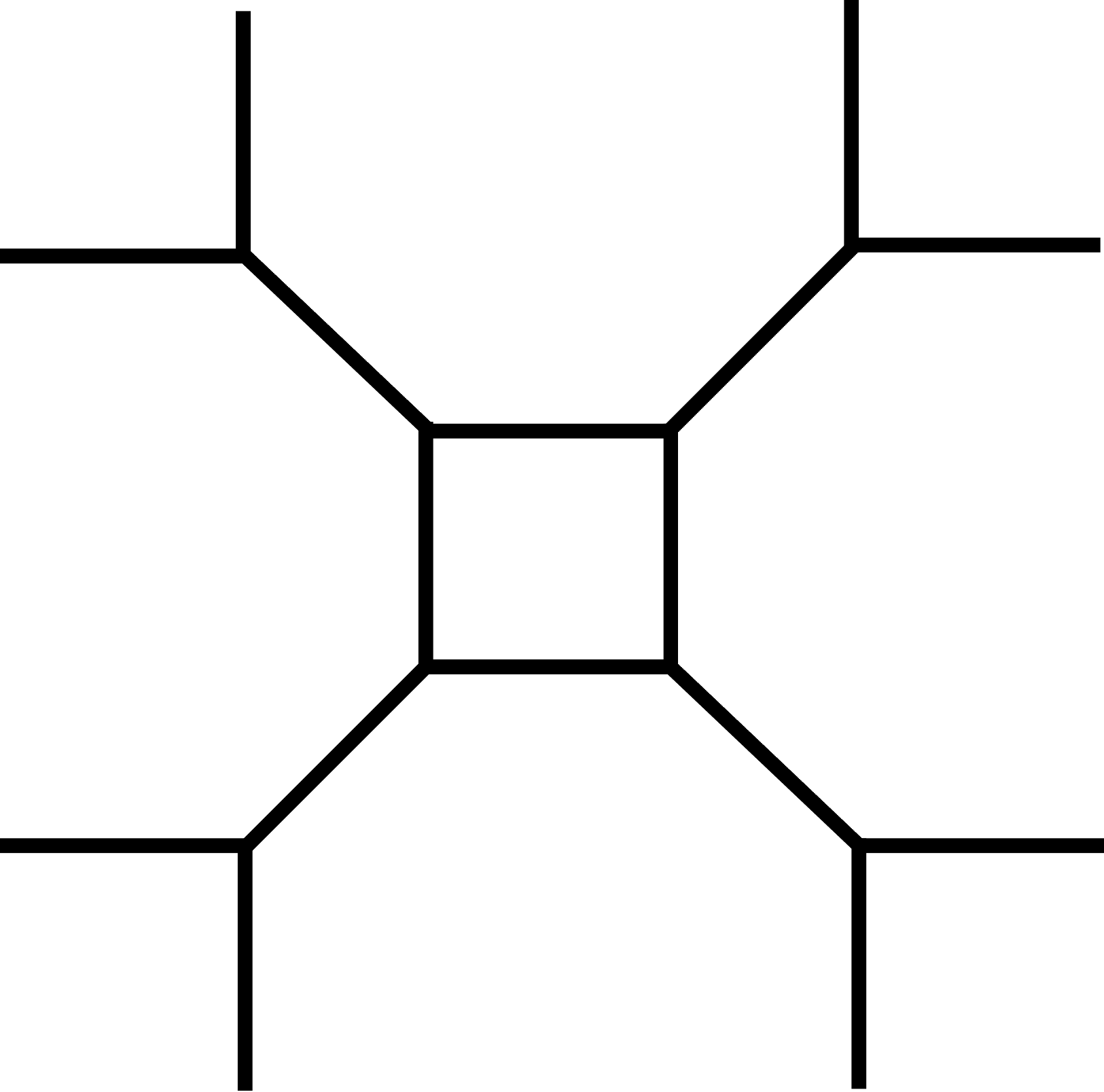}
 \put(-140,140){$Q_{4}$}  
\put(-155,62){$Q_{3}$}   
\put(-60,63){$Q_{2}$}  
\put(-77,139){$Q_{1}$} 
\put(-75,102){$Q_F$} 
\linethickness{1.00pt}
\put(-122,53){\color{blue}\vector(1,0){45}} 
\put(-77,53){\color{blue}\vector(-1,0){45}}     
\put(-105,43){$Q_B$}    
\put(-245,205){$v$}   
\put(20,85){$x$}   
\put(-217,149){$m_4$}  
\put(-217,43){$m_3$}  
\put(3,42){$m_2$}  
\put(3,150){$m_1$}    
\put(-240,98){\vector(1,0){260}} 
\put(-230,0){\vector(0,1){210}} 
\put(-124,119){\color{red}\line(-1,0){8}}
\put(-143,117){$a$} 
\put(-124,77){\color{red}\line(-1,0){8}}
\put(-149,75){$-a$} 
 \end{center}
 \caption{\it The toric graph of the mirror $X_{R;U,z}$ to the local CY $Y_{R;U,z}$.}
 \label{fig:4flavSQCDweb}
\end{figure}
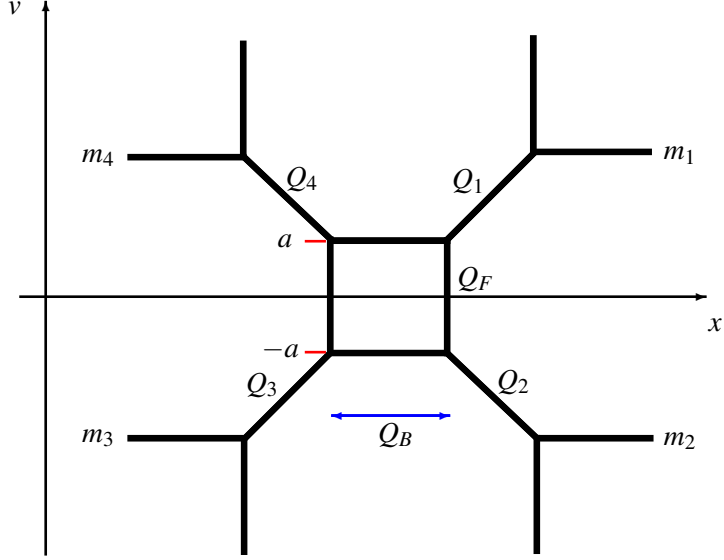

It will later  be useful to recall that 
the family of curves $\Sigma_{u,z}$ can be represented as the limit $R\ra 0$ 
of a certain family of curves $\Sigma_{R;U,z}$ in $\BC^*\times \BC^*$ related 
by  mirror symmetry to the family of toric Calabi-Yau manifolds\footnote{Section  2 in \cite{AKMV} summarises the 
relevant background on  toric geometry in a well-suited form.} having the 
toric graph depicted in Figure \ref{fig:4flavSQCDweb}.
The K\"ahler parameters $t_1,\dots,t_4,t_F, t_B$ of the toric 
Calabi-Yau manifolds will be parameterised through the  variables 
$Q_i=e^{-t_i}$, $i=1,\dots,4$, $Q_F=e^{-t_F}, Q_B=e^{-t_B}$ assigned to the 
edges of the toric graph in Figure \ref{fig:4flavSQCDweb}.

We will consider a certain scaling limit of the K\"ahler parameters which has been
used for the geometric engineering \cite{KKV,KMV} of the four-dimensional, $\CN=2$ supersymmetric 
gauge theory with gauge group $SU(2)$ and four flavors within string theory, see e.g. \cite{HIV} for a review 
discussing this case.
The relevant limit, in the following referred to as four-dimensional (4d) limit, is most easily defined by
parameterising the K\"ahler parameters $t_1,\dots,t_4,t_F, t_B$ as
\begin{equation}
\begin{aligned}
\label{eq:parameterMapSU(2)}
&t_{1}= {R(m_1-a)},\\
&t_{2}= {R(- a-m_2)},
\end{aligned}\qquad
\begin{aligned}
&t_{3}={R(- a-m_3)},\\
&t_{4}={R(m_4-a)}, 
\end{aligned}
 \qquad 
t_{F}= {2Ra} ,
\end{equation}
and sending $R\ra 0$. To simplify the exposition we will  assume that $m_i\in\BR$ for $i=1,\dots,4$. 
In \rf{eq:parameterMapSU(2)} we are anticipating a parameterisation
which will turn out to be useful later. It is based on the fact that the 
K\"ahler parameter associated to 
an edge with equation $rx+sv=c$ and length $l$ is simply given as
$l/\sqrt{r^2+s^2}$. 
Applying this rule to the toric graph  in 
Figure \ref{fig:4flavSQCDweb} 
gives a direct relation between the 
parameters $m_i\in\BR$, $i=1,\dots,4$, in \rf{eq:parameterMapSU(2)} and the 
values of the coordinate $v$ of the corresponding horizontal external edges 
indicated in Figure \ref{fig:4flavSQCDweb}.

Local mirror symmetry \cite{CKYZ} relates this family of toric CY to 
a family of local CY denoted by $\Sigma_{R;U,z}$.
Based on the duality  with brane constructions 
it has been argued in \cite{BPTY}\footnote{It is possible that the following results have been 
derived more directly in the mathematical literature on 
mirror symmetry, but we did not find a reference where 
this has been worked out explicitly for the case of our interest.} 
that 
the curves $\Sigma_{R;U,z}$
can be defined by the equations
\begin{align}\notag
& (w-M_1)(w-M_2) x^2 \\
& \qquad \qquad-\left( \left({ M_1 M_2 }\right)^{\frac{1}{2}}
\left[
\big(1  +zM^{-\frac{1}{2}}\big)w^2
+\big(1  +zM^{+\frac{1}{2}}\big)
\right]
- U  w 
\right) x \label{eq:CurveNf=4}\\
& \qquad  \qquad  \qquad  \qquad  +  z \left( \frac{ M_1  M_2 }{ M_3  M_4 } \right)^{\frac{1}{2}} (w-M_3)(w-M_4) = 0 \, .
\notag\end{align}
We are using the notation $M=M_1M_2M_3M_4$. 
Considering fixed values for $M_1,\dots,M_4$, we will regard the two variables
$z$ and $U$ as parameters for the family of curves $\Sigma_{R;U,z}$.
The parameters $M_1,\dots,M_4,U,z$ of the curve 
defined by the equation \rf{eq:CurveNf=4} are related to the 
K\"ahler parameters  by the mirror map, 
expressing $t_1,\dots,t_4,t_F, t_B$ as periods of the canonical one-form
$ 
\la =\log(w) d\log (x)
$ 
along a suitable set of cycles. 
The rules of local mirror symmetry imply a simple relation between the 
parameters $M_1,\dots,M_4$ in \rf{eq:CurveNf=4} and the parameters
$m_1,\dots,m_4$ introduced via \rf{eq:parameterMapSU(2)}, 
$M_i=e^{-Rm_i}$ for $i=1,\dots,4$. Indeed, it is easy to see that $x\ra \infty$ 
implies that the coordinate $v=-\frac{1}{R}\log(w)$ must approach one of the values
$v=m_1$ or $v=m_2$, and similarly for $x\ra 0$.
The relation between the parameters $U,z$ in \rf{eq:CurveNf=4} and the parameters $t_B$, $t_F=2Ra$
is more complicated.
There exists cycles $\ga_B$ and $\ga_F$ 
on $\Sigma_{R;U,z}$ allowing us to represent the parameters $t_B$ and $t_F$
as the periods $t_B=\int_{\ga_B}\la$ and $t_F=\int_{\ga_F}\la$, respectively.

As discussed in detail in Appendix B of \cite{BPTY}, taking the 
limit $R\ra 0$ of the equation \rf{eq:CurveNf=4} with $w$ being of the form $w=e^{-Rv}$ yields the 
following equation
\begin{equation}
\label{eq:4dSWcurve}
\begin{aligned}
 (v - m_1) (v - m_2) x^2+& \left(-(1+z)v^{2}
+ z (m_1+m_2+m_3+m_4) v + h\right)x\\
& + z (v - m_3) (v - m_4) = 0 ~,
\end{aligned}\end{equation}
with parameter $h$ being related to the higher order terms in the expansion
of $U$ in powers of $R$. This curve can be identified with the curve defined in 
\rf{nf4-curve} by  the change of coordinates $(x,v)\ra (x,y)$ defined by
\begin{equation}
\label{GaiottoShift}
xy = v -  \frac{P_1(x)}{2(x-1)(x-q)} ~,\quad P_1(x) = (m_1+m_2)x^2 -z \, \bar{m}\,x +z (m_3+m_4),
\end{equation}
with $\bar{m}=m_1+m_2+m_3+m_4$, 
bringing the equation for the curve to the form
\begin{equation}
\label{GaiottoCurve}
y^2 = \frac{P_1^2(x) - 4(x-1)(x-z)P_2(x)}{4x^2(x-1)^2(x-z)^2},   \qquad 
 P_2(x) = m_1m_2 x^2  +h\,x  + z m_3m_4  ~.
\end{equation}
This is easily recognised as the curve  \rf{nf4-curve}, with
\begin{align}\label{m-to-a}
& m_4-m_3  = 2a_4    \, , \quad m_4+m_3 = 2a_3\,, \quad   m_1-m_2 = 2a_1   \, , \quad        m_1+m_2 = 2a_2  \, , 
\end{align}
assuming a certain relation between $h$ and $u$ that won't be needed in the following. 

\subsection{Extended K\"ahler moduli space}\label{ExtKaehler}

It will be important for us to notice that only a part of the moduli space of the complex structures
of $\Sigma_{R;U,z}$
is covered by the mirror duals of the toric CY having the toric graph depicted in 
Figure \ref{fig:4flavSQCDweb}. To cover the full moduli space of 
complex structures one will need other toric CY, related to the one considered 
above by flop transitions.  We may introduce an extended K\"ahler moduli space which 
can be described as a collection of chambers representing the K\"ahler moduli spaces
of all toric CY having a mirror dual of the same topological type, joined along walls 
associated to flop transitions.

Our next goal is to describe the chamber structure of the extended K\"ahler moduli 
space in the case $R\ra 0$ of our main interest.   
It is instructive to first analyse the situation in the limit
$z\ra 0$ where $\Sigma_{u,z}$ can be described as the union of 
$\Sigma_{\rm out}$ and 
$\Sigma_{\rm in}$. The curves $\Sigma_{\rm in}$ and $\Sigma_{\rm out}$ are determined by 
the parameters $a^2$, $a_i^2$, $i=1,\dots,4$. We get an unambiguous 
parameterisation assuming $\mathrm{Re}(a)\geq 0$ and $\mathrm{Re}(a_i)\geq 0$, $i=1,\dots,4$.
The equation for $\Sigma_{\rm in}$ can be written as
\begin{equation}\label{Sigma_in}
y^2t^2(t-1)^2=a^2\left(t-\frac{a^2+a_1^2-a_2^2}{2a^2}\right)^2-\frac{D(a)}{4a^2},
\end{equation}
with 
$ 
D(a)=(a+a_1+a_2)(a-a_1-a_2)(a+a_1-a_2)(a-a_1+a_2).
$ 
In the case $a_1>a_2$ 
we see that there exist three chambers, 
\begin{align}
&\mathfrak{C}^{\rm in}_{\1}=\{\,a\in\BC\,;\, \mathrm{Re}(a)\geq 0\,,
a_1-a_2>\mathrm{Re}(a)\,\},\\
&\mathfrak{C}^{\rm in}_{\2}=\{\,a\in\BC\,;\, \mathrm{Re}(a)\geq 0\,,
a_1-a_2<\mathrm{Re}(a)< a_1+a_2\,\},\\
&\mathfrak{C}^{\rm in}_{\3}=\{\,a\in\BC\,;\, \mathrm{Re}(a)\geq 0\,,
\mathrm{Re}(a)> a_1+a_2\,\}.
\end{align}
The boundaries of the chambers correspond to zeros of $D(a)$. Vanishing of $D(a)$ implies that the two branch points of the 
covering $\Sigma_{\rm in}\ra C_{0,3}$ coalesce. 
We may note, on the other hand, that it follows from \rf{eq:parameterMapSU(2)} and \rf{m-to-a} that
$ 
t_{1}= {R(a_2+a_1-a)}$ and 
$t_{2}= {R(a_1- a-a_2)}.
$ 
Vanishing of $D$ is therefore equivalent to the vanishing of a K\"ahler parameter. The case
where $\mathrm{Re}(t_i)>0$ for $i=1,2$ corresponds to the chamber $\mathfrak{C}^{\rm in}_{\1}$.

A similar decomposition into chambers  can be introduced for the parameter space
of $\Sigma_{\rm out}$. Taken together we arrive at a decomposition of the extended K\"ahler 
moduli space of $\Sigma_{u,z}$ for $z\ra 0$ into nine chambers denoted 
$\mathfrak{C}_{\mathfrak{i},\mathfrak{j}}$, with 
$\mathfrak{i}=\1,\2,\3$ labelling the chambers of $\Sigma_{\rm out}$, and
$\mathfrak{j}=\1,\2,\3$ labelling the chambers of $\Sigma_{\rm in}$.

\begin{figure}[t]
 \begin{center}
  \includegraphics[width=90mm,clip]{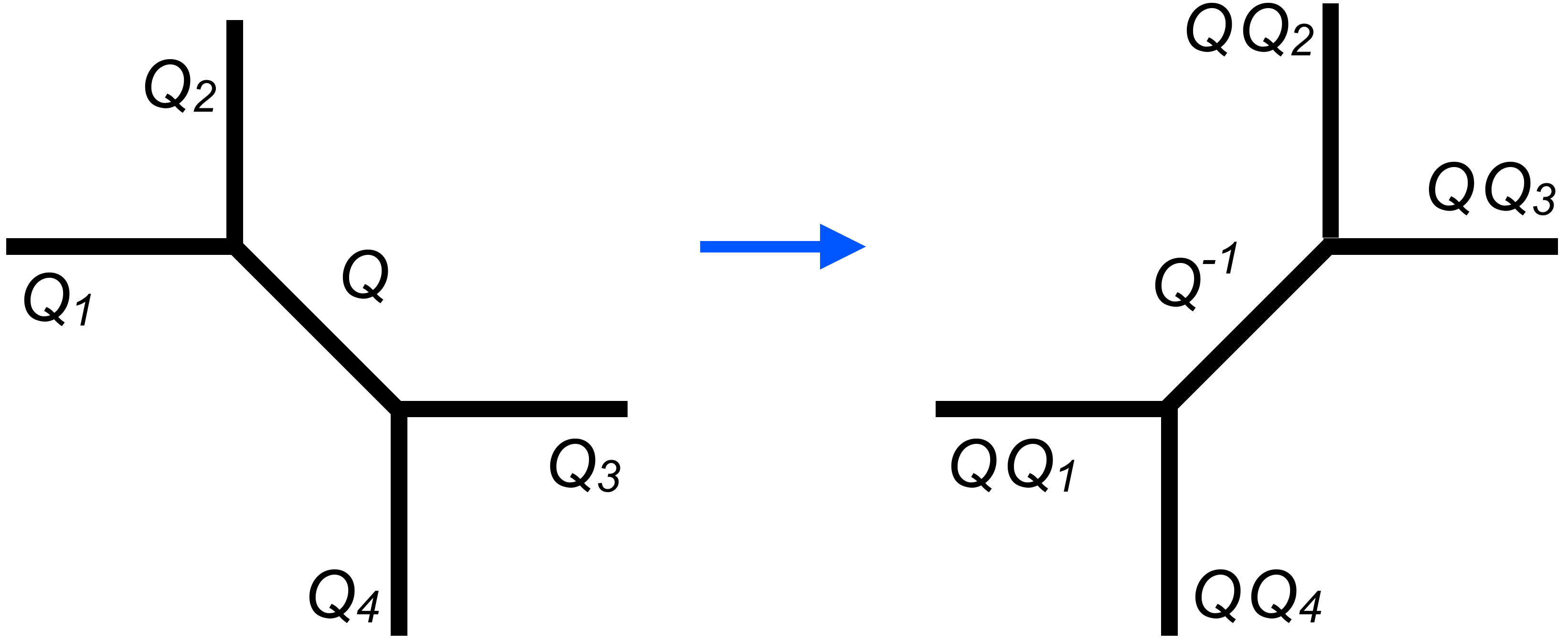}
 \end{center}
 \caption{\it Representation of the flop transition on a subgraph of a toric graph.}
 \label{fig:Flopping}
\end{figure}

The resulting qualitative picture can be expected to hold more generally at least in some neighbourhood
of the boundary component corresponding to the degeneration $z\ra 0$.
The K\"ahler parameters $t_i$, $i=1,\dots,4$ can be represented
as periods of the canonical one-form along cycles surrounding suitable pairs of branch points. 
Coalescence of the branch points implies vanishing of the corresponding periods. 
When one of the periods corresponding to a K\"ahler parameter $t_i$ becomes negative, one can no longer represent 
the mirror of the curves $\Sigma_{R;U,z}$ as the toric CY having the graph in 
Figure \ref{fig:4flavSQCDweb}. The mirror of $\Sigma_{R;U,z}$ may instead be represented
by another toric graph obtained from the one in Figure \ref{fig:4flavSQCDweb} by the local
modification depicted in Figure \ref{fig:Flopping}. This transition is often called a flop. 
In Figure \ref{fig:Flopping} we have also indicated the choice of K\"ahler parameters 
on the toric graph related to the original one by a
flop. For the case at hand it is easy to verify that the rule indicated in 
Figure \ref{fig:Flopping} is necessary to preserve the values of $m_i$ in 
Figure \ref{fig:4flavSQCDweb}.

At least in the case where $z$ is sufficiently small, we expect to get all relevant toric graphs 
by applying flops to the toric CY having the  toric graph depicted in
Figure \ref{fig:4flavSQCDweb}. 


\section{Quantum curves, $\CD$-modules and integrability}\label{sec:q-curves}

\setcounter{equation}{0}



One of the main ideas in \cite{DHSV,DHS} is to regard the relevant free fermion partition functions
as deformations of the chiral free fermion partition functions on the curves $\Sigma$ generated by
turning on  a
B-field proportional to $\la$ on the D6-branes. The deformation induces a non-commutativity of the
coordinates $(x,y)$, turning the curves $\Sigma$ into objects called quantum curves described by certain 
ordinary differential equations. 
We are later going to formulate a precise proposal how to associate a  free fermion partition function
to a quantum curve. In this section we will explain 
what a quantum curve is, 
and why it is natural to allow for 
quantum corrections in the definition of the  quantum curve represented by terms of higher order in $\la$. 

In Subsection \ref{sec:Hitchin} below we will observe that the 
limit $\la\ra 0$ has a natural relation to the Hitchin 
integrable system. 
The relevant quantum corrections are basically determined by 
the requirement to have a consistent deformation of the integrable structure that is present 
at $\la=0$, which will be briefly reviewed in \ref{sec:Hitchin}. A general discussion of the differential equations 
representing the non-commutative deformation of $\Sigma$ is given in Section \ref{sec:d-mod}.
It is observed that the moduli space of holomorphic connections on $C$ is a natural 
one-parameter deformation of the Hitchin system. The moduli space of flat holomorphic connections
has an equivalent representation as the moduli space of the second order differential operators representing 
the quantum curves if one allows quantum corrections in the quantum curve containing  
apparent singularities. The integrable flows of the Hitchin system get ``deformed'' into the 
isomonodromic deformation flows. These flows can be represented as motions of the positions
of the apparent singularities, which is how the $\la$-deformed integrable structure of the 
Hitchin system is represented by quantum corrected quantum curves.  

To simplify the exposition we will mostly
restrict to the case of surfaces $C$ of genus $0$ from now on. It is, however, not hard
to generalise the following discussion to curves $C$ of higher genus.

\subsection{Relation to the Hitchin system}\label{sec:Hitchin}

To motivate our proposal let us revisit the case $\la=0$,  recalling 
that the chiral free fermion partition functions on $\Sigma$ can be represented
as theta functions \cite{AMV}, schematically
\begin{equation}\label{thetasigma}
Z_{\Sigma}(\underline{\vartheta},\mathbf{u})=\sum_{\mathbf{n}}e^{\mathrm{i}\,\mathbf{n}\cdot{\underline{\vartheta}}}
\,e^{\frac{\mathrm{i}}{2}\,\mathbf{n}\cdot\tau^{\Sigma}(\mathbf{u})\cdot \mathbf{n}}e^{\CF_1(\mathbf{u})}.
\end{equation}
The  tuples of integers $\mathbf{n}$ represent the fermion fluxes through cycles of $\Sigma$, and 
$\tau^{\Sigma}(\mathbf{u})$ is the period
matrix of $\Sigma\equiv\Sigma_{\mathbf{u}}$.
The 
variables $\underline{\vartheta}$ in \rf{thetasigma} are naturally interpreted 
as coordinates on the Jacobian
of $\Sigma$ parameterising degree zero line bundles $\CL$ on $\Sigma$.
The free fermion partition function $Z_{\Sigma}(\underline{\vartheta},\mathbf{u})$ is thereby recognised as a function
of the pair of data $(\Sigma,\CL)$. It provides a local description of 
a section of a holomorphic line bundle on 
the Jacobian fibration over the base manifold $\CB$ with 
coordinates $\mathbf{u}$ parameterising the complex structures of $\Sigma$. 

Such Jacobian fibrations naturally arise in the theory of Hitchin systems \cite{Hi} studying 
Higgs pairs $(\mathcal{E},\vf)$ consisting of a holomorphic bundle $\CE$ and 
an element $\vf\in H^0(C,\mathrm{End}(\CE)\otimes K_C)$ modulo gauge transformations. The
integrability of the Hitchin system is realised through the
one-to-one correspondence between Higgs pairs and pairs 
$(\Sigma,\mathcal{L})$, where $\Sigma$ is the spectral curve,
\begin{equation}
\Sigma\,=\,\{\,(x,y)\in T^*C\,;\,\mathrm{det}(y\,\mathrm{id}-\vf(x))=0\,\},
\end{equation} and 
$\mathcal{L}$ is the   line bundle on $\Sigma$ of degree zero having fibres which can be 
identified with the one-dimensional space spanned by an eigenvector
of $\vf$. 
Conversely, given a pair $(\Sigma,\CL)$, where $\Sigma  \subset T^*C$
 is a double cover of $C$, and $\CL$ a holomorphic line bundle on $\Sigma$, one can recover $(\CE , \vf)$ via
$(\CE,\vf) =(\pi_\ast(\CL), \pi_\ast(y))$,
where $\pi$ is the covering map $\Sigma\ra C$, and $\pi_\ast$ is the direct image.

To make this construction more explicit, let us consider the case of holomorphic 
$\mathrm{SL}(2)$-bundles $\CE$, and introduce
a suitably normalised eigenvector $\Phi(x)$ of $\vf(x)$
called Baker-Akhiezer function. It can locally be represented as
\begin{equation}\label{BA-fct}
\Phi(x)=\frac{1}{\vf_0+y}\left(\begin{matrix} \vf_0-y \\ \vf_-
\end{matrix} \right),
\end{equation}
where $y=y(x)$ is the eigenvalue satisfying $y^2=q(x)$, $q(x)=\vf_0^2+\vf_+\vf_-$.
The Baker-Akhiezer function $\Phi(x)$ defined in this way has zeros at the points 
$\hat{x}_k$ projecting to a zero $x_k$ of $\vf_-$ where furthermore $\vf_0=y$, and poles 
at $\check{x}_k=\si(\hat{x}_k)$, with $\si$ being the sheet involution. The divisor 
$\mathbb{D}=\sum_{k}(\hat{x}_k-\check{x}_k)$ characterises the line bundle $\CL$.

To further simplify the exposition let us now restrict attention to the case where the surface $C$
has genus zero $g=0$ with $n$ punctures at $z_1,\dots,z_n$. The Hitchin system will then coincide with 
the Gaudin 
model. The quadratic differential $q(x)$ defining the curve $\Sigma$ has
the form
\begin{equation}\label{q-form}
q(x)=\sum_{r=1}^n\left(\frac{a_r^2}{(x-z_r)^2}+\frac{H_r}{x-z_r}\right).
\end{equation}
Fix a canonical basis $\{\al_1,\dots,\al_{n-3},\be_1,\dots,\be_{n-3}\}$ 
for $H_1(\Sigma)$. The periods of $ydx$ along $\al_k$, $k=1,\dots,n-3$, 
give local coordinates $a^k$ for $\CB$. The Abel-map of the divisor $\mathbb{D}$,
\begin{equation}\label{Abel}
\vartheta_l=\int_\ga\omega_l\,,
\end{equation}
with $\{\omega_l;l=1,\dots,n-3\}$ being a basis for $H^1(\Sigma,K)$ 
such that  $\int_{\al_k}\omega_l=\de_{kl}$,
and $\ga$ in \rf{Abel} 
being a one-dimensional chain\footnote{A formal linear combination
of oriented paths, not necessarily closed, with integral coefficients.} 
such that $\pa \ga=\mathbb{D}$, provides coordinates on 
the Jacobian parameterising the choices of the line bundle $\CL$. 
The coordinates $(\mathbf{a},\underline{\vartheta})$,
$\mathbf{a}=(a_1,\dots,a_{n-3})$, $\underline{\vartheta}=(\vartheta_1,\dots,\vartheta_{n-3})$ are
action-angle coordinates for the Hitchin system. There exists a locally defined
function $\CF(\mathbf{a})$ allowing us to express
the periods $a_k^{\rm\sst D}$  along the dual cycles $\be_k$
as $a_k^{\rm\sst D}=\frac{\pa}{\pa a^k}\CF(\mathbf{a})$.  The  period matrix $\tau^{\Sigma}$
is obtained from $\CF$ as $\tau_{kl}^{\Sigma}=\frac{\pa^2}{\pa a^k\pa a^l}\CF(\mathbf{a})$.

Another useful description of the integrable structure of the Hitchin system uses the 
pairs $(x_k,y_k)$, with $y_k=y(x_k)$ for $k=1,\dots,n-3$ as coordinate functions. This description, often
referred to as the Separation of Variables (SoV) representation\footnote{Going back to \cite{Sk}, applied to 
Hitchin systems in \cite{Hu,GNR,Kr}, and reviewed in \cite[Section 2]{T17b}.}, represents
the phase space as the symmetric product $(T^*C)^{[n-3]}$ with Darboux coordinates $(x_k,y_k)$, $k=1,\dots,n-3$.

It is worth noting that such Jacobian fibrations arise very naturally in the context of local 
CY  of the type considered in this paper. For the case of 
compact base curves $C$ it has been shown in \cite{DDDHP,DDP} that the corresponding Jacobian fibrations
are isomorphic to the intermediate Jacobian fibrations of the associated family $Y_{\Sigma}$ of 
local CY.

\subsection{From quantum curves to $\CD$-modules} \label{sec:d-mod}


In \cite{DHSV,DHS} it is argued that turning on a B-field on the D6-branes induces a non-commutative 
deformation of the algebra of functions on $\Sigma$ described in terms of the coordinates $(x,y)$
by the commutation relations $[x,y]=\mathrm{i}\la$. The deformed algebra of functions can naturally be identified 
with the Weyl algebra of differential operators with generators $x$ and $-\mathrm{i}\la\pa_x$. 
It seems natural to describe the resulting deformation of the curve $\Sigma$ with the help of a deformed 
version of the equation $y^2-q(x)=0$ defining $\Sigma$ which is obtained by replacing $y$ by 
$-\mathrm{i}\la\pa_x$. The equation of the curve $\Sigma$ gets replaced by the differential equation
\begin{equation}\label{quantcurve}
(\la^2\pa_x^2+q(x))\chi=0.  
\end{equation}
A useful framework for making these ideas precise is provided by the theory
of $\CD$-modules.




\subsubsection{$\CD$-modules, differential equations and flat connections}

We will now introduce the basic notions of the theory of 
$\CD$-modules, and later explain why it is 
consistent with the point of view of \cite{DHS} to allow certain
quantum corrections to  the quantum curve obtained by canonical quantisation of the 
equation for the classical curve $\Sigma_{u,z}$.

A $\CD$-module is a sheaf of left modules over the sheaf $\CD_V$ of differential operators on
a smooth complex algebraic variety $V$. For each open subset $U\subset V$ we are given 
a module $\CF(U)$ over $\CD(U)$, the algebra of differential operators on $U$. The various 
modules $\CF(U)$ attached to subsets $U$ satisfy the compatibility conditions defining a sheaf.

An important class of $\CD$-modules is associated to systems of differential equations.
Let $\CG_V$ be a sub-algebra of the algebra $\CD_V$ of  differential operators on $V$, 
generated by commuting differential operators $\CD_i$, $i=1,\dots,m$. To the system
of differential equations
\begin{equation}\label{DEs}
\CD_i\Psi=0,\qquad i=1,\dots,m,
\end{equation}
one may associate the $\CD$-module 
\begin{equation}
\Delta_{\CG_V}^{}:=\CD_V/(\CD_V\cdot \CG_V).
\end{equation}
A solution $\Psi$ of the system \rf{DEs} defines a $\CD$-module homomorphism 
sending $1\in\Delta_{\CG_V}^{}$ to $\Psi$. Conversely, having a $\CD$-module homomorphism 
from $\Delta_{\CG_V}^{}$ to a sheaf $\CF$ one gets a solution $\Psi$ 
to \rf{DEs} with $\Psi\in\CF$ as the image of $1\in\Delta_{\CG_V}^{}$.
The discussion above suggests that we are looking for $\CD$-modules
of this type, with $\CG_V$ being generated by a single
differential operator $\CD_q$ of the form 
$\CD_q=\la^2\pa_x^2+q(x)$.

One may note, on the other hand, that another 
simple type of $\CD$-module is the sheaf of sections of a complex vector bundle $\CE$ on $V$ with a 
holomorphic flat $\lambda$-connection\footnote{A  $\lambda$-connection $\nabla_\la$ satisfies 
$\nabla_\la f s=\la(\pa f)s+f \nabla_\la s$ for smooth functions $f$ and
sections $s$ of $\CE$.} $\nabla_\la$. The $\lambda$-connection $\nabla_\la$, locally represented as
\begin{equation}\label{holoconn}
\nabla_\la=\la\pa_x+\vf(x),\qquad \vf=\left(\begin{matrix} \vf_0 & \;\;\vf_+ \\ \vf_- & -\vf_0\end{matrix}\right),
\end{equation} 
with $\vf_0,\vf_\pm$ holomorphic on $C$, 
defines the action of the 
differential operators in $\CD(U)$ on the sections of $\CE$.  
The $\CD$-modules defined from pairs $(\CE,\nabla_\la)$ can be regarded as
natural $\la$-deformations of the Higgs pairs $(\CE,\vf)$.

Within the moduli space $\CM_{\rm flat}(C)$  of pairs $(\CE,\nabla_\la)$ there is a half-dimensional 
subspace represented by $\lambda$-connections which are gauge equivalent to 
$\lambda$-connections of the form
\begin{equation}\label{oper}
\nabla_{\rm\sst Op}=\la\pa_x+\left(\begin{matrix} 0 & q \\ 1 & 0\end{matrix}\right).
\end{equation}
Flat connections 
of this form are called opers. The horizontality condition $\nabla_{\rm\sst Op}\big(\begin{smallmatrix} \chi_1 \\
\chi_2\end{smallmatrix}\big) =0$ implies that $\chi_2$ solves 
the equation $\CD_q\chi_2=0$, and that $\chi_1=\pa_x\chi_2$. 

Looking for a deformed version of the free fermion partition function associated to quantum 
curves one may note that the $\CD$-modules defined by opers 
only depend on half as many variables as the function 
$Z_{\Sigma}(\underline{\vartheta},\mathbf{u})$ does. The $\CD$-modules associated to 
pairs $(\CE,\nabla_\la)$, on the other hand, depend on just the right number of variables.

\subsubsection{Opers with apparent singularities} 

We are now going to observe that allowing certain quantum corrections in the defining equations
produces quantum curves in a natural one-to-one correspondence to flat connections. To this aim 
we will use the fact that any 
holomorphic connection is gauge equivalent to an oper connection away from certain singularities 
of a very particular type 
which may occur at a collection of points $x_k\in C_{0,n}$, $k=1,\dots,d$.
Given a $\la$-connection of the form 
$\nabla_\la=\la\pa_x+\left(\begin{smallmatrix} \vf_0 & \;\;\vf_+\\ \vf_- & -\vf_0\end{smallmatrix}\right)$ 
it can be  shown by an elementary calculation  that $\nabla_\la$ can be brought to oper form
$\la\pa_x+\left(\begin{smallmatrix} 0 & q_\la \\ 1 & 0\end{smallmatrix}\right)$ 
by means of a gauge transformation $h$,
\begin{equation}\label{meroper}
\nabla_{\rm\sst Op}=h^{-1}\cdot\nabla_\la\cdot h=\la\pa_y+\bigg(\begin{matrix} 0 & q_\la \\ 1 & 0\end{matrix}\bigg),
\end{equation}
which is well-defined on a cover of $C$
branched at the zeros $x_k$, $k=1,\dots,d$, of $\vf_-=\vf_-(x)$.
The resulting formula for the matrix element $q_\la$ is   found to be 
of the form
\begin{equation}\label{opermod}
q_\la(x)=\sum_{r=1}^n\left(\frac{a_r^2}{(x-z_r)^2}+\frac{H_r}{x-z_r}\right)
+\la\sum_{k=1}^{d}\left(\frac{y_k}{x-x_k}
-\frac{3\la}{4(x-x_k)^2}\right).
\end{equation}
Assuming that $\nabla_\la$ is holomorphic on $C_{0,n}$, 
it follows from \rf{meroper} that the monodromy
of $\nabla_{\rm\sst Op}$ around the points $x_k$ is proportional to the identity matrix and therefore
trivial in $\mathrm{PSL}(2,\BC)$. Singularities having this property are called apparent singularities.
Having an apparent singularity at $x=x_k$ is equivalent to the fact that
the parameters $(x_k,y_k)$ introduced in \rf{opermod} satisfy the equations
\begin{equation}\label{appconstr}
\la^2y_k^2+{q}_0^{(k)}=0, \quad k=1, \dots, d,\qquad
q_\la(x)=\sum_{n=-2}(x-x_k)^n q_n^{(k)}.
\end{equation}
Taking into account the constraints \rf{appconstr} and the constraints from regularity at infinity 
it is not hard to see that
for fixed $a_r$, $r=1,\dots,n$, in \rf{opermod}
one gets a family of quadratic differentials $q_\la$ on $C$
depending on $2(n-3)$ independent parameters.

Conversely, if the constraints \rf{appconstr} are satisfied,  and if $d\leq n-3$, there exists a unique 
gauge transformation $h$ holomorphic on a double cover of 
$C_{0,n}\setminus\{x_1,\dots,x_d\}$ with branch points only at $x_1,\dots,x_d$
such that the
connection $\nabla_\la$  defined from  $\nabla_{\rm\sst Op}=\la\pa_x+\left(\begin{smallmatrix} 0 & q_\la \\ 1 & 0\end{smallmatrix}\right)$  by means of \rf{meroper}
is holomorphic on $C_{0,n}$  with first order poles 
only at $x=z_r$. Indeed, by defining 
\begin{equation}
\vf_-(x)=c_0\frac{\prod_{k=1}^{d}(x-x_k)}{\prod_{r=1}^n(x-z_r)},\qquad
\vf_0(x)=\sum_{k=1}^d y_k\left(\prod_{r=1}^n\frac{x_k-z_r}{x-z_r}\right)
\prod_{\substack{l=1\\l\neq k}}^d\frac{x-x_l}{x_k-x_l},
\end{equation}
and using these functions to build
\begin{equation}
h=\bigg(\begin{matrix} 1/\sqrt{\vf_-} & 0\\ 0 & \sqrt{\vf_-}\end{matrix}\bigg)\left(\begin{matrix} 1 & \al\\ 0 & 1\end{matrix}\right),
\quad \al(x)=\frac{\la}{2}\frac{\vf_-'}{\vf_-}-\vf_0(x),
\end{equation}
we find that the connection $\nabla_\lambda$  is holomorphic on $C_{0,n}$.

Allowing quantum corrections containing apparent 
singularities therefore gives us a way to represent all the data characterising a gauge equivalence
class of holomorphic connections in terms of meromorphic opers.
The equivalence between flat 
$\fsl_2$-connections $\nabla_\la$ on $C_{0,n}$ and opers $\nabla_{\rm\sst Op}$ observed above 
can be seen as a
deformation of the Separation of Variables (SOV) for the classical Gaudin model \cite{Sk,DM} 
with deformation parameter $\la$. Comparing with \rf{BA-fct}
we see that the positions $(x_1,\dots,x_{d})$ of the apparent singularities are
directly related to the divisor $\mathbb{D}$ characterising the line bundle $\CL$ 
in the limit $\la\ra 0$.


\subsection{Isomonodromic deformations}\label{sec:isodef}

We are now going to observe that the deformation of the Higgs pairs $(\CE,\vf)$ into 
$\la$-connections leads to a natural deformation of the integrable flows of the Hitchin 
system, given by the isomonodromic deformation flows. It will turn out that this 
integrable structure controls how the free fermion partition function gets deformed 
when $\lambda$ is non-zero.


\subsubsection{Riemann-Hilbert correspondence}\label{sssec:RH}



The Riemann-Hilbert correspondence assigns holomorphic connections to representations $\rho:\pi_1(C)\ra G$
of the fundamental group $\pi_1(C)$ in a group $G$, here taken to be $G=\mathrm{SL}(2,\BC)$.
Considering curves $C$ of genus $0$ 
with a base point $x_0$ one may characterise the representations $\rho$ by the matrices
$M_r$ representing closed curves $\ga_r$ around the punctures $z_r$.   We will consider the cases where
the matrices $M_r$ are diagonalizable, $M_r=C_r^{-1}e^{2\pi \mathrm{i}D_r}C_r^{}$, for a fixed choice of
diagonal matrices $D_r$. 
The Riemann-Hilbert problem is to find a multivalued
analytic
matrix
function $\Psi(x)$ on $C_{0,n}$ such that the monodromy along
$\ga_r$ is represented by 
\begin{equation}
\Psi(\ga_r.x)=\Psi(x)\cdot M_r,
\end{equation}
with $\Psi(\ga_r.x)$ being the analytic continuation of $\Psi(x)$ along $\ga_r$.
The solution to this problem is unique up to left multiplication with
single valued matrix functions. In order to fix this ambiguity we need to
specify the singular behaviour of $\Psi(x)$ at $x=z_r$, leading to the following
refined version of the Riemann-Hilbert problem: 
\begin{quote}
{\it Find a matrix
function $\Psi(x)$ such that (i) $\Psi(x)$ is a multivalued, analytic and invertible
function on $C_{0,n}$ satisfying a normalisation condition, and 
(ii) there exist neighborhoods of $z_k$, $k=1,\dots,n$ where
$\Psi(x)$ can be represented as
\begin{equation}\label{asym}
\Psi(x)\,=\,\hat{Y}^{(k)}(x)\cdot(x-z_k)^{D_k}\cdot C_k\,,
\end{equation}
with $\hat{Y}^{(k)}(x)$  holomorphic and invertible at $x=z_k$,
$C_k\in G$, and $D_k$ being diagonal  matrices for $k=1,\dots,n$.
}\end{quote}
A standard choice of a normalisation condition is to require that 
$\Psi(x_0)=1$ at a fixed point $x_0\in C$. Other  options are to
fix the matrix  $\hat{Y}^{(k)}(z_k)$ appearing in \rf{asym} for one particular value of $k$.
If such a function $\Psi(x)$ exists, it is uniquely determined by the
monodromy data $\mathbf{C}=(C_1,\dots,C_n)$ and the diagonal matrices $\mathbf{D}=(D_1,\dots,D_n)$.
It is known that the solutions to the Riemann-Hilbert problem exist for 
generic representations $\rho:\pi_1(C_{0,n})\ra
G$. 

\subsubsection{Isomonodromic deformations}

We shall now briefly indicate how the Riemann-Hilbert
problem is related to the isomonodromic deformation problem.
Given a solution $\Psi(x)=\Psi(x;\mu,\mathbf{z})$
to the Riemann-Hilbert problem we may define a  connection $A(x)$
as
\begin{equation}
A(x):=\,(\pa_x \Psi(x))\cdot (\Psi(x))^{-1}\,,
\end{equation} 
It follows from ii) that $A(x)$ is a rational function of $x$ which has the form
\begin{equation}
A(x)\,=\,\sum_{r=1}^{n}\frac{A_r(\mu,\mathbf{z})}{x-z_r}\,,
\end{equation}
$\mathbf{z}=(z_1,\dots,z_n)$ being the positions of the punctures.
A variation of  $\mathbf{z}$ for fixed monodromy data $\mu$
leads to a variation of the matrix residues $A_r$. It is not hard to show (see e.g. \cite{BBT}) that 
the resulting variations are described by a nonlinear first order system of partial differential equations
called the Schlesinger equations. The
Schlesinger equations are the Hamiltonian flows defined by
the Hamiltonians and Poisson structure
\begin{equation}\label{Poissonflat}
H_r=\sum_{s\neq r}
\frac{{\rm tr}(A_rA_s)}{z_r-z_s}\,,
\qquad
\{A(x)\,\substack{\otimes\vspace{-0.1cm} \\ ,}\,A(x')\}=
 \left[\frac{\mathcal{P}}{x-x'}\,,A(x)\otimes 1+1\otimes A(x')\right]\!,
\end{equation}
where $\mathcal{P}$ denotes the permutation matrix.


With the help of the equivalence between holomorphic connections and 
meromorphic opers one may describe the isomonodromic 
deformation flows as the flows describing 
 isomonodromic 
deformations of the second order differential operator $\CD_{q_\la}$. It is worth noting that
\begin{itemize} 
\item[(i)] the Hamiltonians $H_r$ generating the isomonodromic deformation flows are related to the 
residues $H_r$ in \rf{opermod} by the gauge transformation from holomorphic connections to opers
with apparent singularities, 
\item[(ii)] the equations \rf{appconstr} are a system of linear equations for the residues $H_r$ in \rf{opermod} 
which can be solved
explicitly to get
$H_r\equiv H_r(\mathbf{x},\mathbf{y};\la)$,  $\mathbf{x}=(x_1,\dots,x_{n-3})$, $\mathbf{y}=(y_1,\dots,y_{n-3})$,
\item[(iii)] the isomonodromic deformation equations can then be represented in Hamiltonian form
as 
\begin{equation}\label{Garnier}
\frac{\pa x_k}{\pa z_r}=\frac{\pa H_r}{\pa y_k},\qquad
\frac{\pa y_k}{\pa z_r}=-\frac{\pa H_r}{\pa x_k},
\end{equation}
\item[(iv)] the coordinates $(\mathbf{x},\mathbf{y})$ are 
Darboux coordinates for the Poisson structure \rf{Poissonflat}, as
equations \rf{Garnier} suggest.
\end{itemize}
The proofs of these statements can be found in \cite{Ok,IKSY,DM}.
In this form it becomes   easy to 
see that the isomonodromic deformation flows turn into flows of the Hitchin integrable system
for $\la\ra 0$, with $ (\mathbf{x},\mathbf{y})$ being the variables in the SOV representation \cite{DM}.
One may recall, in particular, that the variables $x_k$ defining the divisor $\mathbb{D}$ 
are nothing but the zeros of $\vf_-(x)$, and note that
the functions  $H_r(\mathbf{x},\mathbf{y};\la)$
turn into the Hamiltonians of the Hitchin system for $\la\ra 0$.

\subsection{Isomonodromic tau-function}

The isomonodromic tau-function $\mathcal{T}(\mu,\mathbf{z})$ is then defined as the generating function for the
Hamiltonians $H_r$,
\begin{equation}\label{taudef}
H_r\,=\, \pa_{z_r}\log\mathcal{T}(\mu,\mathbf{z})\,.
\end{equation}
It can be shown that the 
integrability of \rf{taudef} is a direct consequence of the Schlesinger 
equations.  Equation \rf{taudef} determines $\mathcal{T}(\mu,\mathbf{z})$ only 
up to addition of a function of the monodromy data. Having fixed this freedom by suitable
supplementary conditions, one may use the Schlesinger equations to determine the 
dependence of $\mathcal{T}(\mu,\mathbf{z})$ on $\mathbf{z}$ via \rf{Poissonflat} and \rf{taudef}.

We will see in the following that the free fermion partition functions we want to 
associate to the $\CD$-modules representing the quantum curves
can be identified with the isomonodromic tau functions 
coming from the Riemann-Hilbert problem characterising the relevant $\CD$-modules.

\section{From quantum curves to free fermion partition functions}\label{Sec:Dmod-tau}

We are now going to explain how to define free fermion partition functions from the 
solutions of the differential equation defining the quantum curve.
This construction generalises the deformed version of the Krichever construction
used in \cite{DHS}. The relation to the theory of infinite Grassmannians and
of the Sato-Segal-Wilson
tau-functions used in \cite{DHS} is explained in Appendix \ref{SW-app}.
The free fermion partition functions defined in this way 
turn out to be closely related to conformal blocks of the free fermion vertex 
operator algebra (VOA). The conformal Ward identities determininig the dependence
of the free fermion partition functions with respect to the complex structure of $C$
are equivalent to the equations defining the isomonodromic tau-functions. 
It will follow that a suitable choice of normalisation factors, which may still depend 
on the monodromy data characterising the equation of the quantum curve 
through the Riemann-Hilbert correspondence, allows us to relate the free fermion
partition functions of our interest to isomonodromic tau-functions.

\subsection{From $\CD$-modules to free fermion states}\label{CDtoFF}


\subsubsection{Free fermions}

The free fermion super VOA is generated by fields $\psi_s(z)$, $\bar{\psi}_s(z)$, $s=1,\dots,N$,
The fields $\psi_s(z)$ will be arranged into a row vector $\psi(z)=(\psi_1(z),\dots,\psi_N(z))$, 
while $\bar\psi(z)$ will be our notation for the column vector with components $\bar{\psi}_s(z)$. 
The modes of $\psi(z)$ and $\bar\psi(z)$, introduced as
\begin{equation}
\psi(z)=\sum_{n\in\BZ}\psi_{n} z^{-n-1}\,,\quad
\bar\psi(z)=\sum_{n\in\BZ} \bar\psi_{n} z^{-n}\,,
\end{equation}
are row and column vectors 
with components $\psi_{s,n}$ and $\bar{\psi}_{s,n}$, respectively,  satisfying 
\begin{equation}
\{\,\psi_{s,n}\,,\,\bar\psi_{t,m}\,\}=\de_{s,t}\de_{n,-m}\,,\qquad
\{\,\psi_{s,n}\,,\,\psi_{t,m}\,\}=0\,,\qquad\{\,\bar\psi_{s,n}\,,\,\bar\psi_{t,m}\,\}=0\,.
\end{equation}
The Fock space $\CF$ is a representation generated from a highest weight vector $\mathfrak{f}_{0}^{}$ 
satisfying  
\begin{equation}
\psi_{s,n}\cdot\mathfrak{f}_{0}^{}=0\,,\quad n\geq 0\,,\qquad
\bar\psi_{s,n}\cdot\mathfrak{f}_{0}^{}=0\,,\quad n>0\,.
\end{equation}
$\CF$ is generated from $\mathfrak{f}_{0}^{}$ by the action of the modes $\psi_{s,n}$, $n<0$, 
and $\bar\psi_{s,m}$, $m\leq 0$. 

We will also consider the conjugate representation $\CF^*$, a {\it right} module generated 
from a highest weight vector $\mathfrak{f}_{0}^{*}$ 
satisfying  
\begin{equation}
\mathfrak{f}_{0}^{*}\cdot\psi_{s,n}=0\,,\quad n< 0\,,\qquad
\mathfrak{f}_{0}^{*}\cdot\bar\psi_{s,n}\,=0\,,\quad n\leq 0\,.
\end{equation}
The Fock space $\CF^*$ is generated from $\mathfrak{f}_{0}^{*}$ by the right action of the modes $\psi_{s,n}$, $n\geq 0$, 
and $\bar\psi_{s,m}$, $m> 0$.  A natural bilinear form $\CF^*\otimes\CF\ra \BC$ is defined
by the expectation value, 
\begin{equation}
\langle \,\mathfrak{f}_0^*\cdot\mathsf{O}_{\mathfrak{f}^*}\,,\,\mathsf{O}_{\mathfrak{f}}\cdot\mathfrak{f}_0\,\rangle= \Omega
(\mathsf{O}_{\mathfrak{f}^*}\mathsf{O}_{\mathfrak{f}}\cdot \mathfrak{f}_0),
\end{equation}
where $\Omega(\mathfrak{f})=c$ if $\mathfrak{f}= c\,\mathfrak{f}_{0}^{}+\sum_{s=1}^N(\sum_{n<0} \psi_{s,n}\mathfrak{f}_{s,n} +
\sum_{m\leq 0}\bar{\psi}_{s,m}\mathfrak{f}_{s,m})$.

\subsubsection{Free fermion states from the Riemann-Hilbert correspondence}\label{FFfromD}

A simple and natural way to characterise a state $\mathfrak{f}\equiv \mathfrak{f}_G\in\CF$ is through the 
matrix $G(x,y)\equiv G_\mathfrak{f}(x,y)$ of two-point  functions  having matrix elements
\begin{equation}\label{FFtwopt}
G_\mathfrak{f}(x,y)_{st}=\big\langle \,\bar{\psi}_s(x)\psi_t(y) \,\big\rangle_{\mathfrak{f}} \equiv
\frac{\langle \,\mathfrak{f}_{0}^{\ast}\,,\,\bar{\psi}_s(x)\psi_t(y)\,\mathfrak{f}\,\rangle}{\langle \,
\mathfrak{f}_{0}^{\ast}\,,\,\mathfrak{f}\,\rangle}.
\end{equation}
Indeed, given a function $G(x,y)$ such that 
\begin{equation}
G(x,y)=\frac{{1}}{x-y}+A(x,y)\,,
\end{equation}
with $A(x,y)$ having an expansion of the form
\begin{equation}\label{A-exp}
A(x,y)=\sum_{l\geq 0} y^{-l-1}\sum_{k>0}x^{-k}A_{kl},
\end{equation}
there exists a state $\mathfrak{f}_G$, unique up to normalisation, such that
its two-point function is given by $G(x,y)$. 
States $\mathfrak{f}_G^{}$ having this property can be constructed as
\begin{equation}\label{CBfromR}
\mathfrak{f}_G^{} 
=N_G^{}\exp\bigg(-\sum_{k>0}\sum_{l\geq 0} \psi_{-k}\cdot A_{kl}\cdot \bar{\psi}_{-l}\bigg) \mathfrak{f}_{0}^{}\,,
\end{equation}
with  matrices $A_{kl}$ 
defined by the expansion \rf{A-exp}, and $N_G\in\BC$ being a normalisation constant.
This can be verified by a straightforward computation. 

We will be mainly interested in two-point functions $G(x,y)$ that have a multi-valued analytic continuation 
with respect to both $x$ and $y$
to the Riemann surfaces $C=C_{0,n}$ with given monodromies. The monodromies describing the analytic
continuation in $x$ are required to act on $G(x,y)$ from the left, while the analytic continuation in $y$ generates
monodromies acting from the right. Consistency with having a pole at $x=y$ with residue being the 
identity matrix requires
\begin{equation}
G(x,\ga_r.y)=G(x,y)\cdot M_r,\qquad G(\ga_r.x,y)=M_r^{-1}\cdot G(x,y). 
\end{equation}
This means that the family of functions $G_x(y):=G(x,y)$ is a solution to 
a generalisation of the Riemann-Hilbert problem formulated above where one allows a first order pole at 
$y=x$, 
and the family $G_y(x):=G(x,y)$ is a solution to a conjugate version of this Riemann-Hilbert problem.
Uniqueness of the solution to the Riemann-Hilbert problem implies that $G(x,y)$ must have the following 
form
\begin{equation}\label{GfromPsi}
G_{\Psi}(x,y)=\frac{(\Psi(x))^{-1}\Psi(y)}{x-y},
\end{equation}
with $\Psi(y)$ being a solution to the Riemann-Hilbert problem  formulated in Section \ref{sssec:RH}. 

The
construction of the fermionic states $\mathfrak{f}_{G}$ described above therefore 
gives us a natural way to assign  fermionic states $\mathfrak{f}_\Psi^{}\equiv \mathfrak{f}_{G_\Psi}$ to 
solutions $\Psi$ of the Riemann-Hilbert problem.


\subsection{Free fermion conformal blocks from $\CD$-modules}\label{FF-RH}

We are now offering
a useful change of perspective by re-interpreting the fermionic states associated to 
$\CD$-modules as free fermion conformal blocks. 
This will allow us to use methods and ideas 
from conformal field theory which will be  useful for the computation of tau-functions.
To this aim we will note that the states $\mathfrak{f}_\Psi^{}\in\CF$ constructed in Section
\ref{FFfromD} are
characterised by a set of Ward identities defined from a solution $\Psi(x)$ of the 
RH problem. Given that conformal blocks can be defined as solutions to such Ward 
identities\footnote{A review of CFT with a very similar perspective can be found  in \cite{T17a}.}
we are led to identify the states $\mathfrak{f}_\Psi^{}\in\CF$ as conformal blocks
for the free fermion VOA.

Let us define the following 
infinite-dimensional spaces of multi-valued  functions on ${C}_{0,n}$,
\begin{equation}\label{RbarRdef}
\begin{aligned}
&\bar{W}_\Psi=\big\{\,\bar{v}(x)\cdot\Psi(x);\;\,\bar{v}(x)
\in\mathbb{C}^N\otimes\mathbb{C}[\mathbb{P}^1\!\setminus\!\{\infty\}]\,\big\}\,,\\
&{{W}}_\Psi=\big\{\,\Psi^{-1}(x)\cdot {v}(x);\;\,{v}(x)\in\mathbb{C}^N\otimes
\mathbb{C}[\mathbb{P}^1\!\setminus\!\{\infty\}]\,\big\}\,,
\end{aligned}
\end{equation}
where $\bar{v}$ and ${v}$ are row and column vectors with $N$ components, respectively, 
and $\mathbb{C}[\mathbb{P}^1\!\setminus\!\{\infty\}]$ is the space of meromorphic functions
on $\mathbb{P}^1$
having poles at $\infty$ only. The elements of the space $\bar{W}_\Psi$ represent solutions of
a generalisation of the  RH problem from Section \ref{sssec:RH}
where the condition of regularity at infinity has been dropped.

Let us next note that the vectors $\mathfrak{f}_\Psi^{}$ defined in \rf{CBfromR} can
be equivalently characterised up to normalisation by the  conditions
\begin{equation}\label{FFWard}
\psi[g]\cdot\mathfrak{f}_\Psi^{}=0\,,\qquad
\bar{\psi}[\bar{f}\,]\cdot\mathfrak{f}_\Psi^{}=0\,,
\end{equation} 
for all $g\in W_\Psi$, $\bar{f}\in\bar{W}_\Psi$, where  the operators $\psi[\bar{f}\,]$ are constructed as 
\begin{equation}\label{psigdef}
\psi[{g}]\,=\,\frac{1}{2\pi\mathrm{i}}\int_{\CC} dz\; \psi(z)\cdot {g}(z)\,,\qquad
\bar{\psi}[\bar{f}\,]\,=\,\frac{1}{2\pi\mathrm{i}}\int_{\CC} dz\; \bar{f}(z)\cdot \bar{\psi}(z)\,,
\end{equation}
with $\CC$ being a circle separating $\infty$ from $z_1,\dots,z_n$.

Indeed,
it can easily be shown that the vector $\mathfrak{f}_\Psi^{}$ is defined uniquely up to 
normalisation by the identities \rf{FFWard}. Let us
note that the columns of $\bar{G}_l(x)$, $l\geq 0$,
and the rows of the matrix-valued functions $G_k(y)$, $k>0$, 
defined through the expansions 
\begin{equation}\label{G-exp}
\frac{(\Psi(x))^{-1}\Psi(y)}{x-y}=\left\{
\begin{aligned}
&\sum_{l\geq 0} y^{-l-1} \bar{G}_l(x),\quad \bar{G}_l(x)=  -x^l\,\mathbf{1}+\sum_{k>0}x^{-k}A_{kl}\,,\\
&\sum_{k> 0} x^{-k} {G}_k(y),\quad {G}_k(y)= y^{k-1}\,\mathbf{1}+\sum_{l\geq 0}y^{-l-1}A_{kl}\,.
\end{aligned}\right.
\end{equation}
generate bases for the spaces $\bar{W}_\Psi$  and ${W}_\Psi$ associated to $\Psi(x)$,
respectively. The conditions \rf{FFWard} are  equivalent to the validity of
\begin{equation}\label{FFWard'}
\bar{\psi}_k^{}\,\mathfrak{f}_\Psi^{}=-\sum_{l\geq 0}(A_{kl}\cdot\bar{\psi}_{-l}^{})\,\mathfrak{f}_\Psi^{},\qquad
{\psi}_l^{}\,\mathfrak{f}_\Psi^{}=\sum_{k> 0}({\psi}_{-k}^{}\cdot A_{kl})\,\mathfrak{f}_\Psi^{},
\end{equation}
for all $k>0$ and all $l\geq 0$.
The identities \rf{FFWard'} can be used to calculate the values of $\langle \mathfrak{v},\mathfrak{f}_\Psi^{}\rangle_\CF$
for $\mathfrak{f}_\Psi^{}\in\CF$ satisfying \rf{FFWard} and arbitrary $\mathfrak{v}\in\CF$ 
in terms of $\langle \mathfrak{f}_{0}^{},\mathfrak{f}_\Psi^{}\rangle_\CF$. This implies that the 
solution to the conditions  \rf{FFWard} is unique up to normalisation. 
It is not hard to check that the
vector $\mathfrak{f}_\Psi^{}$ defined using \rf{G-exp} and \rf{CBfromR} indeed satisfies the identities \rf{FFWard'}.

The definition of  $\mathfrak{f}_\Psi^{}$ through the identities \rf{FFWard} 
is analogous to the definition of Virasoro conformal blocks through the conformal Ward
identities. The uniqueness of $\mathfrak{f}_\Psi^{}$ implies that the space of conformal blocks 
for the free fermionic VOA is one-dimensional.

\subsection{Chiral partition functions as isomonodromic tau-functions}\label{taufromZ}


Out of a representation of the free fermion VOA one may define a representation of the Virasoro algebra
by introducing the energy-momentum tensor as
\begin{equation}\label{FFVir}
T(z)=\frac{1}{2}\lim_{w\ra 0}\sum_{s=1}^N\bigg(\pa_z\psi_s(w)\bar{\psi}_s(z)+\pa_z\bar{\psi}_s(w)\psi_s(z)+\frac{1}{(w-z)^2}\bigg)\,.
\end{equation}
Conformal blocks for the free fermion VOA represent conformal blocks for the 
Virasoro algebra defined via \rf{FFVir}. On the space of 
conformal blocks of the Virasoro
algebra there is a canonical connection \cite{FS}  allowing us to   
represent the variations of a conformal block
induced by variations of the 
complex structure of the underlying Riemann surface $C_{0,n}$ in the form\footnote{This is 
reviewed in [T17a] using a very similar formalism as used in our paper.}
\begin{equation}\label{FSconn}
\pa_{z_r}\mathfrak{f}_\Psi^{}=\mathsf{H}_r\, \mathfrak{f}_\Psi^{},
\end{equation}
with $\mathsf{H}_r$ being suitable linear combinations of the modes of $T(z)$. This connection 
preserves the one-dimensional space of free fermion conformal blocks due to the fact that the adjoint action of the 
Virasoro algebra acts geometrically on the free fermions, transforming them as half-differentials.

The operators $\mathsf{H}_r$ generate a commutative subalgebra of 
the Virasoro algebra, embedded into the Lie algebra 
generated by fermion bilinears via \rf{FFVir}.
Keeping in mind the fact that only the normalisation of $\mathfrak{f}_\Psi^{}$ was left undetermined
by \rf{FFWard} one sees that the equations \rf{FSconn} together with 
\rf{FFWard} can be used to determine $\mathfrak{f}_{\Psi}(\mathbf{z})$
unambiguously in terms of $\mathfrak{f}_{\Psi}(\mathbf{z}_0)$ for any given path
connecting $\mathbf{z}$ and $\mathbf{z}_0$ in $\CM_{0,n}$, the moduli space of complex 
structures on $C_{0,n}$.
Using only the Ward identities one can show that\footnote{The main idea
is simple \cite{Mo}: Consider the expansion of the fermion two point function 
$G_{\Psi}(x,y)$ around $x=y$. Using \rf{G-exp} and $\pa_x\Psi=A\Psi$ one may observe that the
trace part contains ${\mathrm{tr}} A^2(x)$  at order $\CO(x-y)$. The expansion may also
be calculated using the OPE of the fermionic fields where $T(x)$ appears at the same order.
Comparing the resulting expressions yields
\rf{cfblock-tau}. A proof within the formalism used here was outlined in \cite{T17a}.}
\begin{equation}\label{cfblock-tau}
\pa_{z_r}\log\langle\,\mathfrak{f}_{0}^{}\,,\,\mathfrak{f}_\Psi^{}(\mathbf{z})\,\rangle
=
\frac{\langle\,\mathfrak{f}_{0}^{}\,,\SH_r\,\mathfrak{f}_\Psi^{}(\mathbf{z})\,\rangle}{\langle\,\mathfrak{f}_{0}^{}\,,\,\mathfrak{f}_\Psi^{}(\mathbf{z})\,\rangle}
=
H_r(\mu,\mathbf{z}),
\end{equation}
with $H_r$ being the isomonodromic deformation Hamiltonians defined in \rf{Poissonflat}. This means that the isomonodromic tau-function coincides up to 
a function $N(\mu)$ of the monodromy data  with
\begin{equation}\label{tau-ff}
\mathcal{Z}_{\rm ff}(\mu,\mathbf{z})=N(\mu)\CT(\mu,\mathbf{z}),\qquad 
\mathcal{Z}_{\rm ff}(\mu,\mathbf{z}):=
\langle\,\mathfrak{f}_{0}^{}\,,\,\mathfrak{f}_{\Psi}^{}(\mathbf{z})\rangle,
\end{equation} 
relating the isomonodromic tau-functions to free fermion conformal blocks. 

\begin{rem} Starting from a Lagrangian description of the free fermions
on a Riemann surface $C$ one would naturally arrive at a description of the 
free fermion partition functions as determinants of Cauchy-Riemann-operators
on $C$. Such determinants have been studied
for  $C=C_{0,n}$ in  \cite{Pa} where it was shown that they are related to 
the isomonodromic tau-functions. 
This offers an alternative approach to the relation 
between free fermion partition functions and isomonodromic tau-functions
expressed in \rf{tau-ff}. 

A  solution to the  Riemann Hilbert problem has first been constructed using fermionic twist fields
in \cite{SMJ}, and the relation to conformal field 
theory was previously discussed in \cite{Mo}.
\end{rem}

\subsection{Issues to be addressed}

Two points should be noted at this stage: 
First, let us note that the Riemann-Hilbert correspondence relates 
the moduli space $\CM_{\rm flat}(C_{0,n})$ of flat
connections $\pa_y-A(y)$ on $C_{0,n}$ to the character
variety
$\CM_{\rm ch}(C_{0,n})={\rm Hom}(\pi_1(C_{0,n}),{\rm SL}(2,\BC))/{\rm SL}(2,\BC)$.
The definition above therefore defines the tau-function as
a function of two types of data: The variables $\mathbf{z}$ specifying the complex structure 
of $C$, and the monodromy data $M$, represented by the matrices $M_r$ appearing in the
Riemann-Hilbert problem. 
Picking a parameterisation $M_r=M_r(\mu)$, $\mu=(\mu_1,\dots,\mu_{2n-6})$, 
of the monodromy data $M_r$
is equivalent to introducing coordinates $\mu$ for the character variety. Doing this will allow us to 
represent the tau-functions as actual functions $\mathcal{T}(\mu,\mathbf{z})$ 
depending on two types of variables. 
The identification of the tau-function $\mathcal{T}(\mu,\mathbf{z})$ with the free fermion partition function 
$Z_{\rm ff}(\xi,t;\la)$ must therefore involve a map between the variables $(t,\xi)$ and the 
geometric data $(\mu,\mathbf{z})$ that needs to be determined.

Second, the definition above defines the tau-function up to multiplication with 
functions of the monodromy data which do not depend on $\mathbf{z}$.  For the time being we will call
a tau-function {\it any} function $\mathcal{T}(\mu,\mathbf{z})$
satisfying $H_r=\pa_{z_r}\log\mathcal{T}(\mu,\mathbf{z})$, $r=1,\dots,n-3$.
We will later find 
natural ways to fix this ambiguity. Remarkably it will turn out that the choice of 
coordinates $\mu$ for $\CM_{\rm ch}(C_{0,n})$ will determine natural ways for fixing the 
normalisation of $\mathcal{Z}_{\rm ff}(\mu,\mathbf{z})$.

\section{Factorising the tau-functions}\label{sec:factor}
\setcounter{equation}{0}

The definition of the free fermion partition functions given in the previous section, elegant as it may be, is 
not immediately useful for computations. Recently it has  
been shown in \cite{GIL,ILT} how to  compute 
the series expansions for the isomonodromic tau-functions $\CT(\mu,\mathbf{z})$ 
in cross-ratios of the positions $z_r$ explicitly. This result has been re-derived in 
\cite{GL} by a different method which can be seen as a special case of the general 
relations between Riemann-Hilbert factorisation problems and tau-functions discussed in \cite{CGL}. 

In this section we are going  to explain how the existence of 
the combinatorial expansions found  in 
the references above is naturally explained from the theory of free chiral fermions.
The factorisation over a complete set of intermediate states will lead to expressions which 
in the case $C=C_{0,4}$ take the schematic form
\begin{equation}\label{tauexp-schem}
\CT(\si,\kappa;z)=\sum_{n\in\BZ}e^{in\kappa}\CT_n(\si;z).
\end{equation}
This  will allow us to determine the precise relation between the variables $\si,\kappa$ in \rf{tauexp-schem} and 
certain coordinates for the moduli space $\CM_{\rm flat}(C_{0,4})$ of flat $SL(2)$-connections on 
$C_{0,4}$, addressing one of the main issues formulated at the end of  Section
\ref{Sec:Dmod-tau}.

\subsection{Coordinates from factorisation of Riemann-Hilbert problems}\label{coordinates1}

Let us first discuss how the   factorisation of Riemann-Hilbert problems
leads to the definition of coordinates for the space of monodromy data. Within this subsection
we will specialise to the case $N=2$.

\subsubsection{Fenchel-Nielsen type coordinates}

Useful sets of coordinates for
$\CM_{\rm ch}(C_{g,n})$ are  given by the trace functions
$L_{\ga}:=\operatorname{\rm tr}\rho(\ga)$ associated to 
simple closed curves $\ga$ on $C_{g,n}$ \cite{Go}.
Conjugacy classes of irreducible representations of $\pi_1(C_{0,4})$ are uniquely specified by
seven conjugation  invariants
\begin{subequations}
\begin{align}\label{Mk}
&L_k=\operatorname{Tr} M_k=2\cos2\pi \theta_k,\qquad k=1,\ldots,4,\\
&L_s=\operatorname{Tr} M_1 M_2,\qquad L_t=\operatorname{Tr} M_1 M_3,\qquad L_u=\operatorname{Tr} M_2 M_3,
\end{align}
\end{subequations}
generating the algebra of invariant polynomial functions on $\CM_{\rm char}(C_{0,4})$. 
These trace functions satisfy the quartic equation
\begin{align}
 \label{JFR}
& L_1L_2L_3L_4+L_sL_tL_u+L_s^2+L_t^2+L_u^2+L_1^2+L_2^2+L_3^2+L_4^2=\\
 &\nonumber \quad=\left(L_1L_2+L_3L_4\right)L_s+\left(L_1L_3+L_2L_4\right)L_t
+\left(L_2L_3+L_1L_4\right)L_u+4.
\end{align}
For fixed choices of $\theta_1,\ldots,\theta_4$ in \rf{Mk}
one may use
equation \rf{JFR} to describe the character variety as a cubic surface in $\BC^3$.
This surface
admits a parameterisation in terms of coordinates $(\si,\tau)$ of the form
\begin{align}\label{FNcoord}
L_s\,=\,2\cos 2\pi \si\,,\qquad
\begin{aligned}
& (2\sin (2\pi  \si))^2\,L_t\,=\,
C_{t}^+(\si)\,e^{i\kappa}+C_t^0(\si)+C_{t}^-(\si)\,e^{-i\kappa}\,,\\
&  (2\sin (2\pi  \si))^2\,L_u\,=\,
C_{}^+(\si)\,e^{i\kappa}+C_u^0(\si)+C_{}^-(\si)\,e^{-i\kappa}\,,
\end{aligned}
\end{align}
where $C_t^{\pm}(\si)=-C^{\pm}(\si)e^{\pm 2\pi\,\mathrm{i}\,\si}$,
\begin{equation}
\begin{aligned}
C_t^0(\si)&=L_s(L_2L_3+L_1L_4)-{2 }(L_1L_3+L_2L_4)\,\\
C_u^0(\si)&=L_s(L_1L_3+L_2L_4)-{2 }(L_2L_3+L_1L_4).
\end{aligned}
\end{equation}
Equation \rf{JFR} only constrains the product $C_{}^+(\si)C_{}^-(\si)$, leaving the freedom
to trade a redefinition of $\kappa$ in \rf{FNcoord} for a redefinition of $C_{}^+(\si)$ and $C_{}^-(\si)$
which leaves $C_{}^+(\si)C_{}^-(\si)$ unchanged. We will in the rest of this subsection discuss 
natural ways to fix this ambiguity.
The coordinates defined in this way will be called
coordinates of Fenchel-Nielsen type.

\subsubsection{Factorising Riemann-Hilbert problems}\label{RH-factor}

Let us assume $|z|<1$.
We may represent the surfaces $C_{0,4}=\mathbb{P}^1\setminus\{0,z,1,\infty\}$ 
by gluing two  three-punctured spheres $C^{\rm in}$ and 
$C^{\rm out}$. Let us represent both $C^{\rm in}$ and 
$C^{\rm out}$ as $\mathbb{P}^1\setminus\{0,1,\infty\}$, and let 
$A^{\rm in}=\{x\in C^{\rm in} ;|1|<|x|<|z|^{-1}\}$
and  $A^{\rm out}=\{x\in C^{\rm out} ;|z|<|x|<1\}$ 
be annuli in $C^{\rm in}$ and 
$C^{\rm out}$, respectively. By identifying points $x$ in $A^{\rm in}$ with points 
$x'$ in $A^{\rm out}$ iff $x'=zx$ one recovers the Riemann surface  $C_{0,4}$
from $C^{\rm in}$ and 
$C^{\rm out}$.

Having
represented the Riemann surface $C_{0,4}$ by means of the gluing construction 
there is an obvious way to define Riemann-Hilbert problems for $C^{\rm in}$
and $C^{\rm out}$ using the matrices $M_1,M_{2}$ and 
$M_{3},M_4$, respectively. 
A solution $\Psi(x)$ to the Riemann-Hilbert problem on $C_{0,4}$ allows us to 
define solutions $\Psi^{\rm in}(x)$ and $\Psi^{\rm out}(x)$ 
to the corresponding 
Riemann-Hilbert problems on the open surfaces  
$D^{\rm in}=\{x\in \BC;|x|<|z|^{-1}\}$ and $D^{\rm out}=\{x\in\mathbb{P}^1;|x|>|z|\}$
in an obvious way, setting $\Psi^{\rm out}(x)=\Psi(x)T^{\rm in}$ on $D^{\rm out}$ and
$\Psi^{\rm in}(x)=\Psi(zx)T^{\rm out}$ on $D^{\rm in}$, with $T^{\rm in}, T^{\rm out}
\in\mathrm{SL}(2,\BC)$ being fixed matrices describing a possible change of normalisation 
condition in the definition of the Riemann-Hilbert problems on $C^{\rm in}$ and $C^{\rm out}$.
By choosing $T^{\rm in}$, $T^{\rm out}$ appropriately we can get functions 
$\Psi^{\rm in}(x)$ and $\Psi^{\rm out}(x)$ both having diagonal monodromy along the 
boundary circles of $D^{\rm in}$ and $D^{\rm out}$, respectively. The matrices $T^{\rm in}, T^{\rm out}$
which ensure this condition can only differ by a diagonal matrix, leading to 
a relation of the form 
$ 
\Psi^{\rm in}(x)=\Psi^{\rm out}(zx)T,
$ 
for $x\in A$.

Coordinates for the moduli space of flat connections $\CM_{\rm flat}(C_{0,4})$ 
can then be obtained by choosing a
parameterisation for the two pairs of matrices $(M_1,M_{2})$ and $(M_3,M_4)$, 
and using the parameter $\kappa$ for the family of matrices
$T_\kappa=\mathrm{diag}(e^{\mathrm{i}\kappa/2},e^{-\mathrm{i}\kappa/2})$ 
as a complementary coordinate for $\CM_{\rm flat}(C_{0,4})$. 
An equivalent representation can be  obtained by 
trading a nontrivial choice
of the matrix $T$ for an overall conjugation of $M_{1},M_2$ by $T$. 
It will be convenient to consider $\Psi^{\rm in}_{z,\kappa}(x):=\Psi^{\rm in}(x/z)T^{-1}$ 
instead of  $\Psi^{\rm in}(x)$, which is related to  $\Psi^{\rm out}(x)$ simply as
$ 
\Psi^{\rm in}_{z,\kappa}(x)=\Psi^{\rm out}(x)
$ 
for $x\in A$.

\subsubsection{Coordinates from the gluing construction}

Representing $C=C_{0,4}$ by the gluing construction as described in Section \ref{RH-factor}
one needs the solutions of the Riemann-Hilbert problem for $C^{\rm in}\simeq C_{0,3}$ and $
C^{\rm out}\simeq C_{0,3}$. It is a classical result that the solutions to the Riemann-Hilbert problem
on $C_{0,3}$ can be expressed through the hypergeometric function.
We may, in particular, choose $\Psi^{\rm out}$ as
$\Psi^{\rm out}(x)=\big(\begin{smallmatrix} \chi'_+ & \chi'_- \\ \chi_+ & \chi_- \end{smallmatrix}\big)$, with
\begin{equation}\label{Psioutdef}
\begin{aligned}
\chi_{\ep}(x)=\nu_\ep^{\rm out}\, x^{\frac{1}{2}+\ep(\si-\frac{1}{2})}(1-x)^{\frac{1}{2}+\ep\theta_3}F(A_\ep,B_\ep,C_\ep;x),
\end{aligned}
\end{equation} 
for $\ep=\pm 1$, where $\nu_\ep^{\rm out}$ are normalisation factors 
to be specified later, $F(A,B,C;x)$ is the Gauss hypergeometric function and
\begin{equation}\label{ABC}
\begin{aligned}
A_+=A, \quad A_-=1-A,\\
B_+=B, \quad B_-=1-B,\\
\end{aligned}\qquad
\begin{aligned} &C_+=C,\\ & C_-=2-C,
\end{aligned}
\qquad
\begin{aligned}
A&=\theta_3+\theta_4+\si,\\
B&=\theta_3-\theta_4+\si,
\end{aligned}\qquad C=2\si.
\end{equation}
$\Psi^{\rm in}$, on the other hand, may be chosen as
$
\Psi^{\rm in}=\big(\begin{smallmatrix} \xi'_+ & \xi'_- \\ \xi_+ & \xi_- \end{smallmatrix}\big)$, 
where $\xi_{\ep}(x)$ are obtained from $\chi_\ep(x)$ by the replacements 
$x\ra x^{-1}$, $\theta_4\ra \theta_1$, $\theta_3\ra\theta_2$ and $\ep\ra -\ep$.

The well-known formulae for the monodromies of the hypergeometric function then yield, in particular, formulae
for the monodromy $M_3^{\rm out}$ of $\Psi^{\rm out}(x)$ around $z_3=1$ of the form
\begin{equation}\label{M3out}
M_3^{\rm out}=\left(\begin{matrix} \ast & \mu_3^+ \\ \mu_3^- & \ast\end{matrix}\right),
\quad \mu_3^\ep=-\ep \,\left(\frac{\nu_+^{\rm\sst out}}{\nu_-^{\rm\sst out}}\right)^{-\ep}\!
\frac{2\pi\mathrm{i}\,\Ga(C_\ep)\Ga(C_\ep-1)}
{\Ga(A_\ep)\Ga(B_{\ep})\Ga(C_\ep-A_\ep)\Ga(C_\ep-B_{\ep})}.
\end{equation}
A similar formula gives the monodromy $M_2^{\rm in}$ of $\Psi^{\rm in}(x)$ around $1$. 
Keeping in mind the set-up introduced in Section \ref{RH-factor} it is easy to see that $\mathrm{tr}(M_2M_3)$ gets represented as
\begin{equation}\label{trM2M3}
\mathrm{tr}(M_2M_3)=\mathrm{tr}(T^{-1}M_2^{\rm in}TM_3^{\rm out})=
e^{\mathrm{i}\kappa}\mu_2^-\mu_3^++e^{-\mathrm{i}\kappa}\mu_2^+\mu_3^-+N_0,
\end{equation}
where $N_0$ is $\kappa$-independent, and $T=\mathrm{diag}(e^{\mathrm{i}\kappa/2},e^{-\mathrm{i}\kappa/2})$. 
The parameters $(\si,\kappa)$ introduced in this way 
represent coordinates for $\CM_{\rm flat}(C_{0,4})$ of Fenchel-Nielsen type. From 
equations \rf{M3out} and \rf{trM2M3} it is easy to see, in particular, that the definition 
of the coordinate $\kappa$ is directly linked to the choice of normalisation factors  $\nu_\pm^{\rm out}$, $\nu_\pm^{\rm in}$ 
in the definition of $\Psi^{\rm out}$, $\Psi^{\rm in}$. It is furthermore natural to require 
that the determinants of $\Psi^{\rm out}(x)$ and $\Psi^{\rm in}(x)$ are 
equal to $1$, fixing $\nu_+^{\rm out}\nu_-^{\rm out}$ and 
$\nu_+^{\rm in}\nu_-^{\rm in}$ to be equal to $\frac{1}{1-2\si}$, and  leaving us with one undetermined 
normalisation constant.

Two choices appear to be particularly natural from this point of view. One may, on the one hand, choose 
$\nu_+^{\rm out}=1$, $\nu_-^{\rm in}=1$  
in order to ensure that the coefficients appearing in the series expansions of $\Psi^{\rm in}(x)$
and $\Psi^{\rm out}(x)$ are rational functions of $\si$, $\theta_i$, $i=1,\dots,4$.  In that case we easily
see that $C_{}^{\pm}(\si)=C_{\rm r}^{\pm}(\si)$, with 
\begin{align}\label{Cepdef1}
C_{\rm r}^{\pm}(\si)& = \frac{(2\pi)^2\,\Ga(1\pm(2\si-1))^4}{ 
\prod_{s,s'=\pm 1}\Ga\big(\frac{1}{2}\pm\big(\si-\frac{1}{2}\big)+s\theta_1+s' \theta_2\big)\Ga\big(\frac{1}{2}\pm\big(\si-\frac{1}{2}\big)+s\theta_3+s' \theta_4\big)}.
\end{align}

The normalisation factors $\nu_\pm^{\rm out}$ can alternatively be chosen such that
$\mu_3^+=1$, which gives
\begin{equation}\label{mu2+mu3-}
(2\sin(2\pi\si))^2\mu_3^-=-\prod_{s,s'=\pm 1} \,2\sin\pi(\si+s\theta_1+s' \theta_2).
\end{equation}
Adopting an analogous choice for $\nu_\pm^{\rm in}$ leads to $C_{}^+(\si) = 1$ and 
\begin{align}\label{Cepdef2}
(2\sin(2\pi\si))^4C_{}^{-}(\si)&=\prod_{s,s'=\pm 1} \,2\sin\pi(\si+s\theta_1+s' \theta_2)\,2\sin\pi(\si+s\theta_3+s' \theta_4)\\
&= (L_s^2+L_1^2+L_2^2-L_sL_1L_2-4)(L_s^2+L_3^2+L_4^2-L_sL_3L_3-4).
\notag\end{align}
It is worth noting that $C_{}^{\pm}(\si)$ are rational in $L_s$ in this parameterisation.

\subsection{Factorisation of free fermion conformal blocks}

We had previously observed that the free fermion state $\mathfrak{f}_\Psi$ associated with the solution $\Psi$ of
the Riemann-Hilbert problem on $C$ defines a conformal block of the free fermion vertex
algebra on $C$. A standard construction in conformal field theory allows us to represent 
conformal blocks on Riemann surfaces $C$ obtained by gluing two surfaces 
$C^{\rm in}$ and $C^{\rm out}$ in terms of the conformal blocks associated to 
$C^{\rm in}$ and $C^{\rm out}$, respectively. 
Adapting this construction to our case will allow us to represent  the 
free fermion partition functions as overlaps
of the form
\begin{equation}\label{taufactor}
\mathcal{Z}_{\rm ff}(\mu,{z})=
{\big\langle \,\mathfrak{f}_{\rm out}^{\ast}\,,\,
\mathfrak{f}_{\rm in}^{}\,\big\rangle_{\CF}^{}},
\end{equation}
where $\mathfrak{f}_{\rm out}$, $\mathfrak{f}_{\rm in}$ are states in the free fermion Fock space 
defined by factorising the RH problem along a contour $\ga$ separating 
$C$ into two open surfaces $C^{\rm out}$ and $C^{\rm in}$ as described in Section \ref{RH-factor}.
The representation \rf{taufactor} for $\mathcal{Z}_{\rm ff}(\mu,{z})$
can be used to calculate  the free fermion partition functions more explicitly.

\subsubsection{Twisted representations}

As a further preparation we will need to generalise the construction from Section \ref{FF-RH} a bit. 
We will need twisted representations $\CF_\si$ of the free fermion algebra 
labelled by
a tuple $\si=(\si_1,\dots,\si_N)\in\BC^N$ where the fermions have 
non-trivial monodromy around $x=0$,
\begin{equation}
\psi_t(x)=\sum_{n\in\BZ}\psi_{t,n} x^{-n-1+\si_t}\,,\quad
\bar\psi_s(x)=\sum_{n\in\BZ} \bar\psi_{s,n} x^{-n-\si_s}\,,
\end{equation}
with $s,t=1,\dots,N$. The twist fields describing such representations can be conveniently 
described by means of bosonisation. To this aim let us introduce $N$ free bosonic fields, 
\begin{equation}
\phi_s(x)=\mathsf{q}_s+\mathsf{p}_s \log x+\mathrm{i}\sum_{n\neq 0}\frac{1}{n}a_{s,n}x^{-n},
\end{equation} 
$s=1,\dots,N$, having modes satisfying the commutation relations
\begin{equation}
[\mathsf{q}_r,\mathsf{p}_s]=\frac{\mathrm{i}}{2}\de_{r,s},\qquad [a_{r,m},a_{s,n}]=\frac{m}{2}\de_{r,s}\de_{n,-m}.
\end{equation}
We will consider Fock space representation $\CV_{\mathbf{p}}$  labelled by
a tuple $\mathbf{p}=(p_1,\dots,p_N)$ generated from vectors 
${v}_{\mathbf{p}}$ satisfying
\begin{equation}
a_{n,s}\,{v}_{\mathbf{p}}=0,\quad n>0, \qquad \mathsf{p}_s \,{v}_{\mathbf{p}}=p_s\,
{v}_{\mathbf{p}},\qquad
e^{2\mathrm{i}\delta \mathsf{q}_s}{v}_{\mathbf{p}}={v}_{\mathbf{p}-\delta \mathbf{e}_s},
\end{equation}
for all $s=1,\dots,N$, with $\mathbf{e}_s$ being the unit vector having $1$ at the $s$-th component, and $\de\in\BR$.

The direct sum of Fock spaces 
\begin{equation}
\CF_{\si}=\bigoplus_{\mathbf{n}\in\frac{1}{2}\BZ^{N}}\CV_{\si+\mathbf{n}},
\end{equation} 
is a representation of the free fermion VOA generated 
by the fields
\begin{equation}
\psi_s(x)=:e^{\mathrm{i}\phi_s(x)}:\,,\qquad
\bar{\psi}_s(x)=:e^{-\mathrm{i}\phi_s(x)}:\,,
\end{equation}
from the vector $\mathfrak{f}_{\si}\equiv {v}_{\mathbf{\si}}$ satisfying the usual highest weight conditions.
As before we may introduce a conjugate right module $\CF_{\si}^*$. The spaces $\CF_{\si}^*$ 
and $\CF_\si$ are naturally paired by the 
bilinear form 
$\langle .,.\rangle_{\CF_\si}^{}:\CF_{\si}^*\otimes\CF_\si\ra \BC$ defined in the same way as 
previously done for
$\si=0$. 

\subsubsection{Representing conformal blocks within twisted representations}

The construction of free fermion states corresponding to the solutions of the Riemann-Hilbert 
problem described in Section \rf{FFfromD} can now easily be generalised to the cases where 
one of the points at which $\Psi(x)$ can be singular is equal to $0$ or $\infty$. 
We will look for a state  $\mathfrak{f}_{\Psi,\si}^{}\in\CF$ characterised through the 
matrix $G_\Psi(x,y)$ of two-point  functions  with matrix elements
\begin{equation}\label{FFtwopt2}
G_\Psi(x,y)_{st}=\big\langle \,\bar{\psi}_s(x)\psi_t(y) \,\big\rangle_{\Psi} \equiv
\frac{\langle \,\mathfrak{f}_{\si}^{}\,,\,\bar{\psi}_s(x)\psi_t(y)\,\mathfrak{f}_{\Psi,\si}^{}\,\rangle_{\CF_\si}^{}}{\langle \,
\mathfrak{f}_{\si}^{}\,,\,\mathfrak{f}_{\Psi,\si}^{}\,\rangle_{\CF_\si}^{}}.
\end{equation}
However, in order to apply \rf{CBfromR} and \rf{A-exp} 
we now need to use a modified form of the relation between the two-point function 
and the function $A(x,y)$,  taking into account that $\Psi(x)= \Phi(1/x)x^D$ near $x=\infty$,
with $D$ being the the diagonal matrix $D=\mathrm{diag}(\si_1,\dots,\si_N)$,
and $\Phi(x)$ regular at $x=0$. It follows 
that $A(x,y)$ can be introduced via
\begin{equation}
G_{\Psi}(x,y)=\frac{(\Psi(x))^{-1}\Psi(y)}{x-y}=x^{-D}\left(\frac{{1}}{x-y}+A(x,y)\right)y^{D}\,,
\end{equation}
 In a similar way one may define a state $\mathfrak{f}_{\Psi,\si}^*\in\CF_\si^*$ such that
\begin{equation}\label{FFtwopt3}
G_\Psi^{}(x,y)_{st}=\big\langle \,\bar{\psi}_s(x)\psi_t(y) \,\big\rangle_{\Psi}^{} \equiv
\frac{\langle \,\mathfrak{f}_{\Psi,\si}^{*}\,,\,\bar{\psi}_s(x)\psi_t(y)\,\mathfrak{f}_\si^{}\,\rangle_{\CF_\si}^{}}{\langle \,
\mathfrak{f}_{\Psi,\si}^{*}\,,\,\mathfrak{f}_\si^{}\,\rangle_{\CF_\si}^{}}.
\end{equation}
The states $\mathfrak{f}_{\Psi,\si}^{}$ and 
$\mathfrak{f}_{\Psi,\si}^*$ are as before defined uniquely  up to normalisation.

\subsubsection{Factorisation of free fermion conformal blocks}

Using these constructions, and referring back to the factorisation of the Riemann-Hilbert problem 
described in Section \ref{RH-factor}, we can now associate a state 
$\mathfrak{f}_{\rm in}^{}\equiv \mathfrak{f}_{\rm in}^{}(z,\kappa)\in\CF_{\si}$ to 
$\Psi^{\rm in}_{z,\kappa}$, and a state 
$\mathfrak{f}_{\rm out}^*\in\CF_{\si}^*$ to $\Psi^{\rm out}$. Using
the variable $z$ as coordinate for $\CM_{0,4}$ in the case $C=C_{0,4}$ 
one may, on the other hand, use \rf{FSconn} to define the family of states $\mathfrak{f}_{\Psi}^{}({z})$ up to 
a $z$-independent normalisation factor. 
We claim that  $\mathfrak{f}_{\Psi}^{}({z})$ can be normalised  in such a way that we have
\begin{equation}\label{taufactor2}
\mathcal{Z}_{\rm ff}(\mu,\mathbf{z})=
\big\langle\,\mathfrak{f}_{0}^{}\,,\,\mathfrak{f}_{\Psi}^{}({z})\big\rangle_{\CF}^{}=
{\big\langle \,\mathfrak{f}_{\rm out}^{\ast}\,,\,
\mathfrak{f}_{\rm in}^{}(z,\kappa)\,\big\rangle_{\CF_\si}^{}}.
\end{equation}
In \cite[Appendix G]{CLT} it is explained how the relation \rf{taufactor2} can be derived using 
ideas from conformal field theory. It basically represents the free fermion conformal block $\mathfrak{f}_{\Psi}^{}({z})$ 
by the 
gluing construction from CFT associated to the decomposition of $C$ into $C^{\rm in}$ and $C^{\rm out}$ 
described in Section \ref{RH-factor}. It is well-known (see e.g. \cite{T17a}) that the gluing construction defines families of conformal
blocks satisfying \rf{FSconn}. It follows from \rf{taufactor2} and \rf{tau-ff} that
\begin{equation}\label{matel-tau}
{\big\langle \,\mathfrak{f}_{\rm out}^{\ast}\,,\,
\mathfrak{f}_{\rm in}^{}(z,\kappa)\,\big\rangle_{\CF_\si}^{}}=N(\mu)
\mathcal{T}(\mu;z),
\end{equation}
with $\mathcal{T}(\mu;z)$ being the isomonodromic tau-function.

\subsection{Factorisation expansions}\label{sec:fact-exp}

It is furthermore explained in Appendix \ref{SW-app}
how to represent the matrix element occurring in \rf{taufactor2} in terms of the Fredholm determinant 
\begin{equation}\label{matel-Fred}
\mathcal{T}\big(\si,\kappa\,;\,\underline{\theta}\,;\,z\big):=\frac{\big\langle \,\mathfrak{f}_{\rm out}^{\ast}\,,\,
\mathfrak{f}_{\rm in}^{}\,\big\rangle_{\CF_\si}^{}}
{\big\langle \,\mathfrak{f}_{\rm out}^{\ast}\,,\,\mathfrak{f}_0^{}\,\big\rangle_{\CF_\si}^{}
\big\langle \,\mathfrak{f}_0^{}\,,\,\mathfrak{f}_{\rm in}^{}\,\big\rangle_{\CF_\si}^{}}
=
\mathrm{det}(1+\mathsf{A}^{\rm out}\mathsf{A}^{\rm in}),
\end{equation}
with $\mathsf{A}^{\rm in}$ being the operator represented by the matrices
${A}^{\rm in}_{kl}$ defined from $\Psi^{\rm in}_{q,\kappa}$ by first defining $A^{\rm in}(x,y)$
from 
\begin{equation}\label{Psiin-Ain}
\frac{(\Psi^{\rm in}_{q,\kappa}(x))^{-1}\Psi^{\rm in}_{q,\kappa}(y)}{x-y}=
x^{-D}\left(\frac{{1}}{x-y}+A^{\rm in}(x,y)\right)y^{D}\,,
\end{equation}
and then expanding $A^{\rm in}(x,y)$ in a double series of the form \rf{G-exp}. 
The operator $\mathsf{A}^{\rm out}$ is defined in an analogous way.
According to \rf{matel-tau} one may identify the function
$\mathcal{T}(\si,\kappa;\underline{\theta};z)$ as the isomonodromic tau-function
defined with a specific choice of normalisation condition.
Representing ${\big\langle \,\mathfrak{f}_{\rm out}^{\ast}\,,\,
\mathfrak{f}_{\rm in}^{}\,\big\rangle_{\CF_\si}^{}}$
in terms of a Fredholm determinant makes it manifest, in particular, that 
$\mathcal{Z}_{\rm ff}(\mu,\mathbf{z})$ is mathematically well-defined.

Standard identities for determinants allow us to express $\mathrm{det}(1+\mathsf{A}^{\rm out}\mathsf{A}^{\rm in})$
as sum over products of sub-determinants of the infinite matrices formed out of the matrices 
${A}^{\rm in}_{kl}$ 
and ${A}^{\rm out}_{kl}$, respectively, see \cite{CGL} or Appendix \ref{App:Fredholm} for more details.
In this way it is not hard to see that in the case $C=C_{0,4}$ equation
\rf{matel-Fred}  yields series expansions of the following form:
\begin{equation}\label{det-exp}
\mathcal{T}\big(\si,\kappa\,;\,\underline{\theta}\,;\,z\big)
=
\sum_{n\in\BZ} \,e^{in\kappa}\sum_{m=0}^\infty z^{m}\,
\CR_{n,m}(\si,\underline{\theta}),
\end{equation}
where $\underline{\theta}=(\theta_1,\dots,\theta_4)$.
To understand this structure it may be useful to recall that the matrix elements ${A}^{\rm in}_{kl}$ 
of $\mathsf{A}^{\rm in}$ 
are  $2\times 2$-matrices in the case $N=2$ of our interest. 
It easily follows from the discussion in Section \ref{RH-factor} together with \rf{Psiin-Ain} 
that the dependence of 
the  $2\times 2$-matrices ${A}^{\rm in}_{kl}$ on $\kappa$ is for all $k,l$
given by the same factors $e^{\pm \mathrm{i}\kappa}$ in 
the off-diagonal matrix elements of ${A}^{\rm in}_{kl}$. It follows easily that the summation 
index $n$ simply counts the difference of numbers of upper- and lower off-diagonal elements of 
matrices ${A}^{\rm in}_{kl}$ in the sub-determinants appearing in the expansion of 
$\mathrm{det}(1+\mathsf{A}^{\rm out}\mathsf{A}^{\rm in})$.  

One should furthermore note that
$\langle \mathfrak{f}_0^{},\mathfrak{f}_{\rm in}^{}\big\rangle_{\CF_\si}^{}$
has a dependence on $z$ of the form
\begin{equation}
\big\langle \,\mathfrak{f}_0^{}\,,\,\mathfrak{f}_{\rm in}^{}\,\big\rangle_{\CF_\si}^{}=
N_{\rm in}\,z^{\si^2-\theta_1^2-\theta_2^2},
\end{equation}
as follows from the relation between $\mathfrak{f}_{\rm in}^{}$ and 
a conformal block on $C_{0,3}=\mathbb{P}^1\setminus\{\infty,z,0\}$
using the conformal Ward identities, $N_{\rm in}$ being a constant.

The discussion in this section clarifies in particular how the normalisation factors 
$\nu_\ep^{\rm out}$ entering  the definition 
of $\Psi^{\rm out}(x)$,  $\Psi^{\rm in}(x)$ given in Section \ref{coordinates1} via equation \rf{Psioutdef} 
determine unambiguously both (i)  the precise definition of the variable
$\kappa$ in \rf{det-exp},  and (ii) how $\kappa$ is related to the coordinates for 
$\CM_{\rm flat}(C_{0,4})$ defined in Section \ref{coordinates1}. 
A canonical choice is of course $\nu_\ep^{\rm out}=1$ corresponding to the coordinates
$(\si,\kappa)$ defined in Section \ref{coordinates1} using 
\rf{FNcoord} together with formula \rf{Cepdef1} for $C_{}^{\pm}(\si)$. In this case one will 
get an expansion of the form \rf{det-exp} with coefficients $\CR_{n,m}(\si,\underline{\theta})$
which are rational functions of $(\si,\underline{\theta})$. This follows easily 
from the fact that the matrix elements of 
${A}^{\rm in}_{kl}$ 
and ${A}^{\rm out}_{kl}$ are assembled from the power series expansion coefficients of the 
hypergeometric function, which are rational functions of $(\si,\underline{\theta})$.

\begin{rem}
The resulting picture is closely related to the one drawn in \cite{GM,CGL}.
Indeed, using the basic results from the theory of chiral free fermions summarised in Appendix \ref{SW-app}
one may recognise the Fredholm determinants discussed in \cite{CGL} 
as the free fermion matrix elements appearing here.
A more direct proof that the Fredholm determinant on the right of 
\rf{matel-Fred}
is the isomonodromic tau-function can be found in \cite{CGL}. 
The normalisation prescription following from the definition 
\rf{matel-Fred} of the tau-functions is equivalent
to the one used in \cite{ILP}. 
\end{rem}

\section{Representing free fermion partition functions as generalised theta-series}\label{theta-series}
\setcounter{equation}{0}

The results of the last section imply that $\mathcal{Z}_{\rm ff}(\si,\kappa;z):=\big\langle \,\mathfrak{f}_{\rm out}^{\ast}\,,\,
\mathfrak{f}_{\rm in}^{}\,\big\rangle_{\CF_\si}^{}$
can be expanded as\footnote{The dependence on $\underline{\theta}=(\theta_1,\dots,\theta_4)$ is temporarily 
suppressed in the notations.}
\begin{equation}
\mathcal{Z}_{\rm ff}(\si,\kappa;z)=\sum_{n\in\BZ} \,e^{in\kappa}
\CF_{n}(\si,z).
\end{equation}
This expansion has a  form consistent with 
the string duality conjectures discussed in \cite{DHSV} if the Fourier coefficients
$\CF_{n}(\si,z)$  can represented by a function $\CF(\si,z)$ such that
$\CF_{n}(\si,z)=\CF(\si+n,z)$. In that case one
would expect that $\CF(\si,z)$ can be identified with the topological string 
partition function. 

So far we had not fixed a normalisation for the states $\mathfrak{f}_{\rm out}^{\ast}$ and 
$\mathfrak{f}_{\rm in}^{}$, leaving the 
normalisation factors  
$N_{\rm out}=\langle \mathfrak{f}_{\rm out}^{\ast},\mathfrak{f}_0^{}\rangle_{\CF_\si}^{}$
and 
$N_{\rm in}=z^{\theta_1^2+\theta_2^2-\si^2}
\langle \mathfrak{f}_0^{},\mathfrak{f}_{\rm in}^{}\rangle_{\CF_\si}^{}
$ 
entering the relation \rf{matel-Fred}
between free fermion partition partition functions
and Fredholm determinants arbitrary up to now.
Considering generic choices for 
$N_{\rm out}$ and $N_{\rm in}$ we will observe that the 
free fermion partition functions 
$\mathcal{Z}_{\rm ff}(\si,\kappa;z)$ do {\it not} admit
series expansions of the desired form. 

However, we will also see that there exist
a few distinguished choices for $N_{\rm out}\equiv N_{\rm out}(\mu)$
and $N_{\rm in}\equiv N_{\rm in}(\mu)$  such that
the free fermion partition functions 
$\mathcal{Z}_{\rm ff}(\si,\kappa;z)$
admit series expansions of the required form
for suitable choices of coordinates $\mu=\mu(\si,\kappa)$.

\subsection{Explicit form of the factorisation expansion}

Explicit series expansions for the isomonodromic tau functions have first been conjectured in
\cite{GIL}. Proofs of this conjecture were given in 
\cite{ILT}, \cite{BS} and \cite{GL} by rather different methods. The proof closest to the formulation used
in this paper is the one in \cite{GL}. It proceeds by explicit calculation 
of the determinant on the right side of \rf{matel-Fred} using an expansion 
as sum over sub-determinants. After stating 
the result we will discuss some of its features that will be important in the 
following. 

The result of \cite{GIL,ILT,BS,GL} can be written as follows:\footnote{Comparing \rf{ILT-exp1} with
the results of \cite{GIL,ILT,GL} it may be helpful to take the discussion in Sections \ref{rewr-theta} and \ref{alt-theta} 
into account. {Appendix B in \cite{CLT} explains in detail how to derive \rf{ILT-theta} from the main result
of \cite{ILT}}. Formula \rf{ILT-exp1} is equivalent, and more directly related to 
the discussion in Section \ref{sec:factor}.} 
\begin{equation}\label{ILT-exp1}
\mathcal{T}\big(\si,\kappa\,;\,\underline{\theta}\,;\,z\big)=
\sum_{n\in\BZ} \,e^{in\kappa}\,F_n(\,\si\,,\,\underline{\theta}\,)
\CF(\,\si+n\,,\,\underline{\theta}\,;\,z\,),
\end{equation}
using the definitions
\begin{itemize}
\item The variables $\si$ and $\kappa$ are the coordinates for $\CM_{\rm ch}(C_{0,4})$ 
which are defined in Section \ref{coordinates1} using 
\rf{FNcoord} with the normalisation choice giving formula \rf{Cepdef1} for $C_{}^{\pm}(\si)$.
\item 
The functions $F_n(\,\si\,,\,\underline{\theta}\,)$ can be represented as 
\begin{align}
F_n(\,\si\,,\,\underline{\theta}\,)& =\frac
{\prod_{\ep,\ep'=\pm} H_n(\si+\ep\theta_2+\ep'\theta_1)
H_n(\si+\ep\theta_3+\ep'\theta_4)}
{(H_{2n}(2\si))^2},
\notag\end{align}
where $H_n(\si)$, $n\in\BZ$, is the family of functions defined as 
\begin{equation}
H_n(\si)=\frac{G(1+\si+n)}{G(1+\si)(\Ga(\si))^n},
\end{equation}
with $G(p)$ being the Barnes $G$-function satisfying $G(p+1)=\Ga(p)G(p)$. 
Note that $F_n(\si,\underline{\theta})$ are for all $n\in\BZ$  {\it rational} functions of $\si$, as predicted by the discussion
in Section \ref{sec:fact-exp}.
\item
$\CF(\,\si\,,\,\underline{\theta}\,;\,z\,)$ can be represented by a power series of the following form
\begin{equation}\label{Virblock}
\CF(\,\si\,,\,\underline{\theta}\,;\,z\,)=z^{\si^2-\theta_1^2-\theta_2^2}
(1-z)^{2\theta_2\theta_3}\sum_{\xi,\zeta\in\mathbb{Y}}z^{|\xi|+|\zeta|}\CF_{\xi,\zeta}(\si,\underline{\theta}),
\end{equation}
where the summation is extended over pairs $(\xi,\zeta)$ of partitions, 
and $|\xi|$ is the number of boxes in the Young diagram representing the partition $\xi$. The explicit formulae for the 
coefficients $\CF_{\xi,\zeta}(\si,\underline{\theta})$ can be found in \cite{GIL,GL}, where it is also 
observed that they are related to the instanton partition functions in the four-dimensional, $\CN=2$-supersymmetric
$SU(2)$-gauge theory with four flavors.
\end{itemize}
The
normalisations in \rf{ILT-exp1} are fixed such that $\CF_{\emptyset,\emptyset}(\si,\underline{\theta})=1$. 

\subsection{Rewriting as generalised theta series}\label{rewr-theta}

The string dualities discussed in \cite{DHSV} suggest that the relevant fermionic partition 
functions should admit an expansion taking the form \rf{thetaseries0} 
of a generalised theta series. Formula \rf{ILT-exp1} is not of this form, the summand 
depends on the variable $\si$ not only in the combination $\si+n$.
One may note, however, that there are two types of ambiguities involved in the 
definition of the partition functions in general, and in the form of its series 
expansion in particular:
\begin{itemize}
\item
There is generically a normalisation freedom in the definition
of partition functions. While the dependence w.r.t. the variable $z$ is governed by Ward identities, mathematically
expressed in the definition \rf{taudef} of the isomonodromic
tau-function, the normalisation ambiguities leave the freedom
to multiply the tau-function by a function depending only on the monodromy data. 
\item 
The coefficients in the series expansions of tau-functions like \rf{ILT-exp1} depend on the 
precise definition of the coordinate $\kappa$. A change of coordinates from $(\si,\kappa)$ to 
$(\si,\tau)$ with $\tau$ satisfying
$e^{i\kappa}=e^{i\tau}D(\si,\underline{\theta})$ would change the coefficients $F_n$ in 
\rf{ILT-exp1} by a factor of $(D(\si,\underline{\theta}))^n$.
\end{itemize}
By combining these observations we will find a renormalised version of the tau-functions 
which will indeed admit an expansion of generalised theta-series type.

To this aim let us note that the change of variables from the coordinates $(\si,\kappa)$
defined through \rf{FNcoord} with $C^\pm_{\rm r}(\si)$ given in  \rf{Cepdef1} 
to coordinates $(\si,\tau)$  with $C^\pm_{}(\si)$ given in  \rf{Cepdef2}  is such that
\begin{equation}
e^{i\tau}=
e^{i\kappa}\frac{(2\pi)^2(\Ga(2\si))^4}{ 
\prod_{s,s'=\pm 1}\Ga(\si+s\theta_1+s' \theta_2\big)\Ga(\si+s\theta_3+s' \theta_4)}.
\end{equation}
Rewriting \rf{ILT-exp1} in terms of $\tau$ therefore yields the expansion
\begin{equation}\label{ILT-exp1'}
\mathcal{T}\big(\si,\kappa\,;\,\underline{\theta}\,;\,z\big):=\,
\sum_{n\in\BZ} \,e^{in\tau}\,G_n(\,\si\,,\,\underline{\theta}\,)
\CF(\,\si+n\,,\,\underline{\theta}\,;\,z\,),
\end{equation}
where the coefficients $G_n$ can be represented in the form 
\begin{equation}
G_n(\,\si\,,\,\underline{\theta}\,)=\frac{N(\sigma+n,\theta_4,\theta_3)N(\sigma+n,\theta_2,\theta_1)}
{N(\sigma,\theta_4,\theta_3)N(\sigma,\theta_2,\theta_1)},
\end{equation}
with 
\begin{equation}\label{N-Def}
N(\theta_3,\theta_2,\theta_1)\,=\,\frac{\prod_{\ep,\ep'=\pm}G(1+\theta_3+\ep \theta_2+\ep'\theta_1)}{G(1+2\theta_3)G(1+2\theta_2)G(1+2\theta_1)G(1)}.
\end{equation}
The structure of the right hand side of formula \rf{ILT-exp1'} now suggests to define 
\begin{equation}\label{Trenorm}
\mathcal{Z}\big(\si,\tau\,;\,\underline{\theta}\,;\,z\big)=
{N(\sigma,\theta_4,\theta_3)N(\sigma,\theta_2,\theta_1)}\mathcal{T}\big(\si,\kappa\,;\,\underline{\theta}\,;\,z\big),
\end{equation}
which
can indeed be represented in the form of a theta-series. 
\begin{equation}\label{ILT-theta}
\mathcal{Z}\big(\si,\tau\,;\,\underline{\theta}\,;\,z\big)=
\sum_{n\in\BZ} \,e^{in\tau}\,
\CG(\,\si+n\,,\,\underline{\theta}\,;\,z\,),
\end{equation}
with
$
\CG(\si,\underline{\theta};z)=
{N(\sigma,\theta_4,\theta_3)N(\sigma,\theta_2,\theta_1)}
\CF(\si,\underline{\theta};z).
$
Comparing \rf{ILT-theta} with equation \rf{thetaseries0} one may be tempted to identify
the functions $\CG(\si,\underline{\theta};z)$  in \rf{ILT-theta}
as a promising candidate for 
the topological string partition functions.

\subsection{Alternative representations as theta series} \label{alt-theta}


In this subsection we will identify a small family of normalisation conditions 
defining tau-functions
sharing the feature to admit an expansion as a theta-series of the form \rf{thetaseries0}. We will
observe that all normalisation conditions in this class are obtained
from \rf{ILT-theta}
by combining a redefinition of the 
normalising factors $N(\theta_3,\theta_2,\theta_1)$ with a modification of the 
definition of the coordinate $\tau$. Each such choice of 
normalisation thereby corresponds to a particular set of coordinates 
for the space of monodromy data.

To find alternatives to the expansion \rf{ILT-theta}
let us consider 
the possibility to 
replace the functions $N(\theta_3,\theta_2,\theta_1)$  in \rf{Trenorm}
by 
functions $N'(\theta_3,\theta_2,\theta_1)$ such that, for example, 
\begin{equation}\label{M-N}
N(\theta_3,\theta_2,\theta_1)={\prod_{\ep=\pm}S(\theta_3+\ep(\theta_2-\theta_1))}
N'(\theta_3,\theta_2,\theta_1),
\end{equation}
where $S(x)$ is the special function 
$S(x)=(2\pi)^{-x}\frac{G(1+x)}{G(1-x)}.$
Noting that the function $S(x)$ satisfies the functional relation
\begin{equation}
S(x\pm 1)=\mp\left(2\sin\pi x \right)^{\mp 1}S(x)\,,
\end{equation}
we find  the relation
\begin{align}\label{NtoM}
\frac{N(\sigma+n,\theta_4,\theta_3)N(\sigma+n,\theta_2,\theta_1)}
{N(\sigma,\theta_4,\theta_3)N(\sigma,\theta_2,\theta_1)}
&=\frac{N'(\sigma+n,\theta_4,\theta_3)N'(\sigma+n,\theta_2,\theta_1)}
{N'(\sigma,\theta_4,\theta_3)N'(\sigma,\theta_2,\theta_1)}\\
&\times\prod_{\ep=\pm}\Big[2\sin\pi(\si+\ep(\theta_2-\theta_1))2\sin\pi(\si+\ep(\theta_4-\theta_3))\Big]^{-n}.
\notag\end{align}
Introducing a new coordinate 
$\tau'=\tau(\si,\tau)$ which is defined such that
\begin{align}\label{tau'tau}
e^{i\tau}&=e^{i\tau'}\,\prod_{\ep=\pm}\,
{2\sin\pi(\si+\ep(\theta_1-\theta_2))}\,
{2\sin\pi(\si+\ep(\theta_3-\theta_4))},
\end{align}
along with 
\begin{equation}
\mathcal{Z}'\big(\si,\tau'(\si,\tau)\,;\,\underline{\theta}\,;\,z\big)=
\frac
{N(\sigma,\theta_4,\theta_3)N(\sigma,\theta_2,\theta_1)}
{N'(\sigma,\theta_4,\theta_3)N'(\sigma,\theta_2,\theta_1)}\,
\mathcal{Z}\big(\si,\tau\,;\,\underline{\theta}\,;\,z\big)
\end{equation}
we see that $\mathcal{Z}'(\si,\tau';\underline{\theta};z)$ also has a
representation as a generalised theta series, 
\begin{equation}\label{ILT-exp2}
\mathcal{Z}'\big(\si,\tau'\,;\,\underline{\theta}\,;\,z\big)=
\sum_{n\in\BZ} \,e^{in\tau'}\,
\CG'(\,\si+n\,,\,\underline{\theta}\,;\,z\,).
\end{equation}
It is clear the partition functions $\mathcal{Z}'(\si,\tau';\underline{\theta};z)$
and $\mathcal{Z}(\si,\tau;\underline{\theta};z)$ 
differ by a factor only depending on monodromy data, 
We conclude that a change of normalisation of  the partition functions 
correlated with the change of  coordinates $(\si,\tau)\ra(\si,\tau')$ may preserve the 
feature that the partition function can be represented as
a generalised theta-series. 

There are, of course, a few other options similar to  \rf{M-N} one might consider. 
It is natural to restrict attention to redefinitions of the function $N(\theta_3,\theta_2,\theta_1)$ in 
order to preserve a form of the expansion adapted to the pants decomposition it corresponds to. 
By redefinitions similar to \rf{M-N} one can change the sign of the argument of each of the 
four $G$-functions appearing in \rf{N-Def}, giving sixteen options in total.


The  free fermion partition functions defined in the previous 
sections depend on the choice of normalisation factors $\nu_+^{\rm out}$ and $\nu_-^{\rm in}$ for the solutions to the 
Riemann-Hilbert problems on $C^{\rm out}$ and $C^{\rm in}$ introduced in Section \ref{coordinates1}, respectively.
We have  seen that the requirement that the free fermion partition functions admit an 
expansion of generalised theta-series type restricts the normalisation 
freedom considerably. Only very special choices of possibly monodromy-dependent
normalisation factors have this property. However, the requirement to have
a generalised theta series expansion does not fix the normalisation choice uniquely,
there is a fairly small family of choices 
which all yield expansions of theta series type.

\section{Comparison with topological vertex calculations} \label{TopVert}

We will now compare our findings to an alternative computation of the topological 
string partition function which can be done with the help of the topological vertex \cite{AKMV}.
The topological string partition functions have been computed previously for the case of our interest
in \cite{IK03a,IK03b,EK,HIV,IK04}. 
A key issue for our goals is the dependence of the partition functions  on the choice of a chamber. 
The chambers are related flop transitions. The changes of the partition functions associated to flop
transitions were previously studied in \cite{IK04,KM}. We will now summarise 
the relevant results.

The 
partition functions ${Z}^{\rm top}_{\mathfrak{i},\mathfrak{j}}$ associated to the 
chambers $\mathfrak{C}_{\mathfrak{i},\mathfrak{j}}$ 
can be represented the form 
\begin{equation}
\label{Ztop-pertinst}
{Z}^{\rm top}_{\mathfrak{i},\mathfrak{j}}=z^{\si^2-\theta_1^2-\theta_2^2}
{Z}^{\rm out}_{\mathfrak{i}}\, {Z}^{\rm in}_{\mathfrak{j}}\,
Z^{\rm inst}_{},
\end{equation}
  The factor denoted $Z^{\rm inst}$ in \rf{Ztop-pertinst} is known in the literature  as the 
   five dimensional Nekrasov instanton partition function \cite{N}.  This part is 
   independent of the choice of a chamber.
   Of main interest for us are the factors ${Z}^{\rm out}_{\mathfrak{i}}$,  ${Z}^{\rm in}_{\mathfrak{j}}$. 
   In the case $(\mathfrak{i},\mathfrak{j})=(\1,\1)$ corresponding to the toric diagram depicted 
in Figure \ref{fig:4flavSQCDweb}   we find, for example,
   \begin{equation}\label{5d3pt}
   Z^{\rm out}_{\1}  
   =\frac{\mathcal{M}(Q_F)\mathcal{M}(Q_{3}Q_{4}Q_F)}
      {\prod_{i=3}^4\mathcal{M}\big(Q_{i}\big)\mathcal{M}\big(Q_{i}Q_F\big)},
   \qquad
   Z^{\rm in}_{\1}  
   =\frac{\mathcal{M}\big(Q_F\big)\mathcal{M}(Q_{1}Q_{2}Q_F)}
   {\prod_{i=1}^2\mathcal{M}\big(Q_{i}\big)\mathcal{M}\big(Q_{i}Q_F\big)}.
\end{equation}
$\mathcal{M}(Q)$ in \rf{5d3pt} is defined as 
$\mathcal{M}(Q) 
\equiv (Q  q ;q, q)_{\infty}^{-1}
$, with $q=e^{-\la R}$ and $(Q   ;t, q)_{\infty}$ being
\begin{equation}
\label{eq:defcalMeverywhere}
(Q   ;t, q)_{\infty}=
\prod_{i,j=0}^{\infty}(1-Q t^{i} q^j) \quad \text{ for }\; |t|<1,\; | q|<1.
\end{equation}

\begin{figure}[t]
 \begin{center}
  \includegraphics[width=90mm,clip]{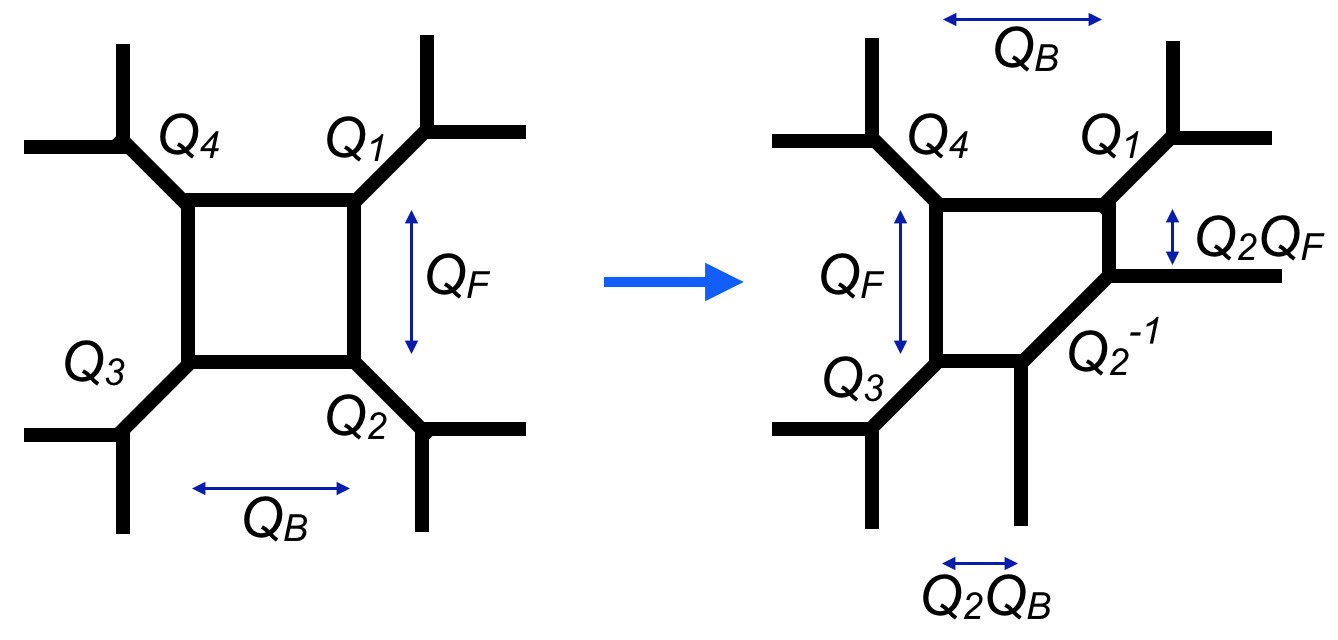}
 \end{center}
 \caption{\it Two toric diagrams related via flopping both engineering  $SU(2)$ superconformal QCD with $N_f=4$ fundamental hypermultiplets.}
 \label{fig:FloppingThe4pointFunction}
\end{figure}

An example for a pair of
toric diagrams related by a flop is depicted in Figure 
\ref{fig:FloppingThe4pointFunction}. The effect of flops on the  partition functions can be 
described
by the relations
\begin{equation}\label{floprel}
\begin{aligned}
{Z}^{\rm in}_{\2} &=  \CR(Q_2)\,Z^{\rm in}_{\1}, \\
{Z}^{\rm in}_{\3} &=  \CR(Q_1)\,Z^{\rm in}_{\2},
\end{aligned}
\qquad 
\begin{aligned}
{Z}^{\rm out}_{\2} &=  \CR(Q_3)\,Z^{\rm in}_{\1}, \\
{Z}^{\rm out}_{\3} &=  \CR(Q_4)\,Z^{\rm in}_{\2},
\end{aligned}
\qquad
\CR(Q)=\frac{\mathcal{M}(Q)}{\mathcal{M}(Q^{-1})}
\,.
\end{equation}
We have thereby  defined $Z^{\rm top}$ as a 
piecewise analytic function
on the union of the chambers $\mathfrak{C}_{ \mathfrak{i}\mathfrak{j}}$.

 
Taking the limit $R\ra 0$ of the functions ${Z}^{\rm top}_{\mathfrak{i},\mathfrak{j}}$ with fixed $m_1,\dots,m_4,a,\la$
is delicate as the functions $\CM(Q)$ diverge in this limit. In order analyse the limit we may use the relation
\begin{equation}\label{M-qBarnes}
\frac{\CM(q^u)}{\CM(1)}=\frac{(q;q)_\infty^{u}(1-q)^{-\frac{1}{2}(u-1)u}}{G_q(1+u)},
\end{equation}
with  $G_q(u)$ being the q-Barnes function which 
has a  limit $\lim_{q\ra 1}G_q(u)=G(u)$ \cite{UN}.  
Note that $(q;q)_\infty$ behaves as $(q;q)_\infty\sim 
e^{-\frac{\pi\mathrm{i}}{12\tau}}
(-i\tau)^{-\frac{1}{2}}$ for $q=e^{2\pi\mathrm{i}\tau}$, $\tau\ra 0$.

However, there exist natural renormalisation prescriptions allowing us to define
meaningful quantities ${\CZ}^{\rm top}_{\mathfrak{i},\mathfrak{j}}$ which can be associated to ${Z}^{\rm top}_{\mathfrak{i},\mathfrak{j}}$ in the limit $R\ra 0$.
Applying \rf{M-qBarnes} for the analysis of the limit of 
${Z}^{\rm top}_{\mathfrak{i},\mathfrak{j}}$ is simplest for $\mathfrak{i}=\mathfrak{j}=2$.
Using \rf{M-qBarnes} one may show that 
\begin{equation}
\CZ^{\rm top}_{\2,\2}=\lim_{R\ra 0} \tilde{Z}^{\rm top}_{\2,\2},\qquad
\tilde{Z}^{\rm top}_{\2,\2}:=\CM(e^{-2Ra_2})\CM(e^{-2Ra_3})(\CM(1))^2\, {Z}^{\rm top}_{\2,\2},
\end{equation}
exists. It follows from the existence of a limit for 
$\tilde{Z}^{\rm top}_{\2,\2}$
that the singular behavior of  ${Z}^{\rm top}_{\2,\2}$ does not depend on the variables 
$\si=-\frac{1}{\la R}\log Q_F$ and $z=Q_B$. While there do exist alternatives for the definition 
of finite quantities from ${Z}^{\rm top}_{\2,\2}$ in the limit $R\ra 0$, it is unnecessary and unnatural to 
introduce extra factors altering the dependence on $\si$ and $z$ in the definition of this limit. A more detailed discussion of these issues can be found in \cite{CPT19}.

This motivates us to define the  four-dimensional limit 
of $Z^{\rm top}_{\2,\2}$ as follows:
\begin{equation}\label{Ztopformula}
\CZ^{\rm top}_{\2,\2}(\,a\,,\,\underline{m}\,;\,z\,)=
\CZ^{\rm out}_{\2}\,\CZ^{\rm in}_{\2}\,
\CF(\,\si\,,\,\underline{\theta}\,;\,z\,),
\end{equation}
where $\CF(\,\si\,,\,\underline{\theta}\,;\,z\,)$ is defined in \rf{Virblock},
the K\"ahler parameters of $X$
are related to 
$\si$ and $\theta_1,\dots,\theta_4$ as
$
\si=a/\la$, $\theta_i=m_i/\la,$ for $i=1,\dots,4
$,
and 
$\CZ^{\rm out}_{\2}=M(\sigma,\theta_4,\theta_3)$,
$\CZ^{\rm in}_{\2}=M(\sigma,\theta_2,\theta_1)$
with
\begin{align}\label{Mdef}
M(\theta_3,\theta_2,\theta_1)=\frac{ 
\prod_{s=\pm}G(1+s\theta_3+\theta_2+\theta_1)G(1+\theta_3+s(\theta_2-\theta_1))}
{G(1+2\theta_3)G(1+2\theta_2)G(1+2\theta_1)G(1)}.
\notag\end{align}

It seems furthermore natural to extend the
the definition of ${\CZ}^{\rm top}_{\mathfrak{i},\mathfrak{j}}$ to all $\mathfrak{i},\mathfrak{j}=\1,\2,\3$
by demanding that we have a natural analog of the relations \rf{floprel} describing 
the effects of flops,
\begin{equation}\label{floprel-4d}
\begin{aligned}
{\CZ}^{\rm in}_{\2} &=  R_\rho(Q_2)\,\CZ^{\rm in}_{\1}, \\
{\CZ}^{\rm in}_{\3} &=  R_\rho(Q_1)\,\CZ^{\rm in}_{\2},
\end{aligned}
\qquad 
\begin{aligned}
{\CZ}^{\rm out}_{\2} &=  R_\rho(Q_3)\,Z^{\rm in}_{\1}, \\
{\CZ}^{\rm out}_{\3} &=  R_\rho(Q_4)\,Z^{\rm in}_{\2},
\end{aligned}
\end{equation}
with $R_\rho(x)=\rho^{x}\frac{G(1-x)}{G(1+x)}$ being a renormalised version of the 
limit $q\ra 1$ of $\CR(q^x)$. The factor $\rho^{x}$ reflects the usual ambiguity in 
the renormalisation of 
$(1-q)^x=
(1-e^{-\la R})^x$ for $R\ra 0$ resulting from the possibility to
redefine  $R\ra \rho R$.

The results display a simple pattern. The normalisation factors $\CZ^{\rm in}$ and $\CZ^{\rm out}$ can 
be one of the functions $N_i=N_i(\vartheta_3,\vartheta_2,\vartheta_1)$,  $i=1,2,3$, defined 
as
\begin{equation}
N_i(\vartheta_3,\vartheta_2,\vartheta_1)=\frac{ 
\prod_{s,s'=\pm}G(1+\vartheta_i+s\vartheta_{i+1}+s'\vartheta_{i+2})}
{G(1+2\vartheta_3)G(1+2\vartheta_2)G(1+2\vartheta_1)G(1)}, \qquad \vartheta_{i+3}\equiv\vartheta_i,
\end{equation}
or $N_s(\vartheta_3,\vartheta_2,\vartheta_1)=M(\vartheta_3,\vartheta_2,\vartheta_1)$. 
The functions $N_i$, $i=1,2,3,s$, are assigned to the region in the parameter space under consideration 
in such a 
way that the arguments in the $G$-functions appearing in the expressions for the functions $N_i$ 
are {\it always positive}.  

Remembering the definition of the chambers 
$\mathfrak{C}_{\mathfrak{i},\mathfrak{j}}$, and taking into account the relations
$\si=a/\la$ and $\theta_i=m_i/\la$ one finds for each
chambers $\mathfrak{C}_{\mathfrak{i},\mathfrak{j}}$ a unique
region defined by the conditions  that all of the arguments
of the functions $N_i$, $i=1,2,3,s$, appearing in the expression for 
${\CZ}^{\rm top}_{\mathfrak{i},\mathfrak{j}}$
are positive for all values of $\la$.  This becomes 
a one-to-one correspondence if one extends the topological vertex results to the cases where 
$a_2>a_1$ and $a_3>a_4$ not considered yet by the necessary modifications. 

Comparing with the discussion in Section \ref{theta-series}
we may finally observe that the partition functions 
characterised by  these normalisation factors all represent Fourier coefficients
in the generalised theta series expansions of suitably normalised free
fermion partition functions. Keeping in mind that there was a 
direct correspondence between the normalisation factors defining such partition functions 
on the one hand, and choices of Darboux coordinates on the other hand, 
makes it natural to ask if there is a simple  relation
between the chambers in the parameter space considered above and the relevant 
choices of Darboux coordinates.

\section{Abelianisation}\label{sec:abel}

In Section \ref{theta-series} we had observed a direct relation between 
choices of coordinates for $\CM_{\rm flat}(C_{0,4})$ and
normalisations of the tau-functions. It was next found that the
theta series coefficients of tau-functions for certain choices of coordinates 
are equal to
the topological string
partition functions for the local CY manifolds of class $\Sigma$. 
These functions change from chamber to chamber in the complex structure 
moduli space. We will now see that there is a natural way 
to describe the dependence on the choice of a chamber. It turns out that the relevant coordinates  
for $\CM_{\rm flat}(C)$ can be defined using a construction introduced in \cite{GMN12} and further 
developed in \cite{HN}. Following \cite{HN} we will refer this construction as abelianisation. We will
begin this section with a  brief review of this construction following \cite{HN,HK}. 

\subsection{Spectral Networks}\label{sec:spectralNet}

The curves $\Sigma$ defined in Subsection \ref{Curves} as covers of Riemann surfaces $C$ were specified by quadratic 
differentials $q=q(x)d^2 x$ in equations \eqref{nf4-curve}-\eqref{C03curve}. 
The \emph{spectral network} $\cW_\theta(q)$ for $q$ is a graph on $C$ having oriented edges called \emph{walls} 
which are defined by $\text{Im}\left(e^{i\theta}\int_a^x\sqrt{q(x')}dx'\right)=0$  and which meet at vertices, the zeros 
$a$ of $q$. Three walls emanate from the branch points where two sheets meet.  It is useful to choose a set of branch 
cuts on $C$ and labels for the sheets of $\Sigma$, such that each wall of the network is labeled by an ordered pair of integers $ij$ 
corresponding to the sheets. Given a positively oriented tangent vector $v$ to the wall and 
$\theta\in\mathbb{R}/2\pi\mathbb{Z}$, the wall carries the label $ji$ if $e^{-i\theta} (y_i-y_j)(v)\in\mathbb{R}_+$ and $ij$ 
if $e^{-i\theta} (y_i-y_j)(v)\in\mathbb{R}_-$. 
For special values of $q$ and $\theta$, two walls $ij$ and $ji$ can overlap and create a double wall. 
When this occurs there exist two possible \emph{resolutions}, which are the infinitesimal ways of displacing the walls with 
respect to each other. When $q$ obeys the Strebel condition $\oint_{\gamma_i}\sqrt{q(x)}dx\in\mathbb{R}_+$ on curves 
$\{\gamma_i\}$ defining a pants decomposition of $C$, the corresponding spectral network is a \emph{Fenchel-Nielsen 
(FN) network} and composed of double walls only. A FN-network defines a pants decomposition of $C$, since its restriction 
to every three-punctured sphere in this decomposition is a FN-network.

\begin{figure}[h]
\centering
\includegraphics[width=\textwidth]{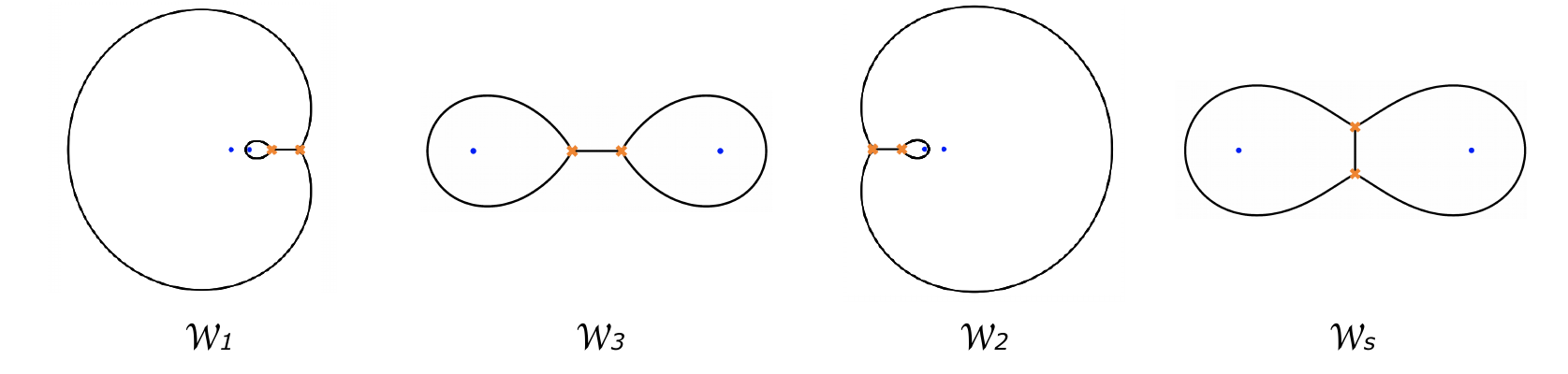}
\caption{\it FN-networks on $C_{0,3}$, with punctures depicted at positions $z=0,1$. These isotopies 
occur in different regions of the parameter space and correspond to the possible choices of chamber $\mathfrak{C}_{\al}$ 
in Section \ref{ExtKaehler}. Network $\cW_1$ corresponds to chamber $\mathfrak{C}^{\rm in}_\1$, while $\cW_3$ 
corresponds to $\mathfrak{C}^{\rm in}_\3$ and $\cW_s$ to $\mathfrak{C}^{\rm in}_\2$. 
The triplets of parameters $\{a_1,a_2,a\}$ in equation \eqref{C03curve} take here the values: 
$\cW_1$) $\{$0.51, 0.32, 0.18$\}$, 
$\cW_3$) $\{$0.49, 0,48, 1$\}$ 
$\cW_2$) $\{$0.32, 0.51, 0.18$\}$ 
and 
$\cW_s$) $\{$0.51, 0.5, 1$\}$. 
}
\label{C03Networks}
\end{figure}

When the base curve is $C_{0,3}=\mathbb{P}^1\backslash\{0,1,\infty\}$, the quadratic differential defining the branched 
cover $\Sigma\to C_{0,3}$ was given in equation \eqref{C03curve}. There exist two inequivalent types of FN-networks: 
$\cW_i$, $i=1,2,3$, are  called type II molecules in Figure \ref{C03Networks} and $\cW_s$ is a type I molecule\footnote{
These figures have been plotted using the Mathematica package \cite{Neitzke:plotter}.}. 
For each of these topologies, there is a choice for the resolution: \emph{British}, where the outer walls of the network 
are oriented clockwise 
or \emph{American}, for counter-clockwise orientation. 
We will here focus on the case where all the parameters $a$ are real. Branch points will either be real, or come in 
complex conjugate pairs. The transitions between different types of molecules occur when two branch points coalesce, 
corresponding to the flop transitions discussed in Section \ref{ExtKaehler}. The branch points $t_{\pm}$ are easily read 
off from equation \rf{Sigma_in} and show that flop transitions occur when $a^2=(a_1+ a_2)^2$ and $a^2=(a_1- a_2)^2$. 
Therefore a molecule changes its isotopy class when $a$ crosses any of the planes $a=\pm a_1\pm a_2$. 
Comparing with Section \ref{sec:curves} one may note that such changes
directly correspond to changes between the chambers in the complex structure moduli space 
defined there. 

On general Riemann surfaces $C$, FN-networks can be defined with respect to pants decompositions 
found by gluing together molecules in the same resolution on the individual pants.

\subsection{$\mathcal{W}$-framed flat connections on $C$} \label{abelflat}

The construction called abelianisation uses the spectral networks
defined by a quadratic differential $q$ to construct a natural one-to-one correspondence
between flat $SL(2)$-connections on $C$ and an (almost-)flat $GL(1)$-connections on the two-fold cover 
$\pi:\Sigma\ra C$ defined by the differential $q$. 
Describing $\Sigma$ as a branched cover of $C$ will then allow us
to define $\mathrm{GL}(1)$-connections $\nabla^{\rm ab}$ on $\Sigma$ from which one can recover all 
flat $\mathrm{SL}(2)$-connections $\nabla$ on $C$ by the construction sketched below.
We are here interested in
flat $SL(2)$-connections $\nabla$ in a complex vector bundle $E$ over a Riemann surface $C$ with fixed 
conjugacy class represented by $D_k = \text{diag}(e^{2\pi i \theta_k}, e^{-2\pi i \theta_k})$ at the $k^\text{th}$ puncture.

\begin{figure}[h]
\centering
\includegraphics[width=1.0\textwidth]{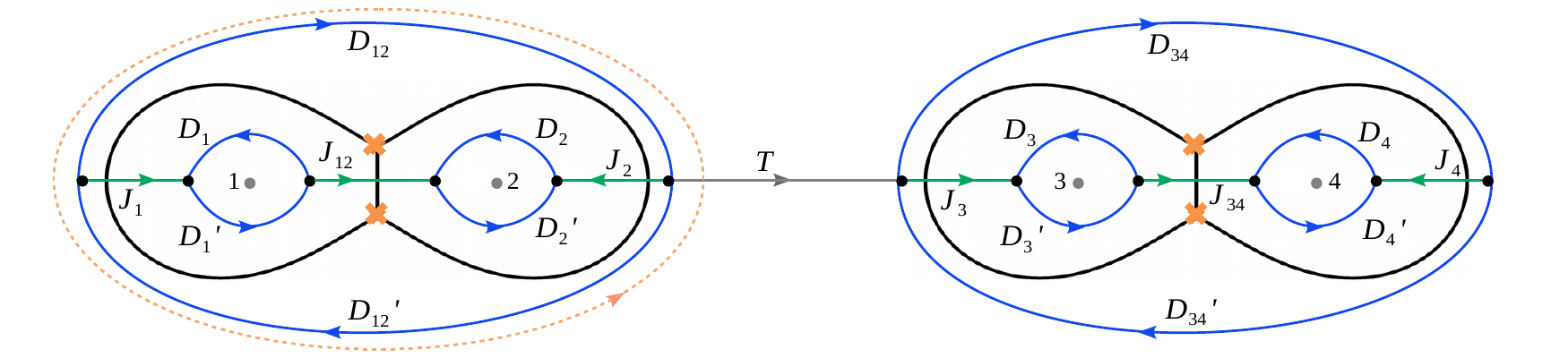}
\caption{\it Fenchel-Nielsen network on the four-punctured sphere.}
\label{C04Networks-IIV4}
\end{figure}

We shall focus on the the cases where the spectral network has type FN. 
A FN-network  $\cW$  decomposes $C$ into annular regions $A_i$ within which 
$\nabla$ can be diagonalised. 
Let $\cG_C$ 
be the set of paths $\wp$ up to homotopy between base points 
marked by black 
dots in the example depicted in Figure \ref{C04Networks-IIV4}
on each side of the walls $w\in\cW$. A path $\wp\in\cG_C$ is ``short" if it does not cross any walls. 

In order to construct $\nabla$ from the connection $\nabla^{\rm ab}$ on $\Sigma$ 
one may start by considering the connection $\pi_\ast\nabla^{\rm ab}$ on the surface $C'$
obtained from $C$ by removing the branch points of $\Sigma$.
Choosing a basis $(s_1,s_2)$ of $E$ at any point on $C\backslash \cW$, the parallel transport of $(s_1,s_2)$ 
with respect to $\pi_\ast\nabla^{\rm ab}$, with $\pi:\Sigma\ra C$ being the covering projection,
along short paths $\wp\subset C\backslash \cW$ is represented by: i)   a matrix $D_\wp=\text{diag}(d_\wp,d_\wp^{-1})$
for $\wp$ not crossing a branch cut
within a pair of pants,  ii)  a matrix $\tilde{D}_{\wp}=D_{\wp}
\big(\begin{smallmatrix} 0 & 1 \\  -1 & 0 \end{smallmatrix}\big)$
for $\wp$ intersecting a branch cut from a simple branch point, and
iii) a matrix $T_\wp=\text{diag}(e^{\pi i\eta},e^{-\pi i\eta})$ 
for $\wp$ traversing an annulus between pairs of pants. The data $d_\wp$ and $\eta$ characterise 
a flat abelian connection $\nabla^{\rm ab}$
on $\Sigma'\setminus \pi^{-1}(\cW)$, with $\Sigma'$ being the complement of the branch points of $\Sigma$. It has been observed in \cite{HN} that 
$\nabla^{\rm ab}$ automatically extends over $\Sigma'$. 

The connection $\pi_\ast\nabla^{\rm ab}$ is almost-flat in the
sense that the holonomy around any branch point is $-1$.
The freedom in the choice of the matrices $D_\wp$  is constrained by the conditions fixing  the holonomies 
around the boundary components. The remaining freedom in the choice of the parameters $d_\wp$
is related to the abelian gauge freedom at the base points. We will describe this in two relevant examples
below.

The abelian connection  $\nabla^{\rm ab}$ can be turned into a non-abelian connection 
$\nabla$ by replacing all products of matrices representing holonomies of $\pi_{\ast}\nabla^{\rm ab}$
by products obtained by splicing in certain triangular matrices for each segment of the path crossing 
a wall. 
Across a single wall, the non-abelian parallel transport of $(s_1,s_2)$ is represented by a triangular jump matrix $J$, whose 
precise form depends on the decoration assigned to the wall \cite{HK}. 
The off-diagonal entries are determined uniquely in 
terms of the matrices $D_\wp$ by the consistency conditions stating
that for every path which is contractible to a turning point 
(marked in orange in Figure \ref{C04Networks-IIV4}), the parallel transport is represented by the identity matrix \cite{HN}.
More details are given below for the cases of our interest. 
 
Note that the map from the path groupoid $\cG_C$ to the corresponding $\mathrm{SL}(2)$ matrices is an anti-homomorphism.
For the composition $\wp=\wp_1\wp_2$ of a path $\wp_1$ from point $i_1$ to $i_2$ with a path $\wp_2$ from $i_2$ to $i_3$ 
one multiplies the holonomy matrices $H_{\wp_1}$ and $H_{\wp_2}$ as $H_\wp=H_{\wp_2}H_{\wp_1}$.

\subsection{Four-punctured sphere -- an example}\label{sec:C04FN}

We will now review how abelianisation works in the case $C=C_{0,4}$, following the previous discussions
in \cite{HN,HK}. Our goal will be to exhibit the residual ambiguities in the definition of 
FN-type coordinates, and
to discuss natural ways to fix them.

\begin{figure}[h]
\centering
\includegraphics[width=1.0\textwidth]{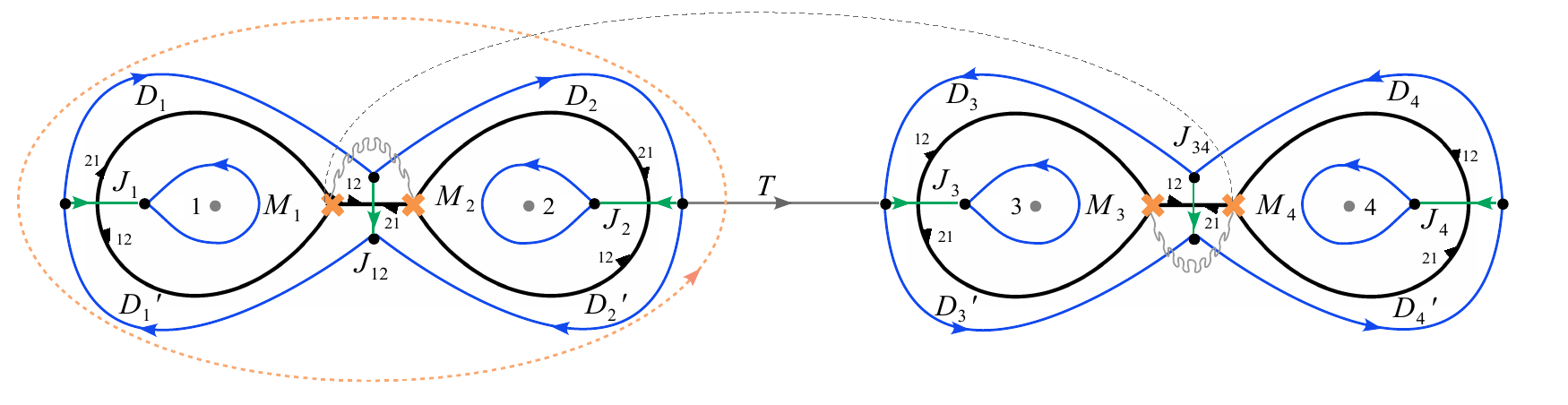}
\caption{\it FN-network on $C_{0,4}$, formed by type-II molecules in the British resolution.}
\label{C04Networks-I-V5}
\end{figure}

As a first example we shall consider the network depicted in Figure \ref{C04Networks-I-V5}.
The $\mathrm{GL}(1)$-connection defined on the cover $\Sigma$ of $C$ is fully characterised
by the matrices $D_\varpi$ associated to the blue paths $\varpi$ in Figure \ref{C04Networks-I-V5}.

Fixing the monodromy around the punctures implies relations of the form
\begin{equation} 
M_i = \left( \begin{matrix} {m}_i & 0 \\ 0 & {m}_i^{-1} \end{matrix} \right) ~, \quad i=1,2,
\quad
D_2 D_1 D_1' D_2' = \left( \begin{matrix} {m}_\alpha^{-1} & 0 \\ 0 & {m}_\alpha \end{matrix} \right)~,
\end{equation}
with ${m}_i=e^{2\pi\mathrm{i}\theta_i}$ for $i=1,2,3,4$ and ${m}_\al=e^{2\pi\mathrm{i}\si}$.
Parameterising the matrices $D_i$, $D_i'$ by complex numbers $d_i$, $d_i'$ for 
$i=1,2,3,4$ leads to the relation $d_1'd_2d_2'/d_1=-{m}_\al^{-1}$, and a similar relation for 
$d_i$, $d_i'$, $i=3,4$. 
 There is an obvious ambiguity in the 
choice of the parameters $d_i$, $i=1,2,3,4$, related to the freedom in the 
choice of the trivialisations at the base points.

This ambiguity affects the definition of FN-type coordinates. Assuming an arbitrary 
choice of the parameters $d_i$, $d_i'$, $i=1,2,3,4$ it is straightforward to find the 
corresponding FN-type coordinates as follows.  In order to 
reconstruct the non-abelian $\mathrm{SL}(2)$-connection $\nabla$ on $C$ from the given 
$\mathrm{GL}(1)$-connection $\nabla^{\rm ab}$ 
one mainly needs to find the 
jump matrices associated to the green paths crossing the walls. 
The jump matrices $J_1$, $J_2$, $J_{12}$ associated to the left part of the network in 
Figure \ref{C04Networks-I-V5} will be parameterised as 
\beq\label{eq:wallXingMatricesC03}
J_i=\left( \begin{matrix} 1 & 0 \\ \tilde{c}_i & 1 \end{matrix} \right)
               \left( \begin{matrix} 1 & c_i \\ 0 & 1 \end{matrix} \right) ~, \quad i=1,2,\quad
J_{12}=\left( \begin{matrix} 1 & c_{12} \\ 0 & 1 \end{matrix} \right)
                \left( \begin{matrix} 1 & 0 \\ \tilde{c}_{12} & 1 \end{matrix} \right)~.
\end{equation}
These matrices are required to 
satisfy the following constraints
\beq 
D_1' J_{12} D_1 J_1^{-1} M_1 J_1 = {\bf 1} ~, \quad 
D_2 J_{12}^{-1} D_2' J_2^{-1} M_2 J_2 = {\bf 1} ~.
\end{equation}
The constraints determine the parameters of the jump matrices uniquely  
in terms of the elements of the $D$-matrices. The resulting expressions are 
\begin{align}
c_1 &= \frac{d_1d_1' ({m}_1 {m}_2 -{m}_\alpha)({m}_1-{m}_2 {m}_\alpha)}{{m}_1 {m}_2 \left(1-{m}_\alpha^2\right)} ~, 
\qquad \nonumber
\tilde{c}_1 = \frac{{m}_1}{d_1d_1'  \left(1- {m}_1^2\right) } \\ 
c_2 &= \frac{d_2 ({m}_1 {m}_2-{m}_\alpha) ({m}_1 {m}_\alpha-{m}_2)}{d_2' {m}_1 {m}_2 \left({m}_\alpha^2-1\right)} 
~, \qquad 
\tilde{c}_2 = \frac{d_2' {m}_2}{d_2\left(1- {m}_2^2\right)} \nonumber\\
c_{12} &= \frac{d_2 d_2' {m}_\alpha \left( {m}_1 {m}_\alpha \left(1+{m}_2^2\right) - {m}_2 \left(1+{m}_1^2\right) \right)}{{m}_1 
{m}_2 \left({m}_\alpha^2-1\right)} ~, \\ 
\tilde{c}_{12} & = \frac{{m}_1(1+{m}_2^2) - {m}_2 {m}_\alpha (1+{m}_1^2) }{d_2 d_2' {m}_1 {m}_2 \left({m}_\alpha^2-1 
\right)} ~. \nonumber
\end{align} 
Determining the jump matrices $J_3$, $J_4$, $J_{34}$ associated to the right part of the network in 
Figure \ref{C04Networks-I-V5} in a similar way, and representing 
the abelian parallel transport along the grey path connecting 
the two pairs of pants by the matrix $T=\text{diag}(e^{\pi i\eta},e^{-\pi i\eta})$,
yields a parameterisation of the non-abelian
$\mathrm{SL}(2)$-connections $\nabla$ on $C$ through the data determining $\nabla^{\rm ab}$.
The clockwise monodromy around the punctures 2 and 3 is  
\beq \label{eq:c04monodromy23}
U = T^{-1} M_3^\text{out} T M_2^\text{in}  = T^{-1} J_3^{-1} M_3 J_3 T J_2^{-1} M_2 J_2 ~ , 
\end{equation}
The  matrix $U$ has trace
\beq \label{eq:TraceM23-II}
L_u = P_+^{-1} + N_0 + P_+ N ~, \qquad 
P_+   = \frac{d_2}{d_2'd_3d_3'} e^{2\pi i \eta} ~,
\end{equation}
where the parameters $d_i$, $d_i'$, $i=1,2,3,4$, obey 
$d_1'd_2d_2'/d_1=-e^{-2\pi i \sigma}=d_3d_4d_4'/d_3'$.
The remaining coefficients in equation \eqref{eq:TraceM23-II} are 
\begin{align} \label{eq:N0coeffB}
(2 \sin(2\pi\sigma))^2 N_0 = &-    2\left[\cos 2\pi\theta_1 \cos 2\pi\theta_4 + \cos 2\pi\theta_2 \cos 2\pi\theta_3 \right] \\
&+ 2\cos 2\pi\sigma \left[ \cos 2\pi\theta_1 \cos 2\pi\theta_3 + \cos 2\pi\theta_2 \cos 2\pi\theta_4 \right] ,\nonumber
\end{align}
and 
\beq \label{eq:NcoeffB}
(2 \sin(2\pi\sigma))^4 N = \prod_{s,s'=\pm 1} 2\sin \pi(\sigma+s\theta_1+s'\theta_2) 2\sin \pi(\sigma+s\theta_3+s'\theta_4) ~.
\end{equation}
Equation \rf{eq:TraceM23-II} exhibits clearly how the gauge freedom in the description of the connection 
$\nabla^{\rm ab}$ affects the definition of the coordinate $\eta$.

In order to discuss natural ways for fixing this freedom let us note that the freedom is related
to the fact that the annular region between the two pairs of pants is not simply connected. Introducing 
extra cuts  can produce simply-connected regions within which the 
$\mathrm{GL}(1)$-connection 
can be trivialised. The holonomy of the corresponding 
$\mathrm{SL}(2)$-connection around the annular region $A$
and the parallel transport $T$ from one boundary to the other are the 
data characterising the connection $\nabla$ in $A$.
One may, however, choose to factorise this holonomy around $A$
into as a product over contributions associated to paths crossing the 
cuts decomposing $A$ into simply-connected pieces. {The resulting freedom 
can be represented by the choice of $d_i$, $d_i'$, $i=1,2,3,4$. However, 
this point of view suggests that the remaining freedom in the definition 
of the FN-type coordinate $\eta$ represented by the factor $\frac{d_2}{d_2'd_3d_3'}$ 
in \rf{eq:TraceM23-II} can be written as the exponential  of a linear function in the 
variable $\si$.
}

A few options are clearly distinguished 
by their simplicity. 
One option is to use a single extra cut to decompose $A$, an example is indicated in 
an example Figure \ref{C04Networks-I-V5} by the dashed grey line. 
The discussion in the previous paragraph would 
suggest the  choice $d_1=d_4=-e^{2\pi i \sigma}$, all other $d_i$, $d_i'$, $i=1,2,3,4$ being equal to unity. 
In this case one would simply find $P_+   =  e^{2\pi i \eta}$. Other simple 
options amount to having $d_2=d_2'$ and $d_3=1/d_3'$, also leading to $P_+   =  e^{2\pi i \eta}$.
One should notice that the FN-type coordinates 
$(\si,\eta)$ defined by these choices are related to the coordinates $(\si,\kappa)$ 
appearing in the generalised theta series
\rf{ILT-exp1} by $\kappa=2\pi\eta$. 

\subsection{Four-punctured sphere -- the other cases}

Another interesting case is associated to the network depicted in Figure \ref{C04Networks-II-II}
below. 
\begin{figure}[h]
\centering
\includegraphics[width=1.0\textwidth]{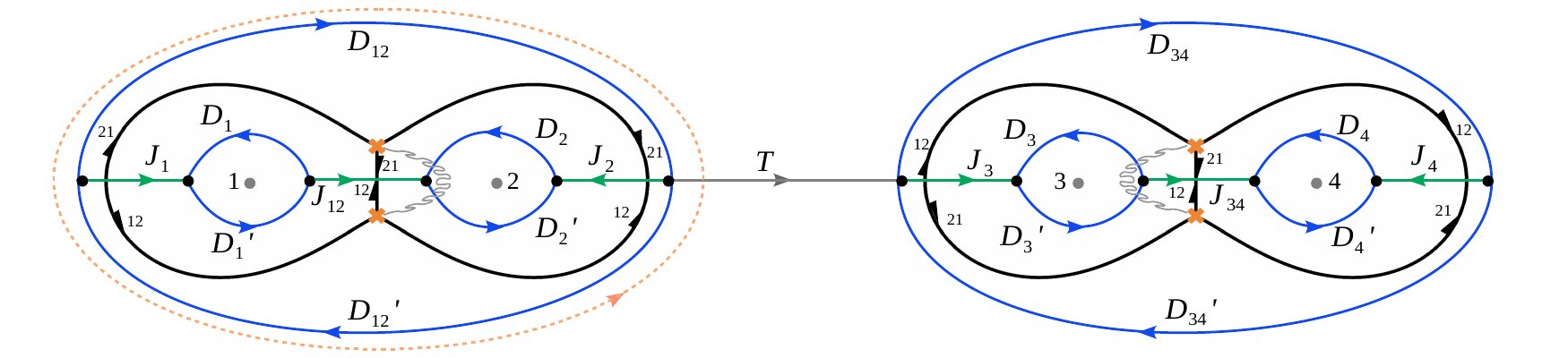}
\caption{\it FN-network on $C_{0,4}$, formed by type-I molecules in the British resolution.}
\label{C04Networks-II-II}
\end{figure}

The $D$-matrices which describe parallel transport along the blue  paths $\wp$ in Figure \ref{C04Networks-II-II} obey 
relations $D_1D_1'=M_1$, $D_2'D_2=M_2$, $D_{12}D_{12}'=S^{-1}$ and similarly for the molecule
on the right side. In this case there are a few natural options for fixing the residual freedom left by these constraints.
A first option is to choose one of the two matrices $D_i$, $D_i'$ to be equal to the identity for $i=1,2,12,3,4$.
However, in there would still be  a discrete family of choices left. Most natural appears to be 
the choice $D_i=D_i'=\sqrt{M}_i$, for $i=1,2,12,3,4$. This is also the case which is most symmetric.

With this choice one may proceed with the construction of the connections $\nabla^{\rm ab}$ and 
$\nabla$ as explained in the previous section. 
 The monodromy around the orange curve is $S=\text{diag}(e^{2\pi i \sigma},e^{-2\pi i \sigma})$ and 
 the clockwise monodromy around the punctures 2 and 3 is  
\beq \label{eq:c04monodromy23}
U = T^{-1} M_3^{{\rm\sst out}} \,T\, M_2^{{\rm\sst in}}  = T^{-1} J_3^{-1} M_3 J_3 T J_2^{-1} M_2 J_2 ~ .
\end{equation}
The matrix $U$ has trace of the form $L_u = \tilde{P}_+^{-1} + N_0 + \tilde{P}_+ N $ with
\begin{align} \label{eq:TraceM23}
& \tilde{P}_+   = - 4 \, \sin\pi(\sigma-\theta_1-\theta_2) \, \sin\pi(\sigma-\theta_3-\theta_4) \, e^{2\pi i \eta} ~,
\notag\end{align}
the coefficients $N_0$ and $N$ being the same as in \rf{eq:N0coeffB} and \rf{eq:NcoeffB}. 
Switching the resolution of the network in Figure \ref{C04Networks-II-II}  is equivalent to replacing $P_+$ by $P_+^{-1}$.



It is then straightforward to treat the remaining cases in a similar way.
In Subsection \ref{sec:spectralNet} we had 
observed a correspondence between flop transitions and changes of topological type of the 
FN-networks. It turns out that a set of simple  rules describes the effect of flop transitions 
on the coordinates defined by 
abelianisation.
In this way 
one finds a simple set of rules for the changes of FN-coordinates induced by flop transitions, 
the result of which is 
summarised in Table \ref{tab:FN-rules}.


%
%

\newcommand{\MIot}{\mathord{\includegraphics[height=3.5ex]{Mo1-12}}}
\newcommand{\MIott}{\mathord{\includegraphics[height=6ex]{Mo1-13}}}
\newcommand{\MItt}{\mathord{\includegraphics[height=6ex]{Mo1-23}}}
\newcommand{\MII}{\mathord{\includegraphics[height=3.5ex]{images/Mo2}}}

\newcommand{\MIotII}{\mathord{\includegraphics[height=6ex]{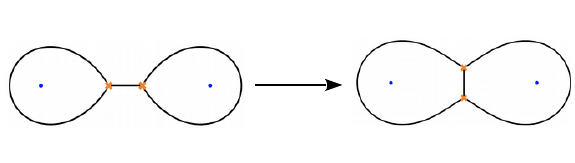}}}
\newcommand{\MIIIott}{\mathord{\includegraphics[height=6ex]{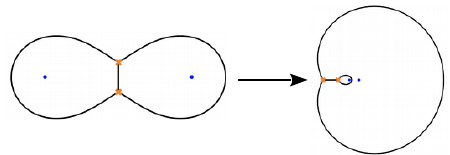}}}
\newcommand{\MIIItt}{\mathord{\includegraphics[height=6ex]{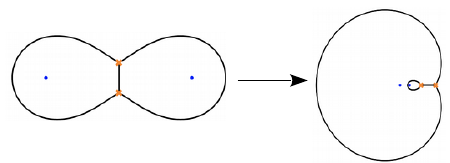}}}

\begin{table}[h!]
\begin{center}
\begin{tabular}{ c || c | c | c  }
Enclosed \\[-4ex] punctures  & $\MIotII$ & $\MIIIott$ & $\MIIItt$  \\ \hline\hline
 (12) & 
       $ 2i\sin\pi(\sigma-\theta_1-\theta_2)
       $  
        & 
      $ 2i\sin\pi(\sigma+\theta_1-\theta_2)
      $  
        &
      $ 2i\sin\pi(\sigma-\theta_1+\theta_2)
      $
        \\ \hline
\end{tabular}
\caption{\it The table collects the changes of FN-type coordinates associated to changes
of the part of the network in the pair of pants containing punctures $1$ and $2$.
The factors given in the lower row table are the ratios between
$e^{2\pi\mathrm{i}\eta_{\mathfrak{j}_2}}$  and $e^{2\pi\mathrm{i}\eta_{\mathfrak{j}_1}}$ 
if $\mathfrak{C}_{\mathfrak{j}_1}$ (resp. $\mathfrak{C}_{\mathfrak{j}_2}$) is the chamber corresponding to 
the network on the left (resp. right) in the upper row. The rules describing changes of the network in the other pair of pants
are obtained by obvious replacements.}
\label{tab:FN-rules} 
\end{center}
\end{table}

\subsection{Comparison}

In this section we have associated FN-type coordinates to 
each chamber $\mathfrak{C}_{\mathfrak{i},\mathfrak{j}}$ with $\mathfrak{i},\mathfrak{j}=\1,\2,\3$.
According to the discussion in Section \ref{theta-series} each of these coordinates defines a
generalised theta series expansion. 
Let us recall that each chamber 
$\mathfrak{C}_{\mathfrak{i},\mathfrak{j}}$ there corresponds to a toric 
diagram from which we can calculate the topological string partitions using the topological vertex. 
It is now straightforward to check that the results of the topological vertex computations agree 
precisely with the coefficient functions in the generalised theta series expansions, chamber by chamber.

{

In the discussions above we had observed some residual freedom in the definition of the FN-type coordinates
left by abelianisation. In the case of Molecule II  discussed in Section \rf{sec:C04FN} this freedom
is
represented by the factor $\frac{d_2}{d_2'd_3d_3'}$ 
in \rf{eq:TraceM23-II}. One should note, however, that we had observed in 
Section \ref{theta-series} that only a rather small subset of the possible choices 
of coordinates $(\si,\eta)$ can appear in series expansions of tau-functions of generalised 
theta series type. The discussions in Section \ref{alt-theta} or \cite[Section 3.3]{CLT}
show that $\frac{d_2}{d_2'd_3d_3'}$ has to be periodic in $\sigma$, severely  restricting 
the left-over freedom. The choice adopted above is distinguished by the property 
that the functions $N_i$, $i=1,2,3,s$ representing the normalisation of the tau-functions 
are {\it real} for real values of the arguments. We believe that the freedom which remains after 
having imposed all these requirements is inessential. This will be further supported by the 
relation with the exact WKB method discussed in \cite{CLT}.
}


\section{Summary and outlook}
\setcounter{equation}{0}

\subsection{The result}\label{sec:summ}

To conclude, let us formulate the resulting picture in a way that suggests various 
generalisations.
Our results amount to a  reconstruction of  the topological string partition functions
from the quantum curve  for certain degenerating families $C_z$ of  base
curves parameterised by a complex number $z$ which controls the degeneration 
occurring for $z\ra 0$.
{This region of the 
moduli space of the complex structures of $\Sigma_{u,z}$ is related by mirror symmetry to 
a region in the extended K\"ahler moduli space which can be represented 
as the  scaling limit of the K\"ahler moduli space of the toric CY discussed in Section \ref{sec:curves}. 
It admits a chamber decomposition 
with chambers separated by walls associated to flop transitions. The sequel \cite{CLT} will
propose a natural extension of this picture to the whole moduli space of quantum curves.}

The quadratic differential $q_\la(x)$ appearing in the equation 
defining the quantum 
curve, 
\begin{equation}\label{q-curveagain}
(\la^2\pa_x^2+q_\la(x))\chi_\pm(x)=0,
\end{equation}
is defined for given monodromy data $\mu=\mu(\si,\tau)$ through the Riemann-Hilbert correspondence.
From the solutions $\chi_\pm$ one may uniquely construct
the tau-functions $\CT(\si,\tau;z)$ as the Fredholm determinants
of an integral operator canonically associated to $\chi_\pm$. 
The tau-functions  $\CT(\si,\tau;z)$
admit an expansion in powers of $z$ around the degeneration point $z=0$.
By multiplying the tau-functions with suitable monodromy-dependent normalisation factors
$N(\si)$ one may
define free fermion partition functions admitting generalised theta series expansions,  
\begin{equation}\label{thetaexp-concl}
\CZ(\si,\tau;z):=N(\si)\CT(\si,\tau;z)=\sum_{n\in\BZ}e^{in\tau}\CG_N(\si+n;z)\,.
\end{equation}
{Abelianisation}
gives a natural way to fix the normalisation factors $N(\si)$ depending
on the choice of a  chamber 
in the extended K\"ahler moduli space of the local CY.
The coefficient functions $\CG_N(\si;z)$ appearing in the expansions \rf{thetaexp-concl}
have been found to be equal to the topological string partition functions, chamber 
by chamber.

It may look surprising that there is an essentially unambiguous way to construct the 
partition functions from the quantum curve. The key ingredients fixing the ambiguities are
(i) integrability controls possible quantum corrections to the quantum curve 
(see Section \ref{sec:q-curves}),
{(ii) abelianisation provides a canonical way to associate  FN-type coordinates
to the parts of the moduli space of quantum curves characterised by real K\"ahler moduli,
chamber by chamber, and
(iii) 
a one-to-one correspondence between choices of FN-type coordinates
and normalisations for the free fermion partition function admitting 
generalised theta series expansions. In the sequel \cite{CLT} the role of abelianisation
will be taken over by the exact WKB expansion for the quantum curve,
which is natural in view of the close relation between exact WKB and abelianisation 
discussed in \cite{HK}. }


The generalisation to the case of $C=C_{0,n}$ is straightforward. 
The variables $(\si,\tau)$ get replaced by tuples $(\underline{\si},\underline{\tau})$ where
$\underline{\si}=(\si_1,\dots,\si_{n-3})$,
and $\underline{\tau}=(\tau_1,\dots,\tau_{n-3})$, and $z$ gets similarly replaced by 
$\mathbf{z}=(z_1,\dots,z_{n-3})$. Cases like higher genus surfaces $C=C_{g,n}$ 
or surfaces with irregular singularities\footnote{See \cite{BGT16,BLMST,BGT17} for similar
results obtained by different approaches in some cases with irregular singularities.} are certainly within reach.
The generalisation to covers of higher degree should be very interesting.

\subsection{Role of integrable structures}

A source of motivation for our proposal has been the relation between the free fermion partition 
function at $\la=0$ \rf{thetasigma},
and the Hitchin integrable system, established by the identification of the variables 
$(\mathbf{a},\underline{\vartheta})$ as action-angle variables of the Hitchin integrable system.
It is shown in \cite{CLT} that \rf{thetasigma}  is recovered in the limit $\la\ra 0$. 
 
It furthermore seems intriguing to observe that the dependence on both $(\underline{\si},\underline{\tau})$ and $\mathbf{z}$
appears to be controlled by the integrable structures of the problem, as can be 
expressed by the  equations
\begin{subequations}
\begin{align}
&{\pa_{z_r}\CT_N(\underline{\si},\underline{\tau};\mathbf{z})}
=H_r \,{\CT_N(\underline{\si},\underline{\tau};\mathbf{z})}, 
\label{isotau-def}\\
&{e^{\pa_{\si_k}}\CT_N(\underline{\si},\underline{\tau};\mathbf{z})}=
e^{-\mathrm{i}\tau_k}\,{\CT_N(\underline{\si},\underline{\tau};\mathbf{z})}.
\label{sigmashift}
\end{align}
\end{subequations}
The factors $H_r$ appearing on the right hand side of \rf{isotau-def} are defined through the 
Riemann-Hilbert correspondence as functions $H_r=H_r(\underline{\si},\underline{\tau};\mathbf{z})$. 
The definition of the coordinate $\tau_k$ appearing on the right of \rf{sigmashift}, on the other hand, is 
unambiguously fixed by using the solutions $\chi_\pm(x)$ obtained by Borel summation in the definition
of coordinates described in Sections \ref{coordinates1}.

While \rf{isotau-def} is the definition of the isomonodromic tau-function through a solution to the 
Schlesinger equations, the difference equations \rf{sigmashift} are associated to 
the integrable structure of $\CM_{\rm flat}(C_{0,n})$ manifested in the Fenchel-Nielsen type coordinates,
allowing one to regard the coordinates $\underline{\si}$ as action-variables, and
$\underline{\tau}$ as angle coordinates, together forming a system of Darboux coordintates
for  the natural symplectic structure on $\CM_{\rm flat}(C_{0,n})$. Equations \rf{sigmashift}
indicate that the integrable structure of $\CM_{\rm flat}(C_{0,n})$ expressed through the Darboux 
coordinates $(\underline{\si},\underline{\tau})$ can be regarded as a deformation of the integrable
structure of the Hitchin system made manifest through the definition of the action-angle coordinates
$(\mathbf{a},\underline{\vartheta})$.

It is clear that equation 
\rf{sigmashift} severely restricts the dependence of $\CT_N(\underline{\si},\underline{\tau};\mathbf{z})$
on $(\underline{\si},\underline{\tau})$, and therefore the choice of the normalisation factors 
left undetermined by the definition \rf{isotau-def} of the isomonodromic tau-function.

\subsection{Perspectives}

Having given a precise {\it analytic} characterisation of the topological string partition 
function may also shed light on what remains to be done to make other approaches 
fully effective. {We will here mention a few possible directions, referring to \cite{CLT} 
for  a discussion of some further directions.}

\subsubsection{Topological recursion}

Topological recursion provides a systematic approach to the expansion of the topological 
string partition functions in powers of $\la$, see \cite{Ey} for a review and further references.
However, it would be good to know
which initial conditions characterise the topological string partition
functions for local CY of class $\Sigma$, and
to what extend one can reconstruct the non-perturbative answer from 
the formal series in $\la$ defined by the topological recursion.

The topological recursion has recently been used in \cite{IKT} to compute the partition functions 
associated to three-punctured spheres $C_{0,3}$. The result agrees with 
the expansions for $\la\ra 0$ of the functions $N_{i}(a_3/\la,a_2/\la,a_1/\la)$, $i=1,2,3,s$, 
defined in Section \ref{TopVert}, as can be checked using the asymptotic expansion of 
the $G$-function for large arguments. The expansions do not depend on the chamber, 
whereas the dependence of  the exact result on the choice of chamber is exactly the one 
described in this paper. This is explained in more detail in \cite[Section 7.3]{CPT19}.

\subsubsection{Matrix models}

Matrix models \cite{DV02,DV09} can potentially give answers for 
the values of the topological string partition functions which are non-perturbative in $\la$ but
restricted to a lattice in the set of allowed K\"ahler parameters defined by the integrality of the numbers of 
integrations.
The precise answer will depend
on the choice of integration contours, in general. Interesting questions are (i) which choice of 
integration contours reproduces the non-perturbative partition functions defined in our paper and
(ii) if there is a canonical way to reconstruct
the full partition functions from the functions on 
the lattices in the set of  K\"ahler parameters defined by the matrix models.
Partial results concerning the first question (i) have been obtained in \cite{CDV}.

\subsubsection{Topological vertex and beyond}

It is not known if the series defined by the topological vertex
formalism are convergent, in general, see however \cite{FM} for recent results 
allowing to prove the convergence in some cases. 
For the theories of class $\Sigma$ one may deduce analyticity of the 
topological string partition functions 
with the help of the   Fredholm determinant representations discussed
in this paper. 

It is worth noting, however, that the class of theories for which the approach 
taken in this paper suggests  an answer includes many cases for which it is not
known how to represent the local CY as limits of toric CY. This will 
be the case for coverings of surfaces $C$ of higher genus and
the so-called Sicilian quivers. It should be possible to generalise 
our approach
to arrive at precise predictions for this class of local CY for which 
not much seems to be known at present.

\subsubsection{$\mathcal{N} = 1$ theories}

Going beyond the various applications of topological string theory 
to the study of $\CN=2$ supersymmetric field theories studied in the literature,
there should also be interesting applications to field theories having only $\mathcal{N} = 1$ 
supersymmetry in four dimensions. 
Intriligator and Seiberg have made a first step in this direction by generalizing the Seiberg-Witten 
theory  \cite{Intriligator:1994sm}. Using their work we can characterize the low-energy 
physics of field theories with an abelian Coulomb branch 
by  spectral curves in a way which is somewhat analogous to the cases with $\CN=2$ 
supersymmetry.
It would be very interesting  if the technology developed in this paper 
could be generalized to predict partition functions for $\mathcal{N} = 1$ theories for which only very few tools exist,
see \cite{Coman:2015bqq,Mitev:2017jqj,Bourton:2017pee} for some previous work in this direction.

\bigskip


{\bf Acknowledgements:} 
Important steps of this work were taken during 
J.T.'s visit at MSRI in the Spring of 2018, where the results have first been presented. 
J.T. would like to 
thank M. Aganagic, A. Okounkov and the MSRI for the hospitality, and 
M. Aganagic, T. Bridgeland, A. Klemm and A. Okounkov for interesting discussions
related to this work.

The work of EP is supported by the German Research Foundation (DFG) via the 
Emmy Noether programm ``Exact results in Gauge theories''.

The work of IC is supported by ERC starting grant H2020 ERC StG No.640159.

\newpage

\appendix

\section{Grassmannians and Sato-Segal-Wilson tau-function}\label{SW-app}
\setcounter{equation}{0}

The construction of free fermion partition function proposed in \cite{DHS} was based on the 
theory of infinite Grassmannians pioneered in \cite{Sa,SW}. We will here compare 
our formulation to the one used in \cite{DHS}. 

\subsection{Grassmannians and tau-functions}\label{SW-tau}

\subsubsection{Infinite Grassmannians}
Let $\CH=L^2(S^1,\mathbb{C}^N)$, where  $S^1$ will often be identified with 
the equator of $\mathbb{P}^1$.  Elements of $\CH$ will be represented as vectors having 
functions on $S^1$ in each of their $N$ components.
We have $\CH=\CH_+\oplus\CH_-$, 
where the functions in $\CH_+$ ($\CH_-$) can be continued analytically inside $S^1$   
(outside of $S^1$ and vanish at infinity). The Segal-Wilson Grassmannian $\mathrm{Gr}(\CH)$ is the set of 
all closed subspaces $W$ of $\CH$ which are the images of embeddings $\sw_+\oplus\sw_-:\CH_+\ra
\CH_+\oplus\CH_-$ 
such that $\sw_+$ is a 
Fredholm operator, and $\sw_-$ is a 
compact operator. 
A subspace $W$ in $\mathrm{Gr}(\CH)$ is spanned by the columns of the 
rectangular matrix $\sw=(\begin{smallmatrix}\sw_+\\  \sw_-\end{smallmatrix})$ called a frame of $W$. 
Frames related by right multiplication with
elements of the group $\CC$ of invertible operators $\sh:\CH_+\ra \CH_+$ differing from the identity $\mathbf{1}$ by 
a trace class operator describe
the same points in $\mathrm{Gr}(\CH)$.

A frame is called admissible if
$\sw_+-\mathbf{1}$ is a trace-class operator on $\CH_+$. We restrict attention to subspaces $W$
having admissible frames.
A frame for a space $W$ can be transformed 
into an admissible frame if $\sw_+$ is invertible,
\begin{equation}
\left(\begin{matrix} \sw_+ \\
\sw_-
\end{matrix}\right)=
\left(\begin{matrix} \mathbf{1} \\
\sw_-^{}\sw_+^{-1}
\end{matrix}\right)\cdot\sw_+
=:
\left(\begin{matrix} \mathbf{1} \\
\mathsf{A}
\end{matrix}\right)\cdot\sw_+.
\end{equation}
This means that the space $W$ can be described as the graph of 
the operator $\mathsf{A}:\CH_+\ra\CH_-$,
$\mathsf{A}:=\sw_-^{}\sw_+^{-1}$. Such a space $W$ is called transverse to $\CH_-$.

A natural line bundle on 
the Segal-Wilson Grassmannian $\mathrm{Gr}(\CH)$ is the dual of  the determinant 
bundle $\mathrm{Det}^*$, which can be represented by pairs $(\sw,\la)$ with 
$(\sw,\la)$ and $(\sw',\la')$ considered to be equivalent iff
$\sw'=\sw\mst$ and $\la'=\la\mathrm{det}(\mst)$ for $\mst\in\CC$. 
$\mathrm{Det}^*$ has a canonical section $\si$ represented by the pairs $(\sw,\mathrm{det}(\sw_+))$.

\subsubsection{Definition of Sato-Segal-Wilson tau-functions}

Let $\Ga$ be the group of continuous maps $S^1\ra\mathrm{GL}(N)$, regarded as multiplication 
operators on $\CH$. The subgroups $\Ga_+$ and $\Ga_-$ are represented by real analytic functions $f$
which extend holomorphically inside the unit circle satisfying $f(0)=1$ and 
outside of the unit circle with $f(\infty)=1$, respectively. 
Noting that the multiplication by 
$g^{-1}\in\Ga_+$ is represented on $\CH=\CH_+\oplus\CH_-$ by a matrix of the form 
$
\left(\begin{smallmatrix} \mathsf{a} & \mathsf{b}\\
0&\mathsf{d}
\end{smallmatrix}\right)
$
one may define the tau-function $\tau_W^{}(g)$ for $W$  transverse
to $\CH_-$ as a function on $\Ga_+$ by setting
\begin{equation}
\tau_W^{}(g)=\frac{\si(g^{-1}W)}{g^{-1}\si(W)},
\end{equation}
where $g^{-1}\si$ is the natural action of $g^{-1}$ on sections of $\mathrm{Det}^*$, 
see \cite{SW} for details. It is not hard to see that the function $\tau_W$ can be represented 
as the Fredholm determinant
\begin{equation}\label{SSW-tau}
\tau_W^{}(g)=\mathrm{det}_{\CH_+}^{}(\mathsf{1}+\mathsf{B}_g\mathsf{A}),\qquad \mathsf{B}_g=\mathsf{a}^{-1}\mathsf{b}.
\end{equation}

This construction can be applied in particular to the case $W=\Psi^{-1}\cdot \CH_+$, with $\Psi$ being the 
fundamental solution matrix of holonomic 
$\CD$-modules which can be analytically continued outside of $S^1$ and is twice differentiable on $S^1$.
In this way we get a natural way to associate points in $\mathrm{Gr}(\CH)$  to $\CD$-modules.

\subsection{Free fermion states associated to points in the infinite Grassmannian}

We are now going to show that the tau-function defined in \rf{SSW-tau} can be represented as a
matrix element in the fermionic Fock space. This connection was part of the motivation 
for the proposal made in \cite{DHS} that the free fermion partition functions relevant for topological string
theory can be defined using the above-mentioned connection between $\CD$-modules and
points  in $\mathrm{Gr}(\CH)$. The approach described in the main text realises these ideas for the case of our interest. 

The spaces $W\in\mathrm{Gr}(\CH)$ we are interested in  are spanned by the columns
of $\sw=(\begin{smallmatrix}\sw_+\\  \sw_-\end{smallmatrix})$ with $\sw_+$ invertible.
Such spaces are specified by the
graphs of the operators $\SA=\sw_-^{}\sw_+^{-1}$.
Let  the operator $\SA:\CH_+\ra\CH_-$ be represented by  matrices $A_{kl}$ with respect to the bases $\CB_+$ and $\CB_-$
for the Hilbert spaces $\CH_+$ and $\CH_-$, respectively,
where $\CB_+=\{{e}_sz^{l},s=1,\dots,N;l=0,1,\dots\}$,
and $\CB_-=\{{e}_sz^{-k},s=1,\dots,N;k=1,2\dots\}$, with $\{e_s,s=1,\dots,N\}$ being  the 
canonical basis of $\BC^N$. 
We may then define a vector $\mathfrak{f}_A^{}\in\CF$ as
\begin{align}\label{fA-def}
&\mathfrak{f}_A^{}:=\mathsf{U}_A^{}\cdot\mathfrak{f}_{0}^{},\qquad \mathsf{U}_A=
\exp\bigg(-\sum_{k>0}\sum_{l\geq 0} \psi_{-k}\cdot A_{kl}\cdot \bar{\psi}_{-l}\bigg) .
\end{align}

A function  $\Psi(y):S^1\ra \mathrm{GL}(N,\BC)$  
analytic outside of $S^1$ and twice-differentiable on $S^1$
defines a multiplication operator on $\CH$ 
allowing us to define the operator 
\begin{equation}
\mathsf{A}_\Psi=\Pi_-\,\Psi^{-1}\,\Pi_+\,\Psi\,\Pi_+,
\end{equation}
where $\Pi_\pm:\CH\ra\CH_\pm$ are the canonical projections.
The operator $\Pi_-\,\Psi^{-1}\,\Pi_+$ is trace-class \cite[Propositon 2.3]{SW}, from which 
it follows that $\mathsf{A}_\Psi$ is trace-class as well. 
One may represent $\SA_\Psi$ as an integral 
operator 
\begin{equation}\label{Ain-intop}
(\mathsf{A}_\Psi f)(x)=\frac{\mathsf{i}}{2\pi } \int_{\CC}{dy}\,\frac{(\Psi(x))^{-1}\Psi(y)}{x-y}f(y).
\end{equation}
From \rf{Ain-intop} it is clear that the matrices representing  $\mathsf{A}_\Psi$ with respect to this basis
are defined using   \rf{GfromPsi} and \rf{A-exp}. We recover the construction used in Section \ref{CDtoFF}
to define free fermion states $\mathfrak{f}_\Psi\in\CF$ from solutions to the Riemann-Hilbert problem.

An operator $\SB:\CH_-\ra \CH_+$ represented by matrices $B_{lk}$ can in a similar way be used to define
\begin{align}\label{fB-def}
&
\mathfrak{f}_B^{\ast}:= \mathfrak{f}_{0}^{\ast}\cdot\mathsf{V}_B^{},\qquad
\mathsf{V}_B^{}=\exp\bigg(-\sum_{l\geq 0}\sum_{k>0} \bar{\psi}_{l}\cdot B_{lk}\cdot {\psi}_{k}\bigg).
\end{align}
One may again associate
such operators in particular to functions $\Psi(y)$ analytic inside of $S^1$
and twice differentiable on $S^1$. Representing the multiplication operator $(\Psi^{\rm out})^{-1}$ with 
respect to $\CH\simeq \CH_+\oplus\CH_-$  in the form 
$\left(\begin{smallmatrix} \mathsf{a} & \mathsf{b}\\
0&\mathsf{d}\end{smallmatrix}\right)$, allows us to define a trace-class operator
 $\SB_{\Psi}:\CH_-\ra \CH_+$, $\SB_{\Psi}=\mathsf{a}^{-1}\mathsf{b}$
 and a state $\mathfrak{f}_\Psi^*\in\CF^*$ in a way which is analogous to the 
 definition of $\SA_{\Psi}$ and $\mathfrak{f}_\Psi$
 given above.
 
The Fredholm determinants representing the Sato-Segal-Wilson tau-functions 
$\tau_W^{}(g)$ via \rf{SSW-tau}
can now be represented as matrix elements in the free fermion Fock-space,
 \begin{equation}\label{SWtau-Fock}
\mathrm{det}_{\CH_+}^{}(\mathsf{1}+\mathsf{B}_g\mathsf{A})=\langle\,\mathfrak{f}_{B_g}^\ast\,,\,\mathfrak{f}_A^{}\,\rangle,
\end{equation}
the next subsection contains  a self-contained proof of the identity \rf{SWtau-Fock}.
 
\subsection{Determinant representation of fermionic matrix elements}\label{App:Fredholm}

Our goal in this subsection is to prove the identity
\begin{equation}\label{matel-det}
\mathrm{det}_{\CH_+}^{}(\mathsf{1}+\mathsf{B}\mathsf{A})=\langle\,\mathfrak{f}_B^\ast\,,\,\mathfrak{f}_A^{}\,\rangle,
\end{equation}
where  $\mathsf{A}:\CH_+\ra\CH_-$ and $\mathsf{B}:\CH_-\ra\CH_+$ are trace-class operators, and 
$\mathfrak{f}_A^{}\in\CF$ and $\mathfrak{f}_B^{\ast}\in\CF^*$ are the states defined from  $\SA$ and $\SB$
in \rf{fA-def} and \rf{fB-def}, respectively. Identities like \rf{matel-det} are probably known, but we did not find a 
convenient reference for the proof.

It will be useful to 
represent the elements of $\CH=L^2(S^1,\mathbb{C}^N)$ using an isomorphism 
$\CH\simeq L^2(S^1,\mathbb{C})$ called blending. Introducing 
the canonical  basis $(e_1,\dots,e_N)$ of $\BC^N$, one may map 
\begin{align}
&L^2(S^1,\mathbb{C}^N)\ni f(x)=\sum_{n\in\BZ}\sum_{k=1}^N f_{n,k}x^{-n}e_k, \quad\mapsto
\quad g(x)=\sum_{m\in\BZ}g_m x^{-m}\in L^2(S^1,\mathbb{C}),\notag\\ 
& \text{where}\quad g_{nN+k}:=f_{n,k}, \quad k=1,\dots,N,\quad
n\in\BZ.
\label{blending1}\end{align}
Let us next notice that 
\begin{equation}
\mathrm{det}_{\CH_+}^{}(\mathsf{1}+\mathsf{B}\mathsf{A})=
\mathrm{det}_{\CH}^{}(1+\mathsf{D}), \qquad \mathsf{D}=
\bigg(\begin{matrix} 0 & -\mathsf{B} \\ \mathsf{A} & \;\,0\end{matrix}
\bigg).
\end{equation}
Using the blending isomorphism we may represent $\mathsf{D}$ as a $\BZ\times\BZ$-matrix $D$.
The determinant $\mathrm{det}_{\CH}^{}(1+\mathsf{D})$ can then be expanded as
\begin{equation}\label{detexp}
\mathrm{det}_{\CH}^{}(1+\mathsf{D})=\sum_{{\mathbb{S}\subset \BZ\,,\, |\mathbb{S}|<\infty}}
\mathrm{det}({D}_\mathbb{S}),
\end{equation}
where ${D}_\mathbb{S}$ is the matrix obtained from ${D}$ by deleting all 
rows and columns in $\mathbb{Z}\setminus \mathbb{S}$. 

We may further decompose 
$\mathbb{S}\subset\mathbb{Z}$ into two sets $\mathbb{P}=\mathbb{S}\cap\mathbb{Z}^{\geq 0}$
and $\mathbb{H}=\mathbb{S}\cap\mathbb{Z}^{< 0}$. 
The block structure of $\mathsf{D}$ implies that $\mathbb{P}$ and $\mathbb{H}$ have the same cardinality.
Using the blending isomorphism we may represent $\mathsf{A}$ as a 
$\BZ^{\geq 0}\times\BZ^{<0}$-matrix $A^b$, and $\mathsf{B}$ as a 
$\BZ^{< 0}\times\BZ^{\geq 0}$-matrix $B^b$.
Due to the block structure of $\mathsf{D}$ one may factorise $\mathrm{det}({D}_\mathbb{S})$
as
\begin{equation}\label{detfactor}
\mathrm{det}({D}_\mathbb{S})=
\mathrm{det}({B}_{\mathbb{H}\mathbb{P}})
\mathrm{det}({A}_{\mathbb{P}\mathbb{H}}),
\end{equation}
where ${A}_{\mathbb{P}\mathbb{H}}$ is obtained from $A^b$ by deleting all rows with indices 
not contained in $\mathbb{P}$ and all columns having indices not in $\mathbb{H}$, with
${B}_{\mathbb{H}\mathbb{P}}$ defined in an analogous way.

The formula following by inserting \rf{detfactor} into \rf{detexp} can be directly compared to the 
representation of  $\langle\,\mathfrak{f}_B^{*}\,,\,\mathfrak{f}_A^{}\,\rangle$ 
in terms of the expansion 
\begin{equation}\label{factorexp0}
\langle\,\mathfrak{f}_B^{*}\,,\,\mathfrak{f}_A^{}\,\rangle=
\sum_{\imath\in\CI}\langle\,\mathfrak{f}_B^{*}\,,\,\mathfrak{f}_{\imath}^{}\,\rangle
\langle\,\mathfrak{f}_{\imath}^*\,,\,\mathfrak{f}_A^{}\,\rangle,
\end{equation}
with $\{\mathfrak{f}_{\imath}^{};\imath\in\CI\}$ and 
$\{\mathfrak{f}_{\imath}^{*};\imath\in\CI\}$ being bases for $\CF$ and $\CF^*$, respectively,
such that $\langle\,\mathfrak{f}_\jmath^{*}\,,\,\mathfrak{f}_{\imath}^{}\,\rangle=\de_{\imath,\jmath}$.

The blending isomorphism relates N-component vectors $\psi(z)$, $\bar{\psi}(z)$ on a punctured 
disc to fermions $\phi(z)$,  $\bar{\phi}(z)$ on an N-fold cover of the punctured disc with 
modes being related as $\phi_{nN+s-1}=\psi_{s,n}$, $\bar\phi_{nN+s}=\bar{\psi}_{s,n}$. 
The Fock-space $\CF$ thereby gets an
alternative representation as Fock space of a single species of free fermions. 
A useful pair of dual bases for $\CF$ and $\CF^*$ can be generated from the vectors 
\begin{equation}
\mathfrak{f}_{\mathbb{P}\mathbb{H}}^*=
\mathfrak{f}_0^*\;\phi_{n_p}\dots\phi_{n_1}\bar{\phi}_{m_1}\dots\bar{\phi}_{m_h},
\qquad
\mathfrak{f}_{\mathbb{H}\mathbb{P}}^{}=
\phi_{-m_h}\dots\phi_{-m_1}\bar{\phi}_{-n_1}\dots\bar{\phi}_{-n_p}\,\mathfrak{f}_0^{}\,,
\end{equation}
associated to the finite sets $\mathbb{P}=\{n_1,\dots,n_p\}\subset \BZ^{\geq 0}$
and $\mathbb{H}=\{-m_1,\dots,-m_q\}\subset \BZ^{<0}$.

It remains to prove  the identities 
\begin{subequations}
\begin{align}\label{matel-det-a}
&\langle\,\mathfrak{f}_{\mathbb{P}\mathbb{H}}^*\,,\,\mathfrak{f}_A^{}\,\rangle
=(-)^{p}\,\mathrm{det}({A}_{\mathbb{P}\mathbb{H}}),
\\
&\langle\,\mathfrak{f}_B^{*}\,,\,\mathfrak{f}_{\mathbb{H}\mathbb{P}}^{}\,\rangle
=(-)^{p}\,\mathrm{det}({B}_{\mathbb{H}\mathbb{P}}).
\label{matel-det-b}\end{align}
\end{subequations}
To prove \rf{matel-det-a} one may use the identities
\begin{align}\label{WI-rep}
\bar\phi_{k}\,\mathfrak{f}_A^{}=-\sum_{l\geq 0}A_{kl}^b\bar{\phi}_{-l}\,\mathfrak{f}_A^{},
\end{align}
following directly from the definition of $\mathfrak{f}_A^{}$, 
allowing us to calculate
\begin{align*}
\langle\,\mathfrak{f}_{\mathbb{P}\mathbb{H}}^*\,,\,\mathfrak{f}_A^{}\,\rangle&=
\langle\, \mathfrak{f}_0^*\,,\,\phi_{n_p}\dots\phi_{n_1}\bar{\phi}_{m_1}\dots\bar{\phi}_{m_p}\, \mathfrak{f}_A\rangle \\[1ex]
&\!\!\!\overset{\rf{WI-rep}}{=}-(-)^{p-1}\sum_{m} A_{m_1m}^b
\langle \,\mathfrak{f}_0\,,\,\phi_{n_p}\dots\phi_{n_1}\bar{\phi}_{m_2}\dots\bar{\phi}_{m_p}
\bar{\phi}_{-m}\, \mathfrak{f}_A^{}\rangle
\\
&=-\sum_{l=1}^p A_{m_1n_l}^b(-)^{l-1}\langle \,\mathfrak{f}_0^*\,,\,\phi_{n_p}\dots\phi_{n_{l+1}}\phi_{n_{l-1}}\dots
\phi_{n_1}\bar{\phi}_{m_2}\dots\bar{\phi}_{m_p}
\,\mathfrak{f}_A^{}\rangle
\end{align*}
Using this identity recursively, and comparing the 
result with Laplace's formula for $\mathrm{det}({A}_{\mathbb{P}\mathbb{H}})$
one gets the identity \rf{matel-det-a}. The proof of \rf{matel-det-b} is completely
analogous.


\newpage

\begin{thebibliography}{99}


\bibitem[ACDKV]{ACDKV}
M. Aganagic, M.C.N. Cheng, R. Dijkgraaf, D. Krefl, C. Vafa,
{\it Quantum Geometry of Refined Topological Strings},
JHEP {\bf 1211} (2012) 019,
arXiv:1105.0630.


\bibitem[ADKMV]{ADKMV}
	M. Aganagic, R. Dijkgraaf, A. Klemm, M. Marino, C. Vafa,
{\it Topological strings and integrable hierarchies}, 
Commun. Math. Phys. {\bf 261} (2006) 451--516,
arXiv:hep-th/0312085



\bibitem[AKMV]{AKMV}
	M. Aganagic, A. Klemm, M. Marino, C. Vafa,
{\it The Topological Vertex},
Comm. Math. Phys. {\bf 254} (2005) 425--478.

\bibitem[AMV]{AMV}
Alvarez-Gaume, L., Moore, G., Vafa, C.: 
{\it Theta functions, modular invariance, and strings}.
Commun. Math. Phys. {\bf 106} (1986) 1--40.

\bibitem[BBT]{BBT}
O.Babelon, D. Bernard, M. Talon,
{\it Introduction to Classical Integrable Systems}. 
Cambridge University Press (2003)

\bibitem[BPTY]{BPTY} 
  L.~Bao, E.~Pomoni, M.~Taki and F.~Yagi,
  {\it M5-Branes, Toric Diagrams and Gauge Theory Duality},
  JHEP {\bf 1204} (2012) 105, 
arXiv:1112.5228.
  
  \bibitem[BS]{BS}
  M. A. Bershtein, A. I. Shchechkin, 
  {\it Bilinear Equations on Painlev\'e $\tau$ Functions from CFT},
Comm. Math. Phys. {\bf 339} (2015) 1021--1061.


\bibitem[BGT16]{BGT16}
G. Bonelli, A. Grassi, A. Tanzini,	
{\it Seiberg-Witten theory as a Fermi gas}, 
Lett.Math.Phys. {\bf 107} (2017) 1--30, 
 arXiv:1603.01174.

\bibitem[BGT17]{BGT17}
G. Bonelli, A. Grassi, A. Tanzini,
{\it New results in $\CN=2$ theories from non-perturbative string},
Annales Henri Poincare {\bf 19} (2018) 743--774, 
 arXiv:1704.01517.
 
 \bibitem[BLMST]{BLMST}
 G. Bonelli, O. Lisovyy, K. Maruyoshi, A. Sciarappa, A. Tanzini,
{\it On Painlev\'e/gauge theory correspondence},
Lett. Math. Phys. {\bf 107} (2017) 2359--2413,
arXiv:1612.06235.



\bibitem[BP]{Bourton:2017pee}
  T.~Bourton and E.~Pomoni,
{\it Instanton counting in Class $\mathcal{S}_k$},
  arXiv:1712.01288 [hep-th].

\bibitem[CGL]{CGL}
M. Cafasso, P. Gavrylenko, O. Lisovyy,
{\it Tau functions as Widom constants},
Commun. Math. Phys. 365 (2019) 2, 741--772, 
arXiv:1712.08546.

\bibitem[CDV]{CDV} M.C.N. Cheng, R. Dijkgraaf, C. Vafa, 
{\it Non-Perturbative Topological Strings And Conformal Blocks},
 JHEP {\bf 1109} (2011) 022,  arXiv:1010.4573. 


\bibitem[CKYZ]{CKYZ}
 T. -M. Chiang, A. Klemm, S. -T. Yau, E. Zaslow,
 {\it Local Mirror Symmetry: Calculations and Interpretations},
Adv. Theor. Math. Phys. {\bf 3} (1999) 495--565,
hep-th/9903053.



\bibitem[CLT]{CLT}
  I.~Coman, P.~Longhi, J.~Teschner,
{\it   From quantum curves to topological string partition functions II},
Preprint arXiv:2004.04585.


\bibitem[CPTY]{Coman:2015bqq}
  I.~Coman, E.~Pomoni, M.~Taki and F.~Yagi,
{\it Spectral curves of $ \mathcal{N} $ = 1 theories of class $ {\mathcal{S}}_k $},
  JHEP {\bf 1706} (2017) 136,
  [arXiv:1512.06079 [hep-th]].
  
  \bibitem[CPT19]{CPT19}
  I.~Coman, E.~Pomoni, J.~Teschner,
  {\it Trinion Conformal Blocks from Topological strings},
   J. High Energ. Phys. {\bf 2020}, 78 (2020),
  e-Print: 1906.06351 [hep-th].


\bibitem[DDP]{DDP}
D. E. Diaconescu, R. Donagi, and T. Pantev, 
{\it Intermediate Jacobians and ADE Hitchin systems}, 
Math. Res. Lett. {\bf 14} (2007) 745--756. 

\bibitem[DDDHP]{DDDHP}
D.-E. Diaconescu, R. Dijkgraaf, R. Donagi, C. Hofman, and T. Pantev, 
{\it Geometric transitions and integrable systems}, 
Nuclear Phys. {\bf B 752} (2006) 329--390.

\bibitem[DV02]{DV02}
R. Dijkgraaf, C. Vafa, 
{\it Matrix models, topological strings, and supersymmetric gauge theories}, 
Nucl. Phys. {\bf B 644} (2002) 3, hep-th/0206255.

\bibitem[DV09]{DV09}
R. Dijkgraaf, C. Vafa, 
{\it Toda Theories, Matrix Models, Topological Strings and $N = 2$
Gauge Systems}, arXiv:0909.2453. 

\bibitem[DVV]{DVV}
R. Dijkgraaf, C. Vafa, E. Verlinde,
{\it M-theory and a Topological String Duality},
hep-th/0602087.

\bibitem[DHSV]{DHSV}
R. Dijkgraaf, L. Hollands, P. Sulkowski, C. Vafa,
{\it Supersymmetric Gauge Theories, Intersecting Branes and Free Fermions},
JHEP {\bf 0802} (2008) 106,
arXiv:0709.4446. 

\bibitem[DHS]{DHS}
R. Dijkgraaf, L. Hollands, P. Sulkowski,
{\it Quantum Curves and D-Modules}, 
JHEP {\bf 0911} (2009) 047,
 arXiv:0810.4157. 



\bibitem[DM]{DM} Dubrovin, M. Mazzocco,
{\it Canonical Structure and Symmetries of the Schlesinger Equations},
Comm. Math. Phys. {\bf 271} (2007) 289--373.


\bibitem[EK]{EK} 
  T.~Eguchi and H.~Kanno,
  {\it Topological strings and Nekrasov's formulas},
  JHEP {\bf 0312} (2003) 006,
 hep-th/0310235.
  
  \bibitem[Ey]{Ey}
  B. Eynard,
  {\it A short overview of the "Topological recursion"},
  Proceedings of the ICM 2014,
  arXiv:1412.3286.
  
  \bibitem[FM]{FM}
  G. Felder, M. M\"uller-Lennert,
  {\it Analyticity of Nekrasov Partition Functions},
  Commun. Math. Phys. {\bf 364} (2018) 683--718,
  arXiv:1709.05232.
  
  \bibitem[FS]{FS} D. Friedan, S. Shenker,
  {\it The Analytic Geometry of Two-Dimensional Conformal Field Theory},
  Nucl. Phys. {\bf B281} (1987) 509--545.
  
\bibitem[Ga]{Ga} D. Gaiotto,
{\it  $N=2$ dualities},
JHEP {\bf 1208} (2012) 034,
arXiv:0904.2715.
 
\bibitem[GMN09]{GMN09}
D. Gaiotto, G. W. Moore, and A. Neitzke, 
{\it Wall-crossing, Hitchin systems, and the WKB approximation,} Adv. Math. {\bf 234} (2013) 239--403,
 arXiv:0907.3987.

  

\bibitem[GIL]{GIL}
O. Gamayun, N. Iorgov, O. Lisovyy,
{\em  Conformal field theory of Painlev\'e VI}, 
J. High Energy Phys.~\textbf{10} (2012) 038, 
arXiv:1207.0787.


\bibitem[GL16]{GL}
P. Gavrylenko, O. Lisovyy,
{\it Fredholm determinant and Nekrasov sum representations of isomonodromic tau functions},
Commun. Math. Phys. {\bf 363} (2018) 1--58,
arXiv:1608.00958v2.

\bibitem[GM]{GM} P. Gavrylenko, A. Marshakov,
{\it Free fermions, W-algebras and isomonodromic deformations},
Theor. Math. Phys. {\bf 187} (2016) 649--677,
arXiv:1605.04554.

\bibitem[GMN12]{GMN12}
D. Gaiotto, G. W. Moore, and A. Neitzke, 
{\it Spectral networks}, Annales Henri Poincare {\bf 14}
(2013) 1643--1731, arXiv:1204.4824.

\bibitem[Go]{Go}
W. Goldman, 
{\it Trace Coordinates on Fricke spaces of some simple hyperbolic surfaces}, 
Handbook of Teichm\"uller theory. Vol. II, 611-684, IRMA 
Lect. Math. Theor. Phys., {\bf 13}, Eur. Math. Soc., Z\"urich, 2009

\bibitem[GNR]{GNR}
A. Gorsky, N. Nekrasov, and V. Rubtsov, 
{\it Hilbert Schemes, Separated Variables, and D-Branes,}
Comm. Math. Phys. {\bf 222} (2001) 299--318,
hep-th/9901089.

\bibitem[GHM]{GHM}
A. Grassi, Y. Hatsuda and M. Marino, 
{\it Topological Strings from Quantum Mechanics}, Annales Henri Poincare {\bf 17} (2016) 3177--3235, 
arXiv:1410.3382.

\bibitem[GS]{GS} S. Gukov, P. Sulkowski,
A-polynomial, B-model, and Quantization,
JHEP {\bf 1202} (2012) 070,
arXiv:1108.0002.

\bibitem[Hi]{Hi}
N. J. Hitchin, {\it Stable bundles and integrable systems}, Duke Math. J. {\bf 54} (1987) 91--114.

\bibitem[HK]{HK} 
  L.~Hollands and O.~Kidwai,
  {\it Higher length-twist coordinates, generalized Heun's opers, and twisted superpotentials},
  Adv. Theor. Math. Phys. 22 (2018) 1713--1822,
  arXiv:1710.04438.
  
\bibitem[HN]{HN}
  L.~Hollands and A.~Neitzke,
  {\it Spectral Networks and Fenchel-Nielsen Coordinates},
  Lett.\ Math.\ Phys.\  {\bf 106} (2016) 811--877,
 arXiv:1312.2979.
  
  \bibitem[HIV]{HIV}
  T.J. Hollowood, A. Iqbal, C. Vafa,
{\it Matrix Models, Geometric Engineering and Elliptic Genera},
JHEP {\bf 0803} (2008) 069,
 hep-th/0310272.  

  
  
\bibitem[Hu]{Hu}  J. C. Hurtubise, 
{\it Integrable systems and algebraic surfaces,} 
Duke Math. J. {\bf 83} (1996) 19--50.

\bibitem[IK03a]{IK03a} 
  A.~Iqbal and A.~K.~Kashani-Poor,
  {\it Instanton counting and Chern-Simons theory},
  Adv.\ Theor.\ Math.\ Phys.\  {\bf 7} (2003) 457,
  hep-th/0212279.
  


\bibitem[IK03b]{IK03b} 
  A.~Iqbal and A.~K.~Kashani-Poor,
  {\it SU(N) geometries and topological string amplitudes},
  Adv.\ Theor.\ Math.\ Phys.\  {\bf 10} (2006) 1,
  hep-th/0306032.

\bibitem[IK04]{IK04} 
  A.~Iqbal and A.~K.~Kashani-Poor,
  {\it The Vertex on a strip},
  Adv.\ Theor.\ Math.\ Phys.\  {\bf 10} (2006) 317,
  hep-th/0410174.
  
  \bibitem[IKT]{IKT} 
 K. Iwaki, T. Koike, Y. Takei,
{\it   Voros Coefficients for the Hypergeometric Differential Equations and
Eynard-Orantin's Topological Recursion},
Journal of Integrable Systems {\bf 4} (2019) xyz004,
arXiv:1810.02946.
  
\bibitem[INOV]{INOV} 
A. Iqbal, N. Nekrasov, A. Okounkov, C. Vafa,
{\it Quantum foam and topological strings},
 JHEP {\bf 0804} (2008) 011,
   hep-th/0312022.
   
 
\bibitem[IKSY]{IKSY}
K. Iwasaki, H. Kimura, S. Shimomura, and M. Yoshida. 
{\it From Gauss to Painlev\'e, a Modern Theory of Special Functions}, Volume {\bf E 16}. 
Aspects of Mathematics, 1991.

\bibitem[ILP]{ILP}
A. Its, O. Lisovyy, A. Prokhorov,
{\it Monodromy dependence and connection formulae for isomonodromic tau functions},
Duke Math. Journal {\bf 167} (2018) 1347--1432,
arXiv:1604.03082. 
 
\bibitem[ILT]{ILT} N. Iorgov, O. Lisovyy, J. Teschner,
{\it Isomonodromic Tau-Functions from Liouville Conformal Blocks},
Comm. Math. Phys. {\bf 336} (2015) 671--694.
 
  \bibitem[IS]{Intriligator:1994sm}
  K.~A.~Intriligator and N.~Seiberg,
{\it Phases of N=1 supersymmetric gauge theories in four-dimensions},
  Nucl.\ Phys.\ B {\bf 431} (1994) 551,
    hep-th/9408155.


\bibitem[IN]{IN}
 K.~Iwaki and T.~Nakanishi, {\it Exact WKB analysis and cluster algebras},
 J. Phys. A {\bf 47} (2014) 474009,
 arXiv:1401.7094.
 
\bibitem[Ka]{Ka} A. Kapustin,
{\it Gauge theory, topological strings, and S duality}, 
JHEP {\bf 0409} (2004) 034,  
hep-th/0404041. 
 
\bibitem[KKV]{KKV} 
S. H. Katz, A. Klemm, C. Vafa, 
{\it Geometric engineering of quantum field theories,} Nucl. Phys. {\bf B497} (1997) 173--195, 
hep-th/9609239.

\bibitem[KMV]{KMV}
S. Katz, P. Mayr, C. Vafa, 
{\it Mirror symmetry and exact solution of 4D N = 2 gauge theories. I}, 
Adv. Theor. Math. Phys. {\bf 1} (1997) 53--114, hep-th/9706110.

\bibitem[KM]{KM}
 Y. Konishi, S. Minabe,
{\it Flop invariance of the topological vertex},
 Int. J. Math. {\bf 19} (2008) 27--45,
 arXiv:math/0601352
 
 \bibitem[Kr77a]{Kr77a}
 I. M. Krichever,
 {\it Integration of nonlinear equations by the methods of algebraic geometry}, 
 Funct. Anal. Appl., {\bf 11}:1 (1977), 12--26. 

\bibitem[Kr77b]{Kr77b}
I. M. Krichever, 
{\it Methods of algebraic geometry in the theory of non-linear equations}, 
Russian Math. Surveys {\bf 32} (6) (1977) 185--213.
 
\bibitem[Kr02]{Kr}
I. Krichever, {\it Vector bundles and Lax equations on algebraic curves}, 
Commun. Math. Phys. {\bf 229} (2002) 229--269, 
[hep-th/0108110].

 
 \bibitem[LMN]{LMN} A.S.
Losev, A.V. Marshakov, N.A. Nekrasov, 
{\it Small instantons, little strings and free fermions.} In: Shifman, M., et al. (eds.) From Fields to Strings: 
Circumnavigating Theoretical Physics, {\bf 1}, pp. 581--621. World Science Publication, Singapore (2005),
arXiv:hep-th/0302191.
 

 









  
 
  



\bibitem[MNOP]{MNOP}
D. Maulik, N. Nekrasov, A. Okounkov and R. Pandharipande,
{\it Gromov-Witten theory and Donaldson-Thomas theory, I},
Compositio Math. 142 (2006) 1263--1285.


\bibitem[MS]{MS}  	
M. Manabe, P. Sulkowski,
{\it Quantum curves and conformal field theory}, 
Phys.Rev. {\bf D95} (2017) no.12, 126003,
arXiv:1512.05785.
  
  
  
  
  \bibitem[MPTY]{Mitev:2014jza} 
  V.~Mitev, E.~Pomoni, M.~Taki and F.~Yagi,
  {\it Fiber-Base Duality and Global Symmetry Enhancement},
  JHEP {\bf 1504} (2015) 052,
  arXiv:1411.2450.
 


\bibitem[MP]{Mitev:2017jqj}
  V.~Mitev and E.~Pomoni,
{\it 2D CFT blocks for the 4D class $\mathcal{S}_k$ theories},
  JHEP {\bf 1708} (2017) 009
  doi:10.1007/JHEP08(2017)009
  arXiv:1703.00736.


\bibitem[Mo]{Mo}
G.W. Moore, 
{\it Geometry of the string equations}, 
Comm. Math. Phys. {\bf 133} (1990) 261--304.

 \bibitem[Npl]{Neitzke:plotter}
  swn-plotter.
http://www.ma.utexas.edu/users/neitzke/mathematica/swn-plotter.nb .


\bibitem[N]{N}
N.A. Nekrasov, 
{\it Seiberg--Witten prepotential from instanton counting. }
Adv. Theor. Math. Phys. {\bf 7} (2003) 831--864,
arXiv:hep-th/0206161.



\bibitem[NO]{NO}
N.A. Nekrasov, A. Okounkov, 
{\it Seiberg-Witten Theory and Random Partitions}. 
The Unity of Mathematics, pp. 525--596. Progress in Mathematics, 244, 
Birkh\"auser Boston, Boston (2006),
arXiv:hep-th/0306238.

\bibitem[NRS]{NRS} N. Nekrasov, A. Rosly, S. Shatashvili,
{\em Darboux coordinates, Yang-Yang functional, and gauge theory}, 
Nucl. Phys. Proc. Suppl.~\textbf{216} (2011) 69--93.



\bibitem[Ok]{Ok}
K. Okamoto, 
{\it Isomonodromic and Painlev\'e equations, and the Garnier system,} 
J. Fac. Sci. Univ. Tokyo, Sect. IA Math {\bf 33} (1986) 575--618.

\bibitem[OP]{OP}
A. Okounkov, R. Pandharipande,
{\it The equivariant Gromov-Witten theory of $P^1$},
Annals of Mathematics {\bf 163} (2006), 561--605,
arXiv:math/0207233.

\bibitem[O09]{O09}
A. Okounkov, {\it Noncommutative geometry of random surfaces},
arXiv:0907.2322.

\bibitem[OR]{OR}
A. Okounkov,  E. Rains,
{\it Noncommutative geometry and Painlev\'e equations},
Algebra Number Theory {\bf 9} (2015) 1363--1400,
arXiv:1404.5938.

\bibitem[Pa]{Pa}
J. Palmer, 
{\it Determinants of Cauchy-Riemann operators as $\tau$-functions}, 
Acta Appl. Math. {\bf 18} (1990) 199--223.



\bibitem[Sa]{Sa}
M. Sato, Y. Sato, 
{\it Soliton equations as dynamical systems on infinite-dimensional Grassmann manifold}, 
Nonlinear partial differential equations in applied science (Tokyo 1982), 
North-Holland Math. Stud., vol. {\bf 81}, Amsterdam 1983, pp. 259--271.

\bibitem[SMJ]{SMJ}
M. Sato, T. Miwa, M. Jimbo, 
{\it  Holonomic quantum fields. II -- The Riemann-Hilbert Problem}, 
Publ. RIMS, Kyoto Univ. {\bf 15} (1979) 201-278.

\bibitem[SW]{SW}
G. Segal, G. Wilson, 
{\it Loop groups and equations of KdV type}, Publ. Math. IHES {\bf 61} (1985) 5--65.

\bibitem[Sk]{Sk}
E. Sklyanin, 
{\it Separation of variables in the Gaudin model}, 
J. Soviet Math. {\bf 47} (1989) 2473--2488.


\bibitem[Sm]{Sm} 
I. Smith, {\it Quiver algebras as Fukaya categories},
Geometry $\&$ Topology {\bf 19} (2015) 2557--2617.




\bibitem[T17a]{T17a} J. Teschner,
{\it A guide to conformal field theory}, 
Les Houches Lect. Notes {\bf 106} (2019),
arXiv:1708.00680.

\bibitem[T17b]{T17b} J. Teschner,
{\it Quantisation conditions of the quantum Hitchin system and the real geometric Langlands correspondence}, in:
{\it Geometry and Physics:  A Festschrift in honour of Nigel Hitchin},
Eds. J.E. Andersen, A. Dancer, and O. Garc\'ia-Prada, Oxford University Press 2018,
arXiv:1707.07873.

\bibitem[UN]{UN} K. Ueno, M. Nishizawa,
{\it Multiple Gamma Functions and Multiple q-Gamma Functions},
Publ. RIMS, Kyoto Univ.
{\bf 33} (1997), 813--838.


\end{thebibliography}
\end{document}